\documentclass[trackchanges,twocolumn]{aastex7}
\shorttitle{OCCAM IX: Stellar Diffusion using APOGEE DR17 }
\shortauthors{Souto et al.}
\graphicspath{{./}{images/}}

\begin{document}

\title{The Open Cluster Chemical Abundances and Mapping Survey: IX.  Measuring the Effects of Stellar Diffusion in the Open Clusters NGC 752 and Ruprecht 147 using APOGEE }

\correspondingauthor{Diogo Souto}
\email{diogosouto@academico.ufs.br}

\author[0000-0002-7883-5425]{Diogo Souto}
\affiliation{Departamento de F\'isica, Universidade Federal de Sergipe, Av. Marcelo Deda Chagas, S/N Cep 49.107-230, S\~ao Crist\'ov\~ao, SE, Brazil}
\email{diogosouto@academico.ufs.br}

\author[0000-0003-4019-5167]{Taylor Spoo}
\affiliation{Department of Physics and Astronomy, Texas Christian University, TCU Box 298840 Fort Worth, TX 76129, USA}
\email{...}

\author[0000-0002-0740-8346]{Peter M. Frinchaboy}
\affiliation{Department of Physics and Astronomy, Texas Christian University, TCU Box 298840 Fort Worth, TX 76129, USA}
\affiliation{Maunakea Spectroscopic Explorer, Canada-France-Hawaii-Telescope, Kamuela, HI 96743, USA}
\email{p.frinchaboy@tcu.edu}

\author[0000-0001-6476-0576]{Katia Cunha}
\affiliation{Steward Observatory, University of Arizona, 933 North Cherry Avenue, Tucson, AZ 85721-0065, USA}
\affiliation{Observatório Nacional/MCTIC, R. Gen. José Cristino, 77,  20921-400, Rio de Janeiro, Brazil}
\email{kcunha@arizona.edu}

\author[0000-0002-4818-7885]{Jamie Tayar}
\affiliation{Department of Astronomy, University of Florida, 211 Bryant Space Science Center, Gainesville, FL 32611, USA}
\email{jtayar@ufl.edu}

\author[0009-0008-0081-764X]{Alessa Ibrahim Wiggins}
\affiliation{Department of Physics and Astronomy, Texas Christian University, TCU Box 298840 Fort Worth, TX 76129, USA}
\email{a.ibrahim@tcu.edu}

\author[0000-0001-9738-4829]{Natalie Myers}
\affiliation{Department of Physics and Astronomy, Texas Christian University, TCU Box 298840 Fort Worth, TX 76129, USA}
\email{n.myers@tcu.edu}

\begin{abstract}
A growing understanding of stellar processes that alter surface chemical abundances over time has opened new avenues for using these changes as probes of stellar properties. On the main sequence and near the turnoff, stellar surface abundances are affected by gravitational settling and radiative acceleration, collectively known as atomic diffusion. 
In this work, we use SDSS/APOGEE DR17/DR19 data to investigate atomic diffusion in the open clusters NGC~752 and Ruprecht~147, thereby constraining how these signatures vary with age.
From the analysis of Fe, C, N, Na, Mg, Al, Si, S, K, Ca, Ti, V, Cr, Mn, Co, and Ni, we find significant abundance differences between stars near the turnoff and the cooler main-sequence, where warmer stars are depleted relative to the cooler main-sequence stars at the $\geq1\sigma$ level for all elements available in the analysis.
These abundance differences are consistent with the signatures expected from atomic diffusion and are further supported by comparisons with stellar models that include diffusion. By fitting the observed $T_{\rm eff}$--[Fe/H] patterns with MIST isochrones, we obtain best-fit ages of 1.0~Gyr for NGC~752 and 2.5--3.2~Gyr for Ruprecht~147.
Carbon is a key diagnostic, as it shows both atomic diffusion and extra-mixing signatures associated with first-dredge-up.
For NGC~752, the coolest stars in our sample, which provide the best proxy for the initial cluster composition, yield [Fe/H]$_{\rm CS} = 0.01~\pm~0.02(\pm~0.05)$ dex. For Ruprecht~147, we obtain [Fe/H]$_{\rm CS} = 0.17~\pm~0.00(\pm~0.05)$ dex.
Our findings further constrain atomic diffusion models, suggesting that atomic diffusion affects age estimates of stars near the main-sequence turnoff.
\end{abstract}



\section{Introduction} \label{sec:intro}

Precise measurements of stellar abundances and ages are fundamental tools in Galactic archaeology, allowing us to reconstruct the formation and evolutionary history of the Milky Way from the fossil record preserved in its stars \citep{Freeman2002}. Open clusters are especially valuable for these studies, as their well-constrained ages, distances, and compositions offer critical benchmarks for calibrating stellar models and probing chemical evolution across the Galactic disk \citep{Randich2018,Magrini2023}. However, accurately interpreting observed abundance patterns requires accounting for stellar processes such as atomic diffusion and internal mixing, which can significantly modify surface compositions and introduce systematic biases into abundance-derived ages if neglected \citep{MIST_Dotter2016, MIST_Choi2016, Dotter2017}.

A widely adopted approach for determining stellar ages involves modeling the evolution of luminosity and effective temperature with time to construct theoretical isochrones, which describe stellar populations at specific ages. By matching observed stars to isochrones with the corresponding chemical compositions, precise age estimates can be obtained. Stars near the main-sequence turnoff (MSTO) and along the subgiant branch (SGB) are particularly suitable for this technique, typically yielding ages with uncertainties of $\sim$10–20\% under favorable observational conditions \citep[e.g.,][]{Soderblom2010, Casagrande2016}. Recent large-scale spectroscopic surveys, such as the Apache Point Observatory Galactic Evolution Experiment \citep[APOGEE;][]{dr17}, the Gaia-ESO Survey \citep{gaia_eso}, and the GALactic Archaeology with HERMES survey \citep[GALAH;][]{galah}, have substantially enhanced the applicability of this method. Combined with Gaia astrometry, these surveys enable robust studies of stellar ages throughout the Galaxy, particularly through targeted observations of SGB stars \citep[e.g.,][]{Hayden2022_GALAHchemclock, Maosheng2022_Nature}.

However, standard isochrone fitting typically assumes that a star’s observed surface abundances reflect its initial, birth composition. In practice, this assumption breaks down due to various internal mixing processes that alter the surface chemical composition over time. Stars are dynamic systems, and material can be exchanged between the stellar interior and atmosphere through both convective and non-convective processes. For example, in red giants, it is well established that the first dredge-up (FDU) modifies the surface abundances of carbon and nitrogen, a convective episode that transports material processed via the CNO cycle from the interior to the surface \citep{Karakas2014_DawesReview, Lagarde2012}. 
Atomic diffusion is another well-established physical process that operates more efficiently in stars near the main-sequence turnoff (MSTO) and, to a lesser degree, along the subgiant branch (SGB). First proposed over a century ago \citep{Chapman1917a, Chapman1917b}, atomic diffusion arises from the interplay of gravitational settling, which causes heavier elements to sink below the surface, and radiative levitation, which can counteract this effect by pushing certain ions upward. 
As a result, surface photospheric abundances gradually diverge from the star’s pristine composition. In particular, elements may diffuse out of the convective envelope, leading to surface depletions that persist until mixing processes restore them. During red giant evolution, the deepening of the convective envelope can dredge up material from the interior, reestablishing surface abundances that more closely reflect the stellar interior \citep{Michaud2004}. 
Atomic diffusion is expected to be most effective in the outer layers of main-sequence stars with effective temperatures above 6000 K and in horizontal branch stars with $T_{\rm eff} > 8000$ K \citep{Richer1998,Dotter2017}.

Fortunately, atomic diffusion has been extensively studied both theoretically and observationally \citep[][and references therein]{atomic_diff_book}. Clear signatures of this process have been detected in several globular clusters, including M 92 \citep{Boesgaard1998_M92, King1998_M92}, NGC 6397 \citep{Korn06_NGC6397, Korn07_NGC6297}, and NGC 6752 \citep{Gruyters2013_NGC6752, Gruyters2014_NGC6752}. More recently, evidence for atomic diffusion has also been reported in M 4 \citep{Nordlander2024}, as well as in a number of open clusters, such as M 67 \citep{souto_19, souto_2018, Liu2019_M67, BertelliMotta2018_M67, Gao2018_M67, Onehag2014_M67}, NGC 2420 \citep{Semenova2020_NGC2420}, and Coma Berenices \citep{Souto2021_ComaBer}. 
\citet{Grilo2024} investigated potential signatures of atomic diffusion in the young open cluster Pleiades (age $\sim$100 Myr), but found no statistically significant evidence for diffusion-related abundance variations.
Star clusters provide ideal laboratories for studying atomic diffusion, as they contain stars at various evolutionary stages with shared ages and compositions. By comparing abundance patterns across the main-sequence, turnoff, and red-giant phases, we can assess the signatures of atomic diffusion and evaluate its implications for stellar evolution.

In this study, we investigate the signatures of atomic diffusion across multiple elements from abundances derived using spectra from the SDSS/APOGEE survey.
This work builds on \citet{souto_2018, souto_19}, employing a similar methodology that includes the use of consistent isochrone models and detailed spectroscopic abundance determinations, as well as the APOGEE/ASPCAP (\citealt{GarciaPerez2016}) abundances. Using data from SDSS/APOGEE and applying quality and membership selections based on the OCCAM survey, we expand the analysis to two additional open clusters: NGC 752 and Ruprecht 147.
NGC 752 is a nearby ($\sim$450 pc), intermediate-age open cluster with an estimated age of 1.3 Gyr \citep{cg20} and solar metallicity ([Fe/H] $\sim$ -0.01 dex; \citealt{Reddy2012MNRAS.419.1350R,Lum2019ApJ...878...99L,Carrera2019A&A...623A..80C}). Ruprecht 147 (also known as NGC 6774) is a middle-aged open cluster ($\sim$2.5 Gyr; \citealt{cg20}) located even closer to the Sun ($\sim$250–300 pc), and has a slightly supersolar metallicity of $[{\rm Fe/H}] \sim 0.10$ dex \citep{Bragaglia2018A&A...619A.176B,Beeson2024MNRAS.529.2483B}.
The proximity and well-characterized properties of these clusters make them ideal targets for precise spectroscopic analysis, enabling us to further explore the impact of atomic diffusion.

\section{Data \& Models} \label{sec:data_analysis}

\subsection{SDSS/APOGEE Survey} \label{sdss_data}

In this work, we use data from the Apache Point Observatory Galactic Evolution Experiment surveys \citep[APOGEE-1 \& -2;]{apogee}, which are part of the SDSS-III \& -IV projects \citep{sdss3,sdss4}. Spectroscopic data were obtained in both hemispheres with 2.5m telescopes \citep{sloan_telescope, du_pont} at APO (Apache Point Observatory, New Mexico, USA) and LCO (Las Campanas Observatory, La Serena,  Chile), using the APOGEE-N and -S spectrographs \citep{apogee_inst}, respectively. The targeting selection for APOGEE is described in  \citet{zasowski13,zasowski17}, \citet{ap2n_target}, and \citet{ap2s_target}, with targeting for open clusters being further described in \citet{frinchaboy_13} and \citet{donor_18}.

\subsection{MIST Stellar Evolution Models}

To investigate the significance of any diffusion effects, we need to compare the observed data with stellar evolution models.
In this work, we have selected the MESA Isochrones \& Stellar Tracks \citep[MIST;][]{MIST_Dotter2016, MIST_Choi2016} models that utilizes the stellar evolution code Modules for experiments in Stellar Astrophysics \citep[MESA;][]{MESA_Paxton2011,MESA_Paxton2013,MESA_Paxton2015}, which also includes the effects of atomic diffusion. 
These models cover an age range of $5 \leq\ log(Age) \leq 10.3$ and metallicity range of $-4 \leq [Fe/H] \leq +0.5$ as well as report surface abundances for 19 isotopes of 17 elements from hydrogen to iron. 
Stellar evolutionary tracks were computed with and without radiative acceleration to gauge their influence. 
Each element is treated independently in the calculation of ionization balance, monochromatic opacity, and radiative acceleration. For elements without explicit atomic-diffusion evolution in the isochrones, we adopt the iron abundance as a proxy.
All isochrones used in this study were generated using the MIST Isochrone Interpolation tool\footnote{https://waps.cfa.harvard.edu/MIST/interp\_isos.html}.

\section{Analysis}

Our stellar data are taken from the first data release of the SDSS-V MWM survey \citep[DR19;][]{dr19}. 
A full description of the APOGEE data quality and parameter limitations is presented in \citet{dr17} and Holtzman et al. ({\it in prep}).
For this work, we have selected the SDSS/APOGEE survey to study diffusion in the open clusters NGC 752 and Ruprecht 147. These open clusters were selected because they cover a wide range in stellar class from dwarfs to red giant stars.
Notably, besides the open cluster M67 that was previously studied in \citet{souto_2018, souto_19}, NGC 752 and Ruprecht 147 are the only clusters in the OCCAM sample that include stars spanning the main sequence, turnoff, and red giant branch evolutionary stages. This full evolutionary coverage is essential for probing abundance variations near the turnoff, where atomic diffusion signatures are expected to be most prominent.

\subsection{Cluster Membership and Selected Sample}

To select open cluster members for this study, we used the cluster membership probabilities from the Open Cluster Chemical Analysis and Mapping survey \citep[OCCAM,][]{frinchaboy_13,cunha_15,donor_18,donor_2020,Myers_2022_OCCAM,Otto2026AJ....171...91O_OCCAM}. Specifically, we used the OCCAM proper motion (PM) and radial velocity (RV) membership probabilities from the \citet{Myers_2022_OCCAM} VAC to select members of NGC 752 and Ruprecht 147. We adopted a cut-off RV PROB $>$ 0.1 and RV PROB $>$ 0.1. We checked that the same sample is obtained when using the \cite{Otto2026AJ....171...91O_OCCAM} DR19 results.
We note that we did not use the metallicity-based probability, as it could remove stars with potential signatures for atomic diffusion, since the iron surface abundance would change at and around the turnoff phase. 

We also used radial velocity measurements from different APOGEE observations of the same star to remove possible binaries from our sample. The DR17 parameter VSCATTER provides the scatter around the average radial velocity determined from individual stellar visits; here, we applied a VSCATTER cut of $< 1~km~s^{-1}$. In addition, we selected only stars with reliable stellar parameters and elemental abundance measurements in DR17 by applying the APOGEE/ASPCAP quality bit-wise flags described in \citet{jonsson_2020} and Holtzman et al. ({\textit{in prep.}}). Specifically, we excluded stars flagged by the relevant DR17 stellar-parameter STARFLAG criterion and by the ASPCAP chemistry-quality flag, corresponding to the STARFLAG and ASPCAP flag cuts previously summarized as STARFLAG $\ne 16$ and ASPCAP chemistry flag $\ne 23$. We further required APOGEE spectra with SNR $\geq~100$. To ensure cluster membership, we adopted the OCCAM survey VAC probabilities from \citet{Myers_2022_OCCAM}, requiring both radial-velocity membership probability RV\_PROB $> 0.1$ and proper-motion membership probability PM\_PROB $> 0.1$. After applying these selection criteria, our final sample consists of 35 stars from NGC~752 and 23 stars from Ruprecht~147.

\begin{figure*}
    \centering
    \includegraphics[width=0.87\textwidth]{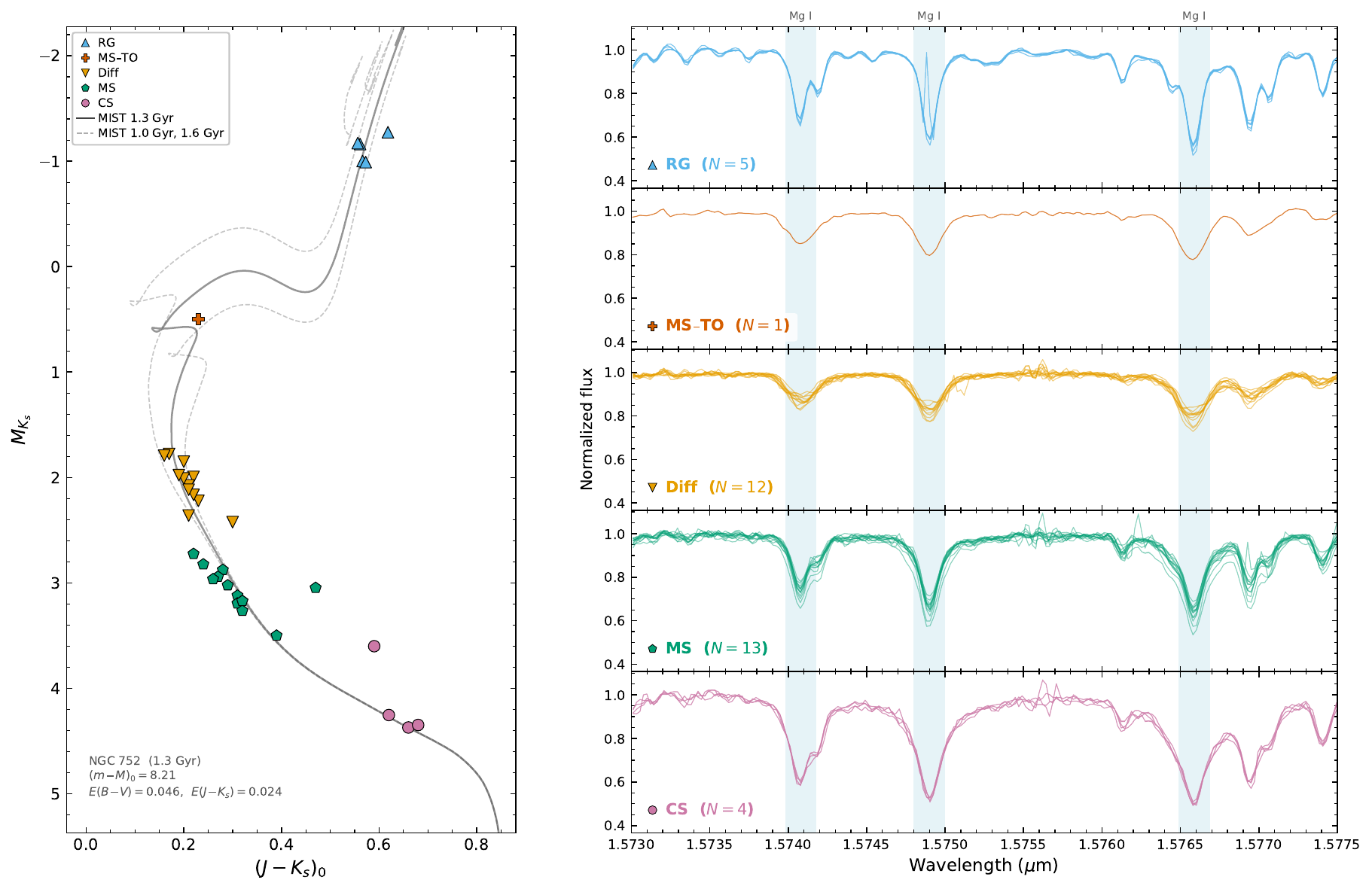}
    \includegraphics[width=0.87\textwidth]{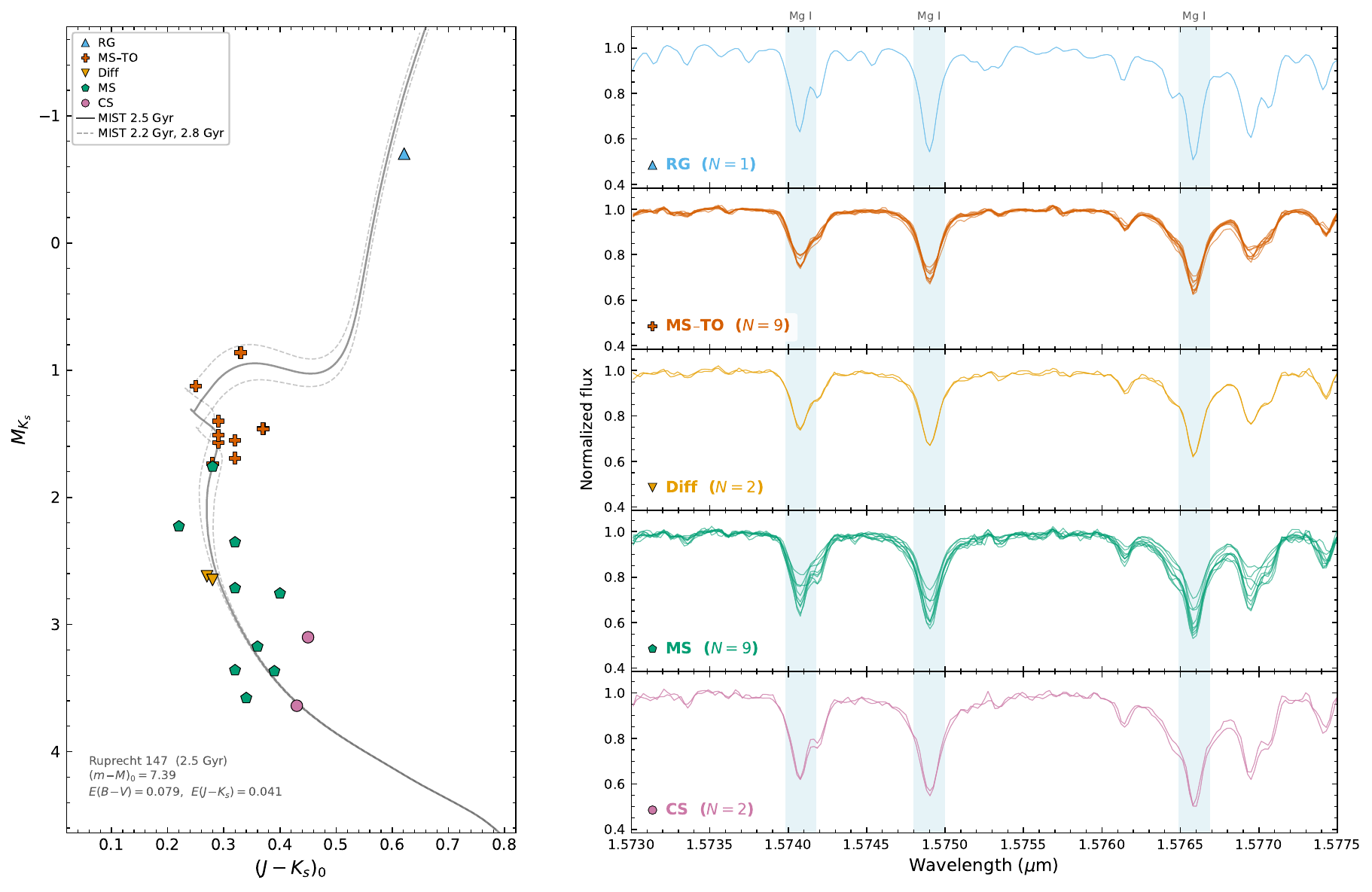}
   \caption{Color--magnitude diagram and APOGEE spectra for stars in NGC~752 and Ruprecht 147 in the top and bottom panels, respectively.
The left panel shows the dereddened $M_{K_s}$ versus $(J-K_s)_0$ diagram. Symbols indicate the evolutionary classes adopted in this work: red giants (RG; blue triangles), main-sequence--turnoff transition stars (MS--TO; orange plus symbols), diffusion-dominated stars (Diff; yellow inverted triangles), main-sequence stars (MS; green pentagons), and cool stars (CS; pink circles). The solid gray curve shows the MIST isochrone for the adopted cluster parameters, and dashed gray curves show MIST isochrones with ages $\pm$0.3~Gyr for comparison.
The right panels show the normalized APOGEE spectra in the wavelength interval $1.5730$--$1.5775~\mu{\rm m}$, separated by evolutionary class. Individual stellar spectra are shown as thin colored curves, and the number of stars in each group is indicated in each panel. The light-blue shaded regions mark the Mg~I features used to illustrate the spectral behavior across the cluster sequence.}
    \label{fig:CMD_Spec}
\end{figure*}

Figure~\ref{fig:CMD_Spec} presents the photometric and spectroscopic properties of our final sample in NGC~752 and Ruprecht~147. The left panels show the dereddened color--magnitude diagrams, $M_{K_s}$ versus $(J-K_s)_0$, while the right panels show a narrow APOGEE H-band spectral region containing three strong Mg~I features. 
The top and bottom rows correspond to NGC~752 and Ruprecht~147, respectively. 
In the CMD panels, we overplot the adopted MIST isochrone for each cluster as a solid gray curve, together with two additional isochrones offset by $\pm 0.3$~Gyr from the adopted cluster age, shown as dashed gray curves. 
The symbols indicate the evolutionary classes adopted in this work, as defined in Section~\ref{sec:mass_trends}: cool stars (CS), main-sequence stars (MS), diffusion-dominated stars (Diff), main-sequence--turnoff transition stars (MS--TO), and red giants (RG). 
The spectroscopic panels show the observed APOGEE spectra for stars in each of these evolutionary classes, ordered from evolved stars at the top to cooler main-sequence stars at the bottom. The shaded regions mark the Mg~I features highlighted in this figure and illustrate how the line strengths vary across the cluster sequences. The stellar parameters and abundances for the full sample are listed in Table~\ref{tab:par_abu}.

\subsection{Atmospheric Parameters: $T_{\rm eff}$ and log $g$} \label{sec:atm_par}

We determined the effective temperatures for the studied stars using the color--$T_{\rm eff}$ relations of \citet{GHB2009_PhotCal}, adopting the calibrations appropriate for dwarf and giant stars. For each star, we computed temperatures from five color indices: $B-V$, $V-J$, $V-H$, $V-K_s$, and $J-K_s$. The optical $B$ and $V$ magnitudes were taken from the UCAC4 catalog \citep{Zacharias2013AJ....145...44Z}, and the near-infrared $J$, $H$, and $K_s$ magnitudes from 2MASS \citep{2MASS_DR_allsky}. We adopted cluster metallicities of [Fe/H] = $-0.04$ dex for NGC~752 and [Fe/H] = $+0.12$ dex for Ruprecht~147 when applying the calibrations. The final photometric $T_{\rm eff}$ was defined as the mean value of the five color indices, and the dispersion among the individual color temperatures was used as an estimate of the internal uncertainties. A detailed comparison of this work's $T_{\rm eff}$ with those from APOGEE/ASPCAP DR17 and DR19 is presented in Appendix \ref{app:teff}.

We determined surface gravities using Equation~\ref{eq:logg_lum}, which follows directly from the Stefan--Boltzmann law combined with the definition of surface gravity.

\begin{equation}
\log g = \log g_{\odot} 
+ \log\left(\frac{M}{M_{\odot}}\right)
+ 4\log\left(\frac{T_{\rm eff}}{T_{{\rm eff},\odot}}\right)
- \log\left(\frac{L}{L_{\odot}}\right)
\label{eq:logg_lum}
\end{equation}

For each star, we adopted the photometric $T_{\rm eff}$ values determined in this work, together with stellar masses and luminosities estimated from MIST isochrones appropriate for each cluster. We assumed an age of 1.3~Gyr and [Fe/H] = $-0.04$ dex for NGC~752, and an age of 2.5~Gyr and [Fe/H] = $+0.12$ dex for Ruprecht~147, following \citet{cg20} ages and metallicities from \cite{Reddy2012MNRAS.419.1350R,Lum2019ApJ...878...99L,Carrera2019A&A...623A..80C,Bragaglia2018A&A...619A.176B,Beeson2024MNRAS.529.2483B}. Stellar masses and luminosities were assigned by comparing each star with the corresponding isochrone, using the adopted $T_{\rm eff}$ and the $K_s$ magnitude as reference quantities, and selecting the closest matching point along the isochrone. We adopted the solar reference values of $\log g_{\odot}=4.438$ dex and $T_{{\rm eff},\odot}=5772$~K, following the IAU recommendations of \citet{Prsa2016AJ....152...41P}.

\subsection{Detailed Abundance Analysis} \label{sec:TS}

In this study, we performed a detailed spectroscopic analysis to determine chemical abundances for all stars in our sample using the effective temperatures and surface gravity values discussed in the previous section. These detailed abundance determinations provide an independent benchmark against which we compare the automated results from the ASPCAP pipeline, as presented in Appendix \ref{app:abundances}.

The spectral analysis was carried out using the BACCHUS code \citep{Masseron2016}, with MARCS models of atmosphere (\citealt{Gustafsson2008}),  and we derive elemental abundances by fitting synthetic spectra to observed lines via a minimum $\chi^2$ criterion with further visual inspection. 
For estimating the microturbulence velocity, we adopted the same procedure as in \citet{souto_2016} and \citet{Smith2013_method}, and the same spectral lines as in \cite{souto_2018} were analyzed to determine individual abundances of the elements: C, N, Na, Mg, Al, Si, S, K, Ca, Ti, V, Cr, Mn, Fe, Co, and Ni.
It is important to note that for specific elements, such as Ti, V, Cr, and Co, the neutral atomic lines become quite weak as the effective temperature increases. 
This is the case, in particular, for turnoff stars, for which the determination of precise individual abundances becomes more challenging or impossible. 
The chemical abundances obtained from our detailed analysis are presented in Table \ref{tab:par_abu}.

The estimated abundance uncertainties here are the same as those reported in Table 4 of \cite{souto_2018}, as our analysis methodology and atmospheric parameter range for the sample are very similar.
The abundances of Fe consistently exhibit the lowest total uncertainties (typically $<$0.07 dex) across all evolutionary stages (main-sequence, turnoff, and red giants), making it a robust indicator for comparative chemical studies. 
The abundances of carbon and calcium also display relatively small uncertainties, below 0.07 dex in most cases, although the uncertainties are slightly higher for red giants. In contrast, silicon and magnesium abundances show increased sensitivities, particularly to effective temperature and microturbulent velocity, resulting in abundance uncertainties up to $\sim$0.10 dex for turnoff stars.

The mean open cluster Fe abundances and standard deviations obtained from our detailed analysis are $\langle$[Fe/H]$\rangle$ -0.03$\pm$0.04 for NGC 752, and $\langle$[Fe/H]$\rangle$=0.14$\pm$0.04 for Ruprecht 147.
These mean Fe abundances and their associated dispersions are slightly larger than the level of chemical homogeneity reported by \citet{Sinha2024}, who found that most open clusters exhibit internal abundance scatters of $\leq$ 0.02 dex across well-measured elements. Although NGC 752 and Ruprecht 147 were not part of their cluster sample, our slightly larger internal dispersions (0.04 dex) may reflect subtle abundance variations driven by physical processes such as atomic diffusion, particularly given our deliberate selection of stars across a range of evolutionary stages, which we discuss in detail in the next Section.

\subsection{Deviation from the local thermodynamical equilibrium}

Based on the previous non- local thermodynamical equilibrium (NLTE) study in the APOGEE H-band by \citet{Osorio_2020_NLTE}, we do not expect large NLTE abundance corrections for the stars analyzed in this work. For Na, K, and Ca, LTE and NLTE line profiles yield essentially the same abundances in the solar spectrum, while only small differences are found for Ca in Arcturus. Magnesium is the element most affected by NLTE effects, particularly for the strong Mg~I lines near $\lambda$15740--15766~\AA{}, for which line-core differences can become important at APOGEE-like resolution. However, these strong Mg features are excluded from our analysis; instead, we use weaker Mg~I lines that are less affected by the reported NLTE departures, as discussed in Section \ref{sec:mg}. 

\begin{deluxetable*}{llcccccccccc}
\tablecaption{Stellar Parameters and Elemental Abundances\label{tab:par_abu}}
\tabletypesize{\scriptsize}
\tablewidth{0pt}
\tablehead{
\colhead{SDSS\_ID} & \colhead{Cluster} & \colhead{$J_0$} & \colhead{$H_0$} & \colhead{$K_0$} & \colhead{S/N} & \colhead{Phase} & \colhead{$M/M_{\odot}$} & \colhead{$T_{\rm eff}$} & \colhead{$\log g$} & \colhead{$[{\rm Fe/H}]$} & \colhead{\ldots} \\
}
\startdata
116885323 & NGC752 & 10.607 & 10.442 & 10.379 & 183.3 & Diff & 1.28 & 6678 & 4.30 & -0.044 & \ldots \\
116885668 & NGC752 & 7.581 & 7.141 & 7.024 & 414.4 & RG & 1.87 & 4982 & 2.99 & -0.030 & \ldots \\
116885332 & NGC752 & 10.937 & 10.696 & 10.640 & 154.9 & Diff & 1.24 & 6385 & 4.29 & -0.063 & \ldots \\
116886272 & NGC752 & 10.263 & 10.109 & 10.063 & 305.7 & Diff & 1.28 & 6719 & 4.31 & -0.112 & \ldots \\
116886106 & NGC752 & 11.457 & 11.212 & 11.168 & 191.7 & MS & 1.14 & 6190 & 4.38 & 0.034 & \ldots \\
\ldots & \ldots & \ldots & \ldots & \ldots & \ldots & \ldots & \ldots & \ldots & \ldots & \ldots & \ldots \\
\enddata
\tablecomments{This table is published in its entirety in machine-readable format. A portion is shown here for guidance regarding its form and content. The full machine-readable table contains the SDSS identifier, cluster name, dereddened 2MASS photometry ($J_0$, $H_0$, and $K_0$), APOGEE signal-to-noise ratio, evolutionary phase, isochrone-derived stellar mass, $T_{\rm eff}$, $\log g$, the abundance ratios [Fe/H], [C/H], [N/H], [Na/H], [Mg/H], [Al/H], [Si/H], [S/H], [K/H], [Ca/H], [Ti/H], [V/H], [Cr/H], [Mn/H], [Co/H], [Ni/H], the corresponding abundance uncertainties e\_[C/H], e\_[N/H], e\_[Na/H], e\_[Mg/H], e\_[Al/H], e\_[Si/H], e\_[S/H], e\_[K/H], e\_[Ca/H], e\_[Ti/H], e\_[V/H], e\_[Cr/H], e\_[Mn/H], e\_[Fe/H], e\_[Co/H], e\_[Ni/H], the Gaia DR3 source identifier, and the SDSS-IV APOGEE identifier. Blank entries indicate measurements that could not be reliably determined for a given star.}
\end{deluxetable*}

\section{Results \& Discussion}

This work presents atmospheric parameters and abundance results for 58 stars at different evolutionary stages in the NGC 752 and Ruprecht 147 open clusters. The studied elements are: Fe, C, N, Na, Mg, Al, Si, S, K, Ca, Ti, V, Cr, Mn, Co, and Ni.
Our sample is designed to probe abundance trends linked to physical processes operating in stars, such as atomic diffusion and extra mixing, as a signature of the first dredge-up. 
To evaluate the extent of these signatures in the chemical abundances measured for our open cluster samples, we compare our derived elemental abundances with results and predictions from MIST stellar evolution models. 
In Appendices \ref{app:teff} and \ref{app:abundances}, we present a comparison of atmospheric parameters and individual abundances using the APOGEE/ASPCAP pipeline, placing this work's detailed results in context with those derived from automatic pipelines. 

\subsection{Populations as a Function of Stellar Mass}
\label{sec:mass_trends}

\begin{figure*}
    \centering
   { \includegraphics[width=0.47\textwidth]{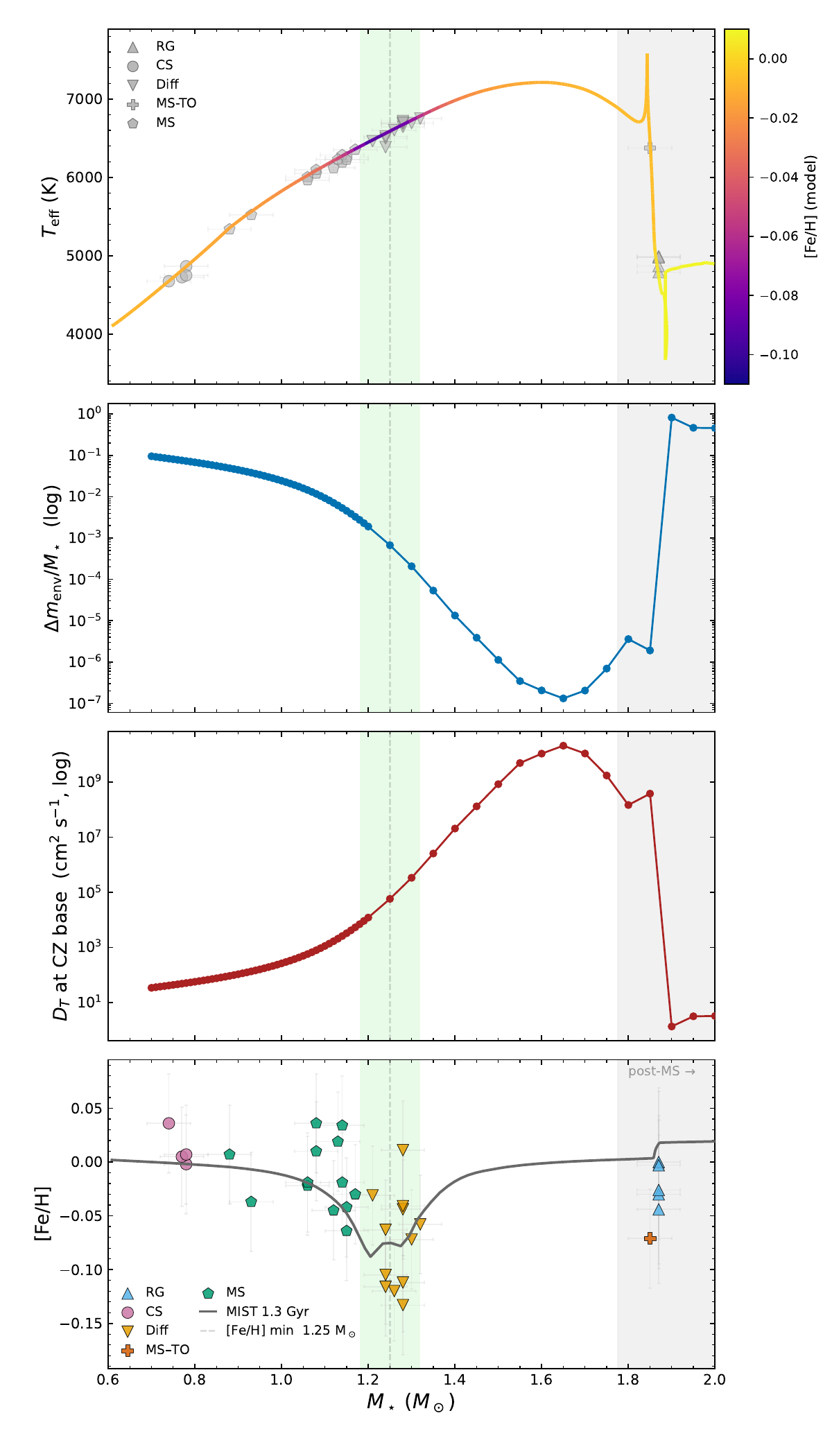}}
   { \includegraphics[width=0.47\textwidth]{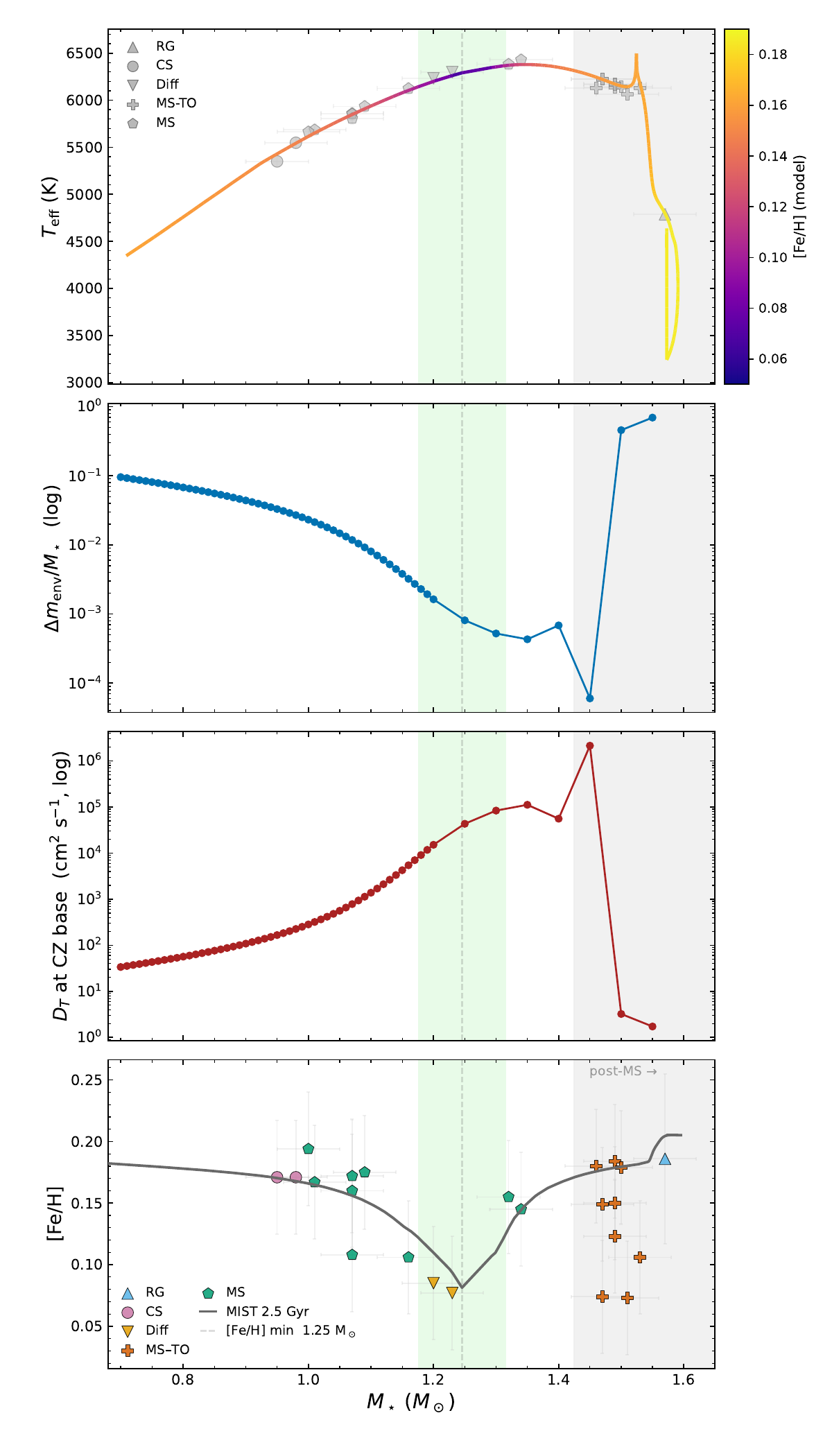}}
    \caption{Atomic diffusion diagnostics as a function of stellar mass for NGC~752 (\textit{left}) and Ruprecht~147 (\textit{right}). In each column, the four panels share the same abscissa (isochrone-derived stellar mass); the green band marks the predicted [Fe/H] minimum ($M_{\star} = 1.25 \pm 0.05M_{\odot}$ for NGC~752), and the gray region marks masses evolved past the main sequence at the cluster age. \textit{First panel:} mass--$T_{\rm eff}$ plane; the solid curve is the MIST isochrone (1.3~Gyr for NGC~752; 2.5~Gyr for Ruprecht~147), color-coded by the model surface [Fe/H], with observed members shown in gray, coded by evolutionary class. \textit{Second panel:} fractional mass of the surface convection zone from MIST tracks interpolated at the cluster age. \textit{Third panel:} turbulent-diffusion coefficient $D_{T}$ evaluated at the base of the convection zone. \textit{Fourth panel:} observed [Fe/H] versus stellar mass, coded by evolutionary class, compared with the MIST prediction (solid gray curve).}
    \label{fig:mass_analysis}
\end{figure*}

The interpretation of abundance trends along a cluster sequence depends critically on the evolutionary status assigned to each star. Evolutionary classes are traditionally defined from the position of each star in the observed $T_{\rm eff}$--$\log g$ plane, or equivalently in the color--magnitude diagram. 
The atomic diffusion observational signature, which this work aims to seek, is not controlled by the evolutionary label itself, but by the structure of the star. 
The abundance surface depletion produced by diffusion depends primarily on the mass of the surface convection zone and on the additional mixing processes that counteract gravitational settling \citep{Thoul1994ApJ...421..828T, Dotter2017}, both of which are functions of stellar mass. 
At a fixed near-solar composition and an age typical of open clusters (0.6--3.0 Gyr), the maximum depletion occurs at an approximately age-independent mass of $M_{\star}~\simeq~1.25M_{\odot}$, whereas the cluster turnoff mass decreases with age. 
Therefore, depending on the cluster age, the stars carrying the strongest diffusion signal may be genuine turnoff stars, as in M67 \citep{souto_2018,Liu2019_M67}, or they may be upper-main-sequence stars located well below the turnoff, as in Coma Berenices \citep{Souto2021_ComaBer}. Because this distinction determines how the abundance trends in this work should be interpreted, we begin by discussing the mass domain and use it to define the evolutionary classes adopted throughout the remainder of the paper.

Figure~\ref{fig:mass_analysis} presents the mass-based diagnostics for NGC~752 (left) and Ruprecht~147 (right). In each column, four vertically stacked panels share a common abscissa, the isochrone-derived stellar mass. 
From top to bottom, the panels show: (i) the mass--$T_{\rm eff}$ plane, with the MIST isochrone \citep{MIST_Choi2016, Dotter2017} color-coded by the model surface [Fe/H] at each point along the sequence; (ii) the fractional mass of the surface convection zone, $\Delta m_{\rm env}/M_{\star}$, obtained from MIST evolutionary tracks of fixed mass and interpolated to the cluster age (1.3~Gyr for NGC~752 and 2.5~Gyr for Ruprecht~147; \citealt{cg20}); (iii) the turbulent-diffusion coefficient, $D_{T}$, adopted in the MIST models (Equation~\ref{eq:DT}); and (iv) the observed [Fe/H] as a function of stellar mass, compared with the model prediction. 
In all panels, the green shaded band marks the mass range of maximum predicted depletion ($M_{\star} = 1.25 \pm 0.05\,M_{\odot}$ for NGC~752; $M_{\star} \simeq 1.2$--$1.25\,M_{\odot}$ for Ruprecht~147), while the gray shaded region marks masses that have already evolved off the main sequence at the cluster age.

The two middle panels identify the structural origin of the atomic diffusion depletion signal. 
The fractional convective-envelope mass, shown in the second panel, decreases monotonically by roughly four to six orders of magnitude between $\sim$0.7 and $\sim$1.65\,M$_{\odot}$. 
A thinner convective envelope contains less mass over which gravitationally settled material can be diluted; therefore, for a similar settling flux, the surface abundance depletion increases with stellar mass. 
However, the convective-envelope mass alone cannot explain the full shape of the predicted (and observed) abundance pattern. It accounts for the increasing depletion toward higher masses, but it does not explain why the depletion reaches a minimum and then becomes weaker again at higher masses. 
This turnover point is produced by the turbulent mixing (D$_{T}$) included in the MIST models \citep{Dotter2017}, following \citet{Proffitt1991ApJ...380..238P} as modified by \citet{VandenBerg2012ApJ...755...15V}:

\begin{equation}
D_{T} \;=\; D_{0}
\left( \frac{\rho_{\rm CZ}}{\rho} \right)^{3}
\left( \frac{M_{\rm CZ}}{M_{\star}} \right)^{-3/2},
\label{eq:DT}
\end{equation}
with $D_{0} = 1~{\rm cm^{2}s^{-1}}$, calibrated on the metal-poor globular cluster NGC~6397. Evaluated at the base of the convection zone, where $\rho = \rho_{\rm CZ}$, the turbulent-diffusion coefficient scales as $D_{T} \propto (M_{\rm CZ}/M_{\star})^{-3/2}$ and increases steeply as the convective envelope becomes thinner. 
In the models shown here, $D_T$ rises from $\sim$50~${\rm cm^{2}s^{-1}}$ at $0.7M_{\odot}$ to $\gtrsim 10^{7}~{\rm cm^{2}s^{-1}}$ at $1.4M_{\odot}$. At higher masses before the red giant branch (RGB), where the convective envelope becomes extremely shallow or effectively disappears ($M_{\star} \gtrsim 1.45M_{\odot}$), the computed values should be regarded as lower limits, shown as open triangles in Figure \ref{fig:mass_analysis}. 

The predicted minimum [Fe/H] marks the transition between two competing regimes. 
On the low-mass side, surface depletion is limited by convective dilution: the deeper convective envelope mixes the settled material over a larger mass range. On the high-mass side, turbulent mixing becomes sufficiently efficient to counteract gravitational settling and restore the surface abundances toward their initial values. 
The mass at which these two effects balance ($M_{\star} \simeq 1.25M_{\odot}$), at near-solar composition, is set primarily by the stellar structure and by the adopted turbulent-mixing prescription, and is only weakly sensitive to age over the range considered here for the two studied open clusters.

This nearly fixed abundance depletion as a function of mass, in contrast to the age-dependent turnoff mass, is central to the evolutionary interpretation of our sample. 
In NGC~752, at an age of 1.3~Gyr, the cluster turnoff occurs at $M_{\star} \gtrsim 1.78M_{\odot}$, whereas the stars showing the strongest depletion are concentrated around $M_{\star} \approx 1.2$--$1.35M_{\odot}$. 
In this regime, these are upper-main-sequence stars, and the abundance minimum in NGC~752 does not coincide with the cluster turnoff itself. 

In Ruprecht~147, the older cluster age shifts the turnoff to lower masses, around $M_{\star} \simeq 1.5M_{\odot}$. This is much closer to the mass range where the MIST models predict the strongest surface abundance depletion, near $M_{\star} \simeq 1.2$--$1.3M_{\odot}$, than in the younger NGC~752 cluster. 
Therefore, in Ruprecht~147, the stars classified as MS--TO are sampling the region immediately above the predicted abundance minimum, where the models begin to recover toward the initial cluster abundance. 
In other words, the diffusion-sensitive mass range and the observed turnoff region are closer to overlapping in Ruprecht~147, while they are clearly separated in NGC~752. 
It also explains why the depletion signature occurs at a lower effective temperature than in NGC~752 (see Figure \ref{fig:mass_analysis}, top panel).

Guided by this framework, we define five evolutionary classes, which are adopted consistently throughout the remainder of this paper:

\begin{itemize}
\item \textit{CS} (cool stars): the coolest unevolved members of the sample ($M_{\star} \lesssim 0.8M_{\odot}$ for NGC 752 and $M_{\star} \lesssim 1.0M_{\odot}$ for Ruprecht 147). These stars have deep convective envelopes ($\Delta m_{\rm env}/M_{\star} \sim 10^{-1}$), which efficiently dilute the effects of gravitational settling. Their surface abundances are therefore expected to remain closest to the initial cluster composition.

\item \textit{MS} (main sequence): unevolved main-sequence stars with $0.8 \lesssim M_{\star}/M_{\odot} \lesssim 1.2$. In this mass range, the convective envelope becomes progressively thinner as mass increases, but the predicted diffusion-driven depletion remains moderate.

\item \textit{Diff} (diffusion-dominated): stars within the mass interval of maximum predicted surface depletion ($M_{\star} \approx 1.2$--$1.35M_{\odot}$; green band in Figure~\ref{fig:mass_analysis}). These stars are the primary probes of atomic diffusion in this work.

\item \textit{MS--TO} (main-sequence turnoff): stars at, or just beyond, the main-sequence turnoff of each cluster ($M_{\star} \gtrsim 1.78M_{\odot}$ in NGC~752 and $M_{\star} \simeq 1.5M_{\odot}$ in Ruprecht~147).

\item \textit{RG} (red giants): evolved stars on the red-giant branch. In these stars, the deepening convective envelope during the first dredge-up has largely erased the surface-depletion signature of atomic diffusion, restoring the surface abundances toward values close to the initial cluster composition.
\end{itemize}

We expect both the CS and RG samples to provide the closest estimates of the initial cluster composition, because their deep convective envelopes efficiently dilute or erase the surface abundance signatures produced by atomic diffusion. However, RG stars are known to experience abundance changes during the first dredge-up, particularly in C and N, and, to a lesser extent, in Na. We therefore adopt the CS sample as our primary reference for the pristine open-cluster composition. 

The bottom panels of Figure~\ref{fig:mass_analysis} show that, under this classification, the observed [Fe/H]--mass trends closely follow the model predictions in both clusters. Pink circles are the CS sample, green pentagons are the MS, yellow downward triangles are Diff, orange pluses are the MS-TO, and blue upward triangles are the RG sample.
In NGC~752, the Diff-class stars reach ${\rm [Fe/H]} \simeq -0.10$ to $-0.13$~dex, tracing the bottom of the predicted depletion trough. In contrast, the CS and MS stars scatter around the initial cluster composition, while the red giants recover near-initial abundances after first dredge-up.
In Ruprecht~147, a similar V-shaped profile is present, with the stars near the 1.2 M$_{\odot}$ occupying the depletion region.
However, we see some stars near the turnoff remaining more depleted than the MIST isochrone predicts.

We note that the stellar masses are isochrone-derived, and we adopt an uncertainty of $\pm 0.05M_{\odot}$, shown as horizontal error bars in the bottom panels of Figure~\ref{fig:mass_analysis}. 
As an independent check, masses estimated using the empirical calibration of \citet{Torres2010A&ARv..18...67T} are consistent with the isochrone masses within these uncertainties. In the following subsection, we present the element-by-element abundance trends as a function of effective temperature, adopting the evolutionary classes defined above.

\subsection{Abundance Trends as a function of $T_{\rm eff}$}

\begin{figure*}
    \centering
   { \includegraphics[width=0.32\textwidth]{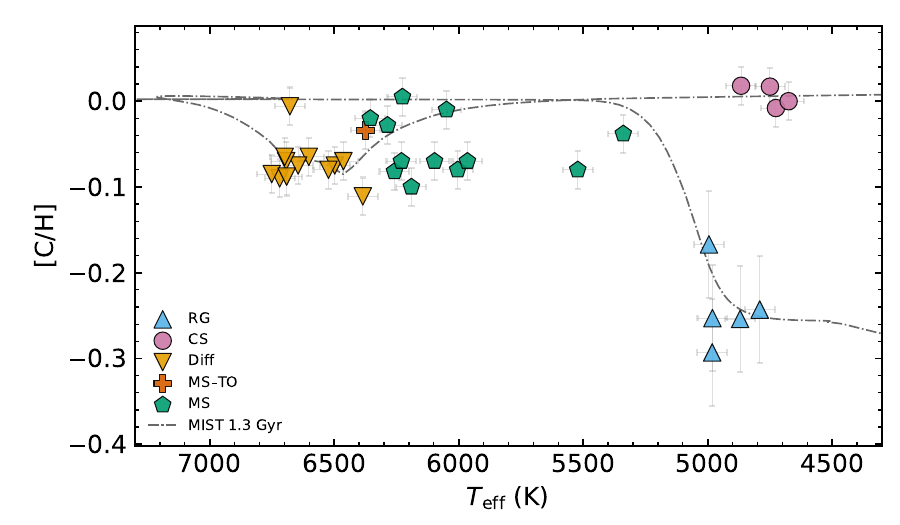}} 
   { \includegraphics[width=0.32\textwidth]{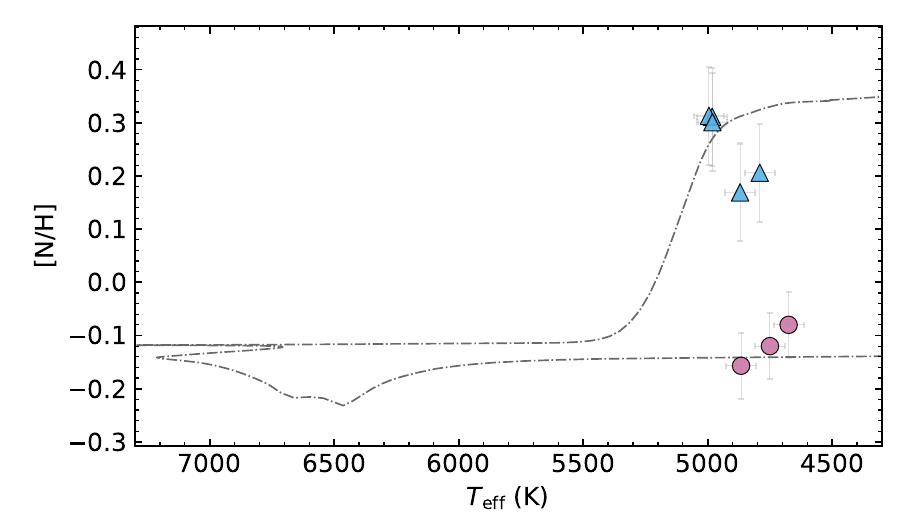}}
   { \includegraphics[width=0.32\textwidth]{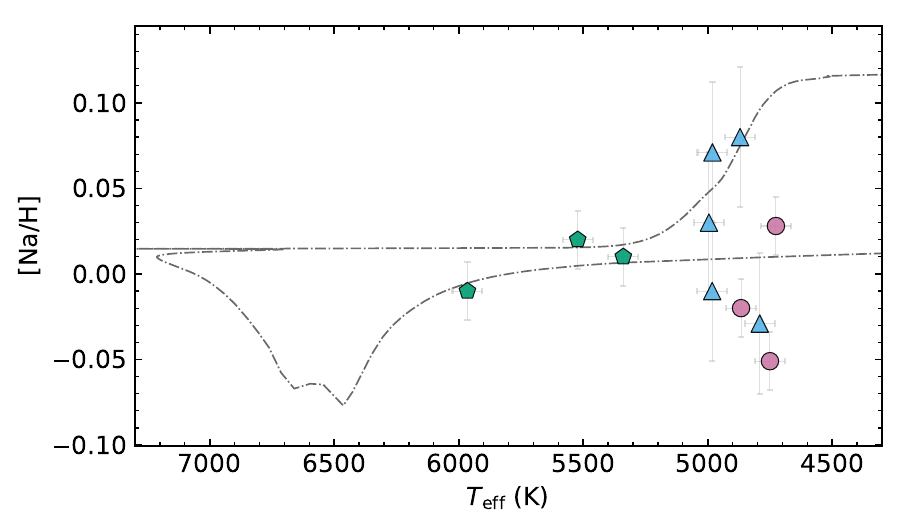}}\\ 
   { \includegraphics[width=0.32\textwidth]{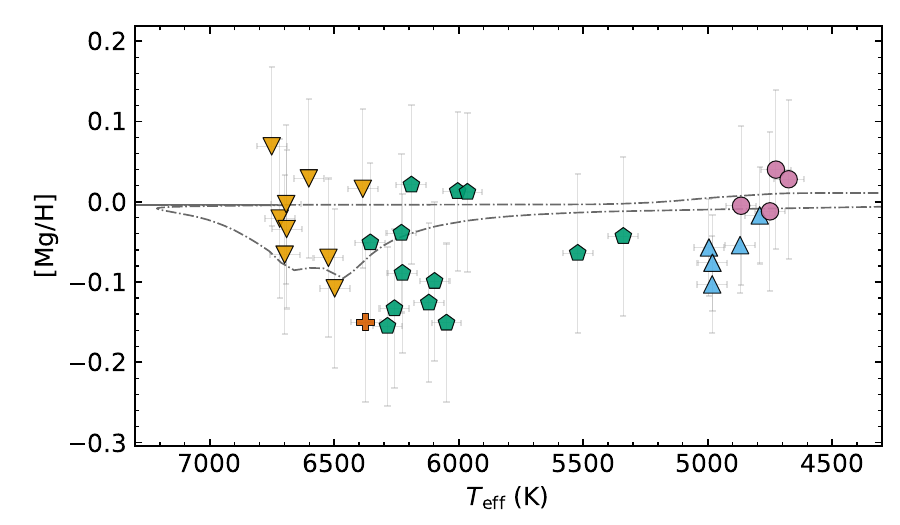}}
   { \includegraphics[width=0.32\textwidth]{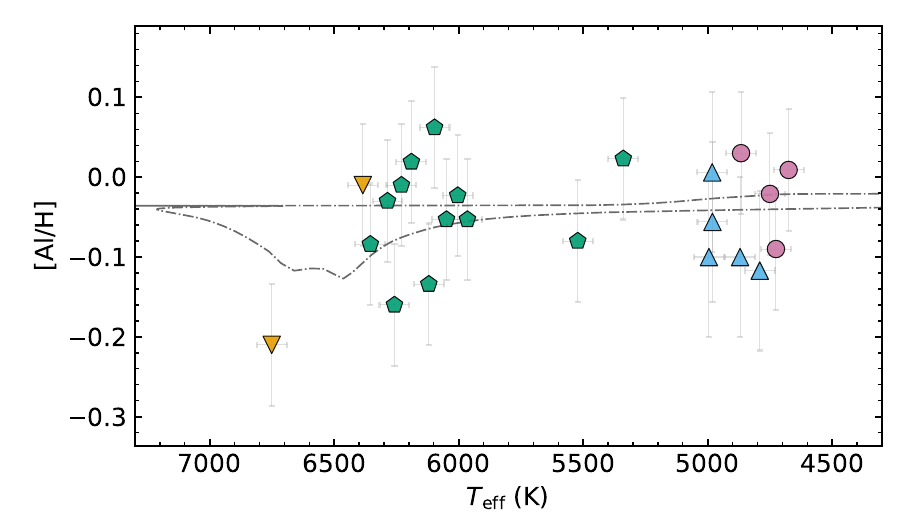}} 
   { \includegraphics[width=0.32\textwidth]{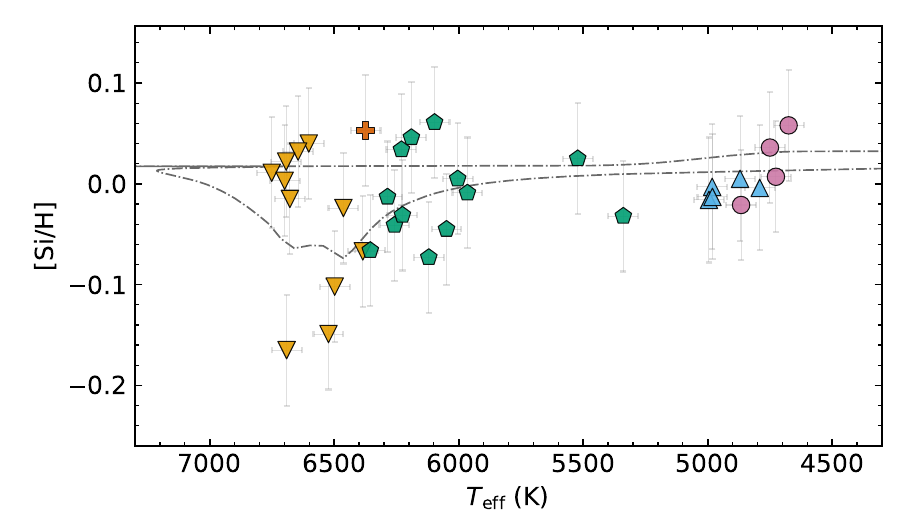}}\\
   { \includegraphics[width=0.32\textwidth]{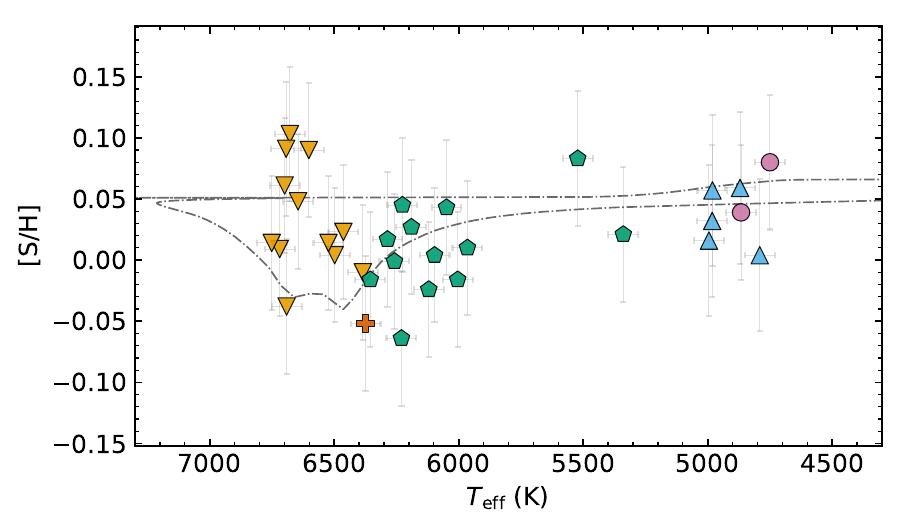}} 
   { \includegraphics[width=0.32\textwidth]{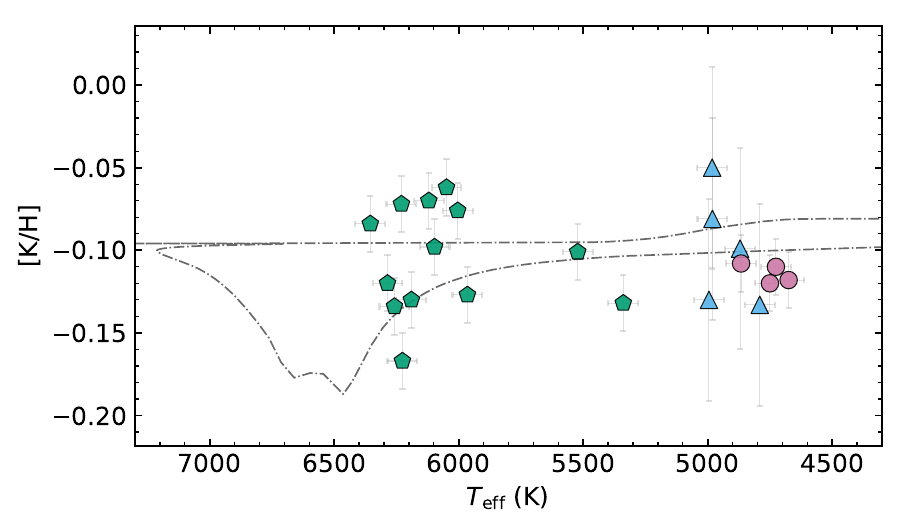}}
   { \includegraphics[width=0.32\textwidth]{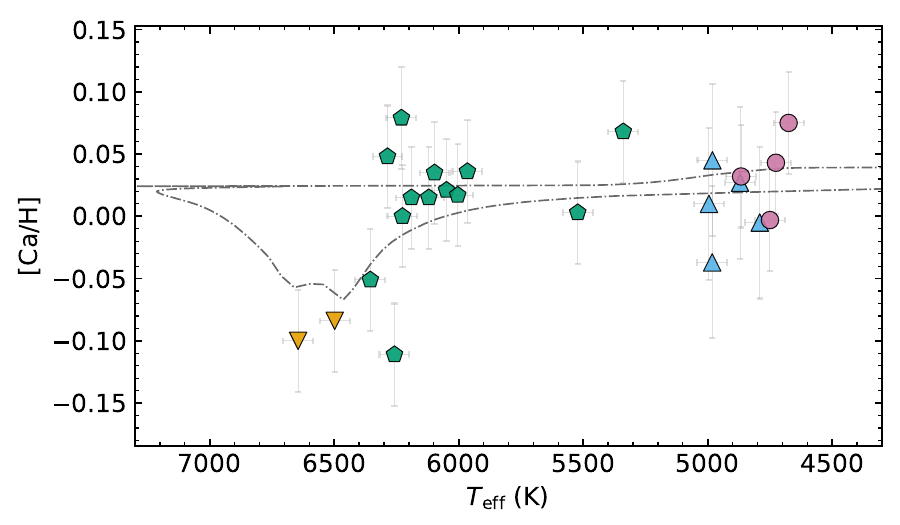}}\\   
   { \includegraphics[width=0.32\textwidth]{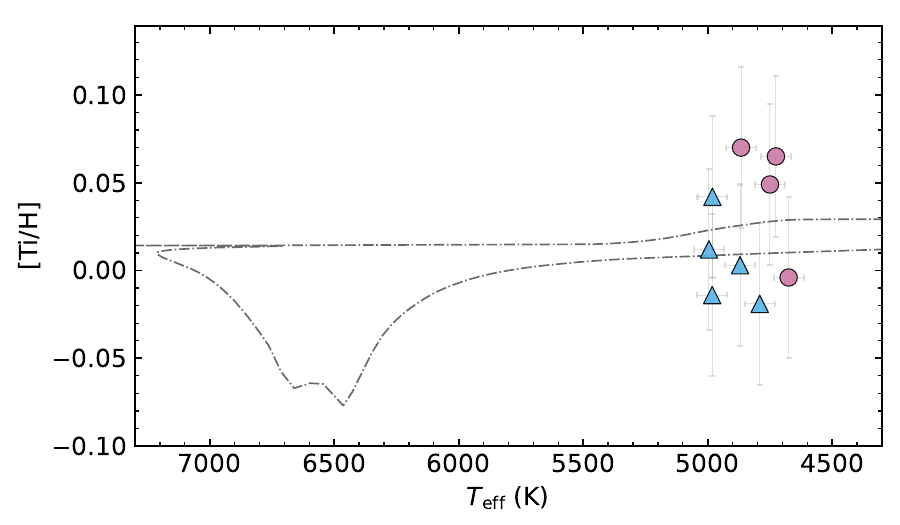}}   
   { \includegraphics[width=0.32\textwidth]{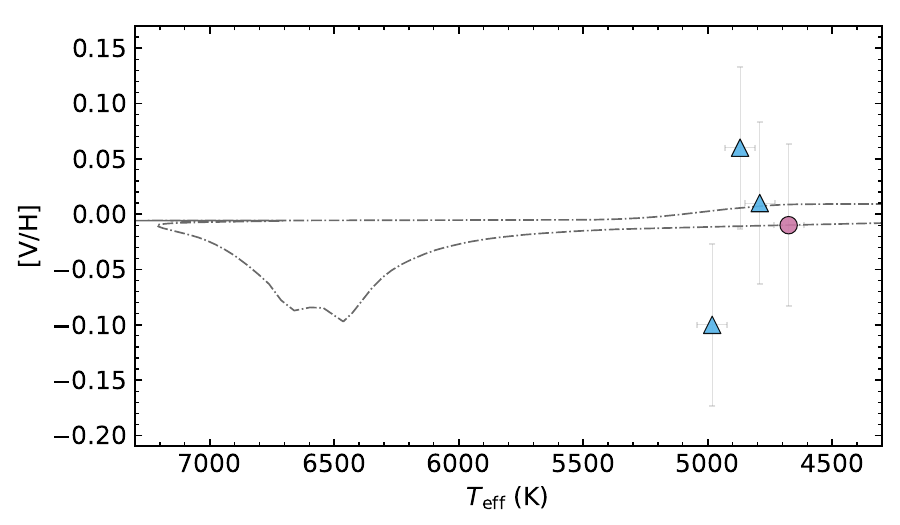}}    
   { \includegraphics[width=0.32\textwidth]{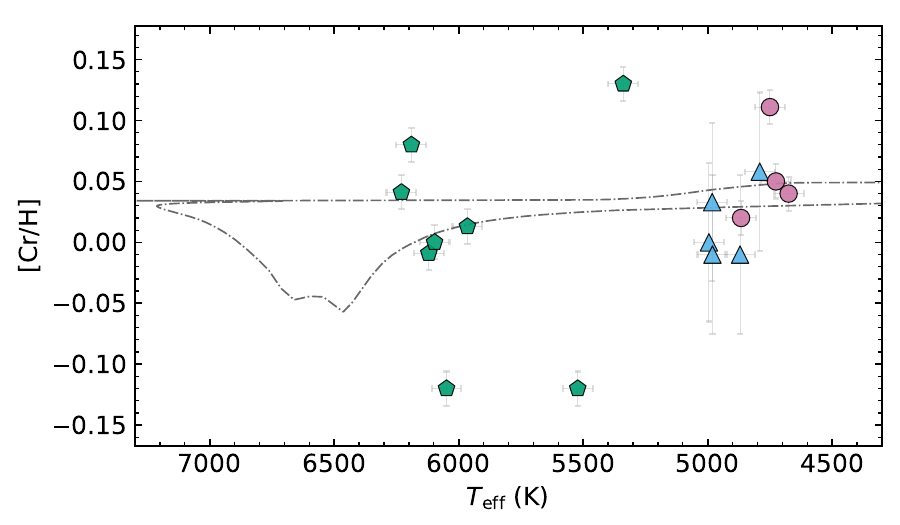}}  \\ 
   { \includegraphics[width=0.32\textwidth]{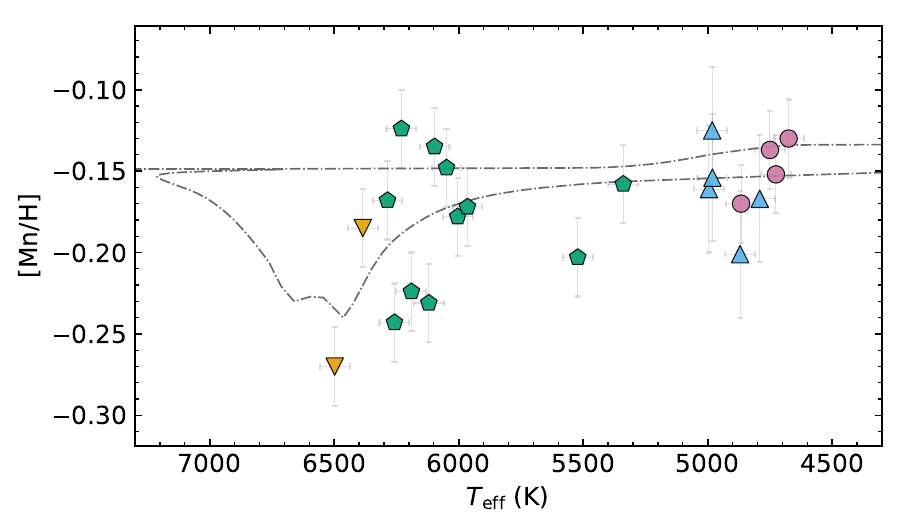}}   
   { \includegraphics[width=0.32\textwidth]{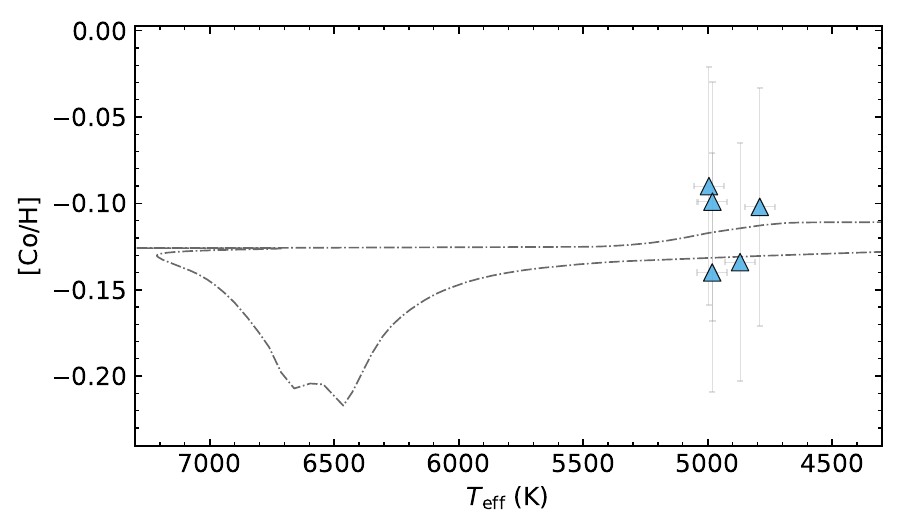}}    
   { \includegraphics[width=0.32\textwidth]{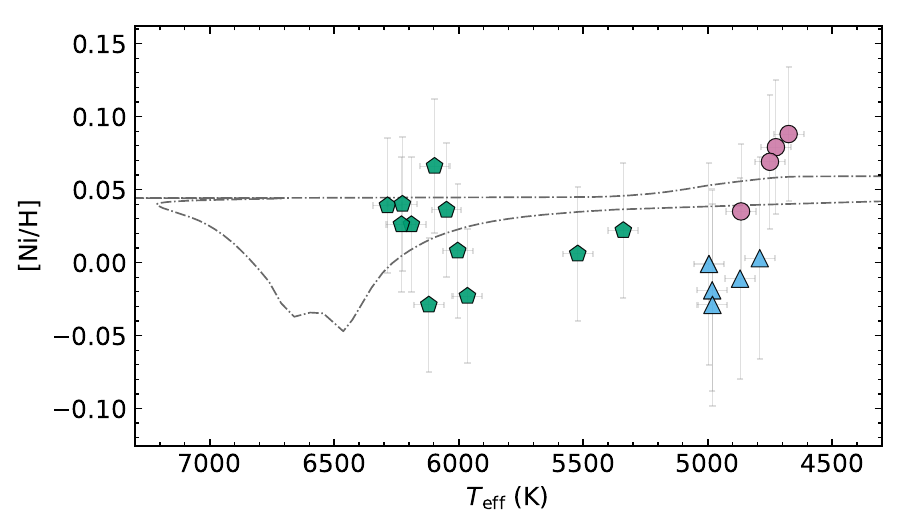}}   
   \caption{Abundances, [X/H], as a function of effective temperature of the studied elements for the open cluster NGC 752. Each panel corresponds to a different chemical species. The different symbols indicate the evolutionary stages: pink triangles represent red giant (RG) stars, green squares indicate main-sequence (MS) stars, yellow circles show stars in the main-sequence turnoff transition region (MS--TO), and blue plus symbols represent turnoff (TO) stars. The dot-dashed curves show the MIST atomic diffusion model \citep{MIST_Choi2016} for an age of 1.3~Gyr and [Fe/H] = 0.00.}
    \label{fig:NGC752_teff_elems}
\end{figure*}

\begin{figure*}
    \centering
   { \includegraphics[width=0.32\textwidth]{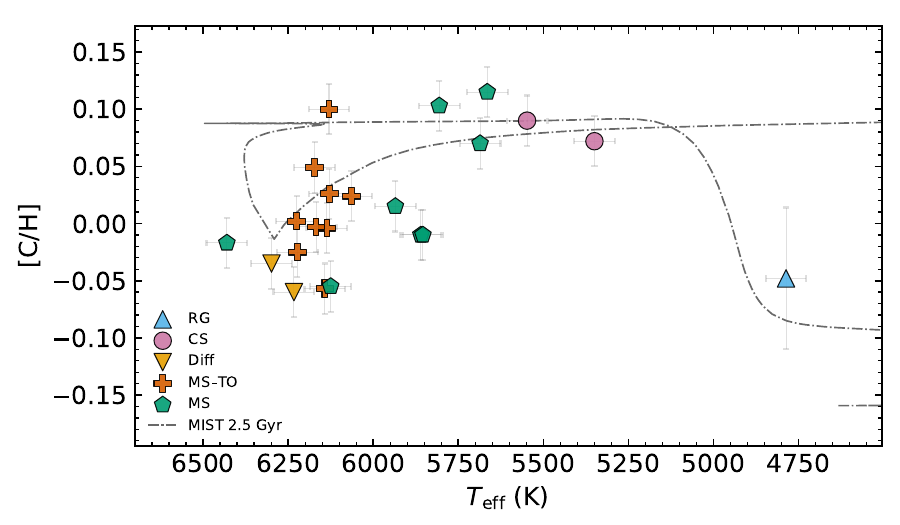}} 
   { \includegraphics[width=0.32\textwidth]{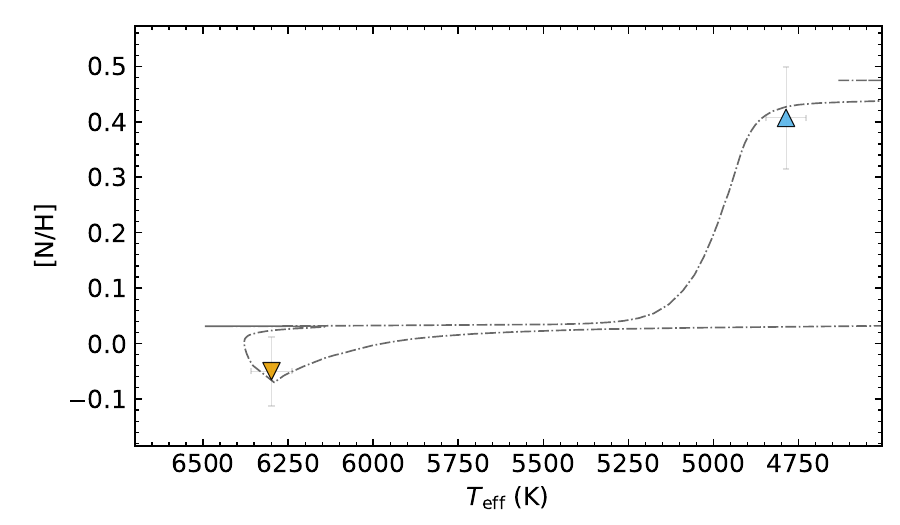}}
   { \includegraphics[width=0.32\textwidth]{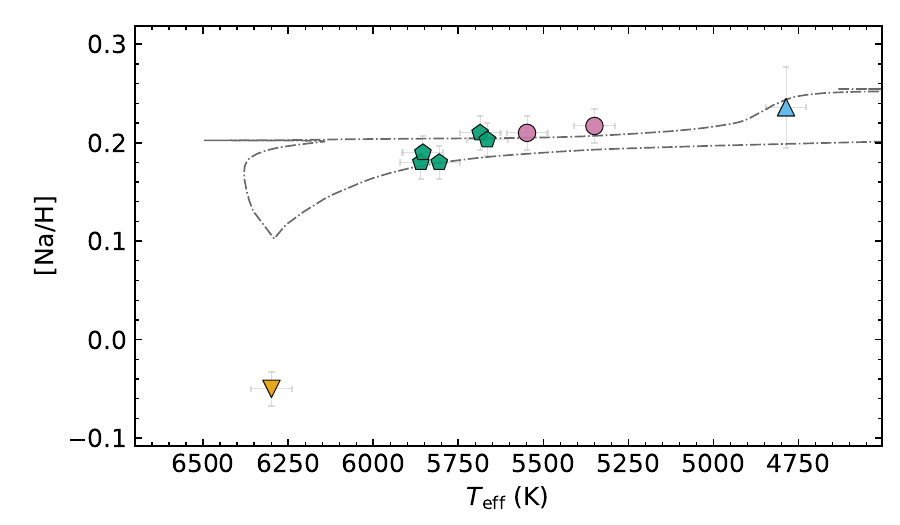}} \\
   { \includegraphics[width=0.32\textwidth]{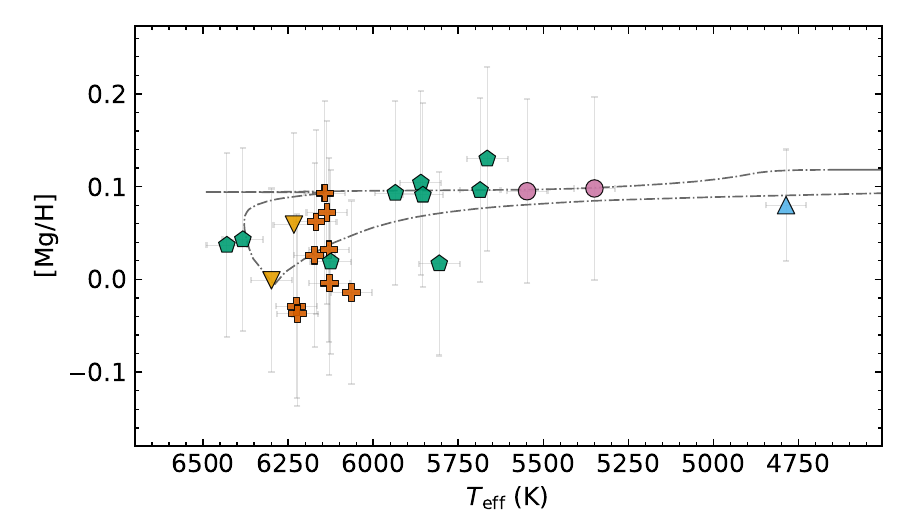}}
   { \includegraphics[width=0.32\textwidth]{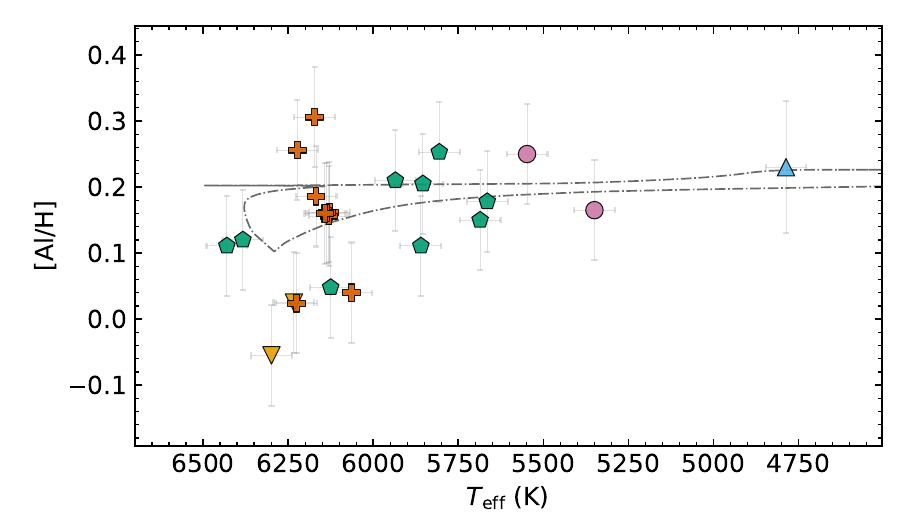}} 
   { \includegraphics[width=0.32\textwidth]{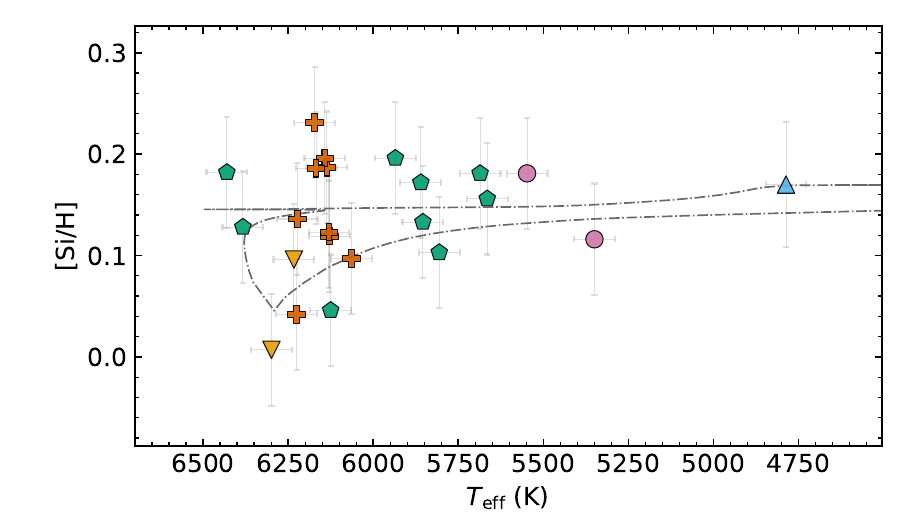}}\\
   { \includegraphics[width=0.32\textwidth]{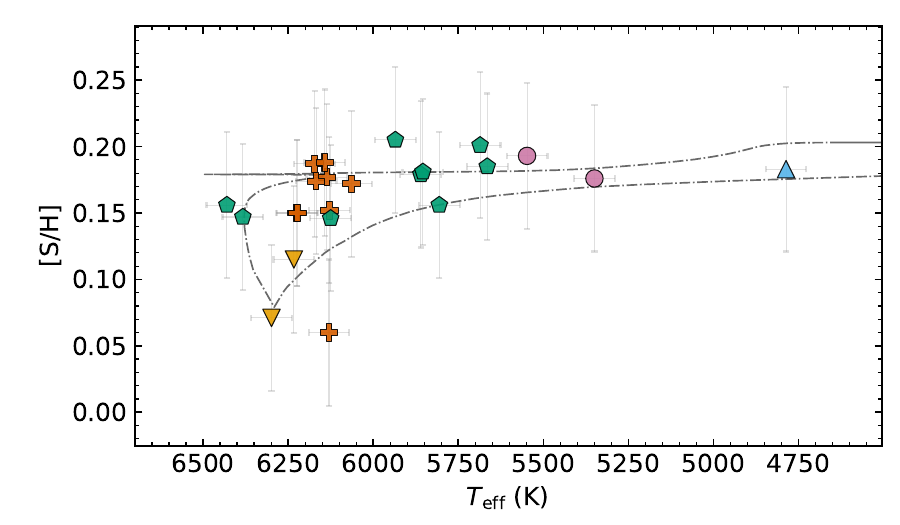}} 
   { \includegraphics[width=0.32\textwidth]{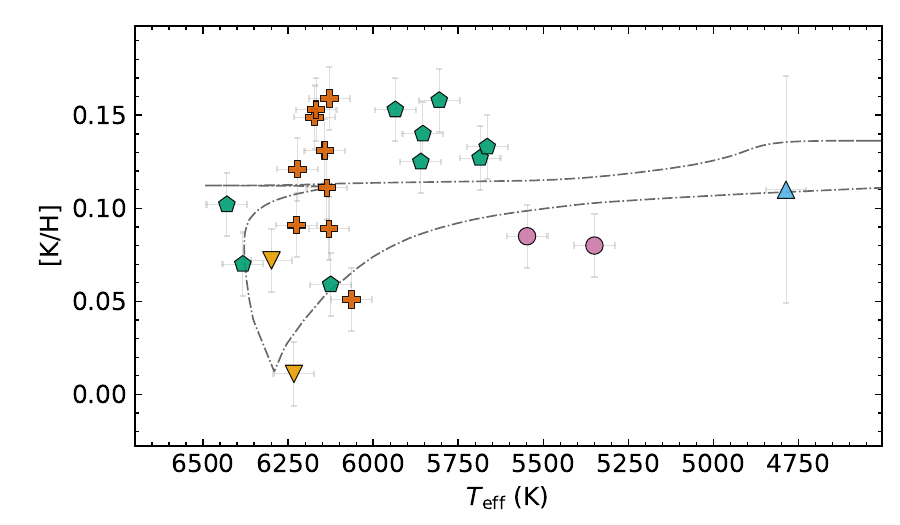}}
   { \includegraphics[width=0.32\textwidth]{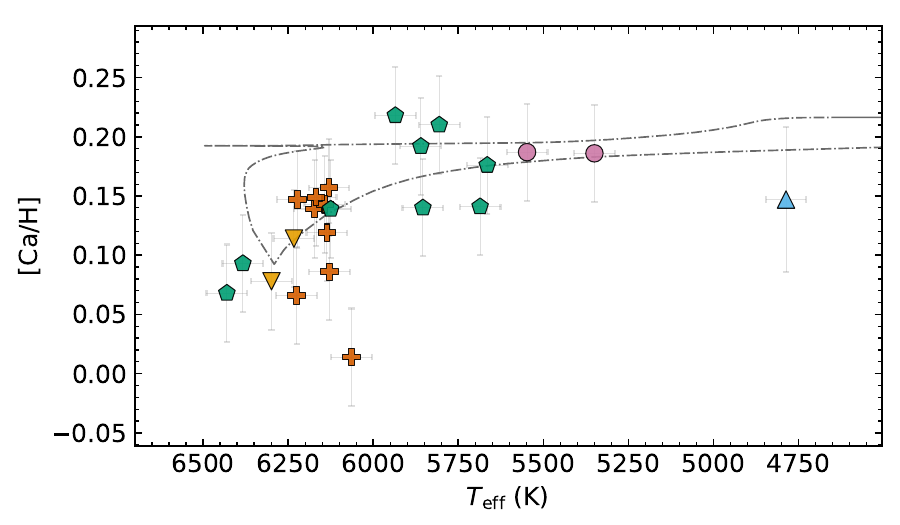}}\\
   { \includegraphics[width=0.32\textwidth]{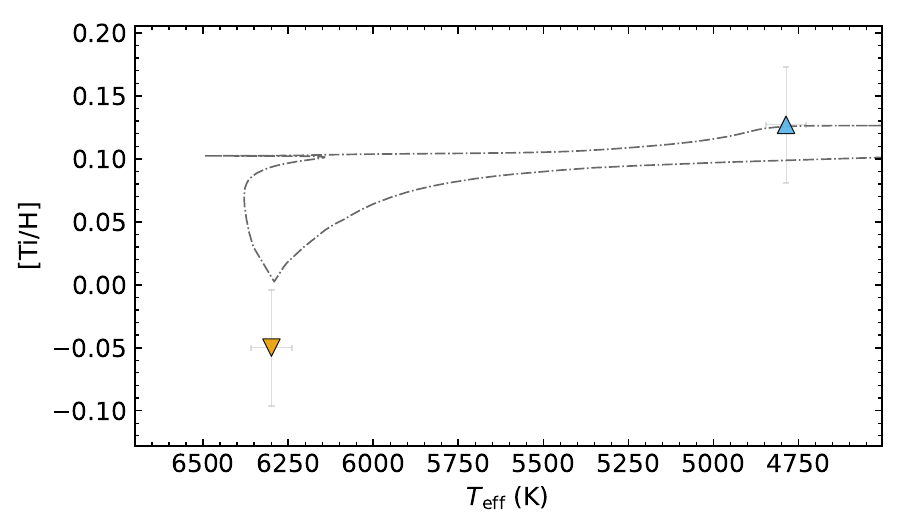}} 
   { \includegraphics[width=0.32\textwidth]{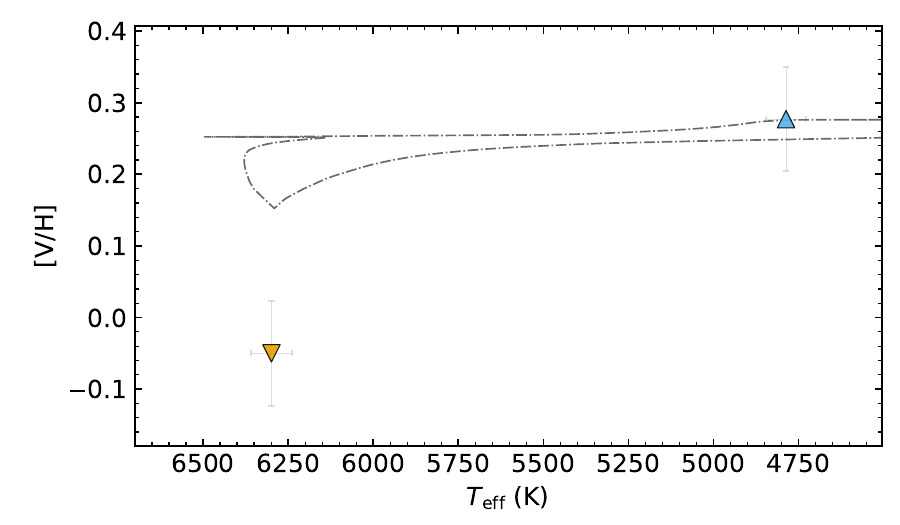}}
   { \includegraphics[width=0.32\textwidth]{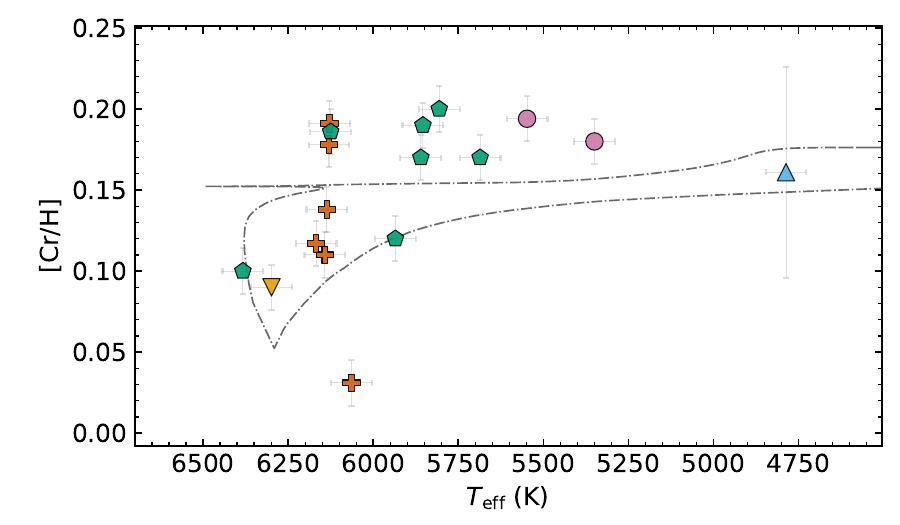}}\\
   { \includegraphics[width=0.32\textwidth]{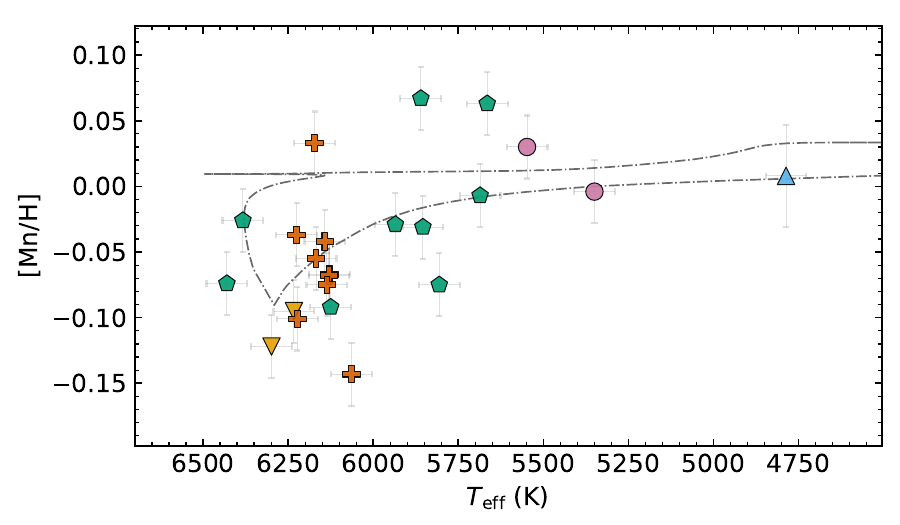}} 
   { \includegraphics[width=0.32\textwidth]{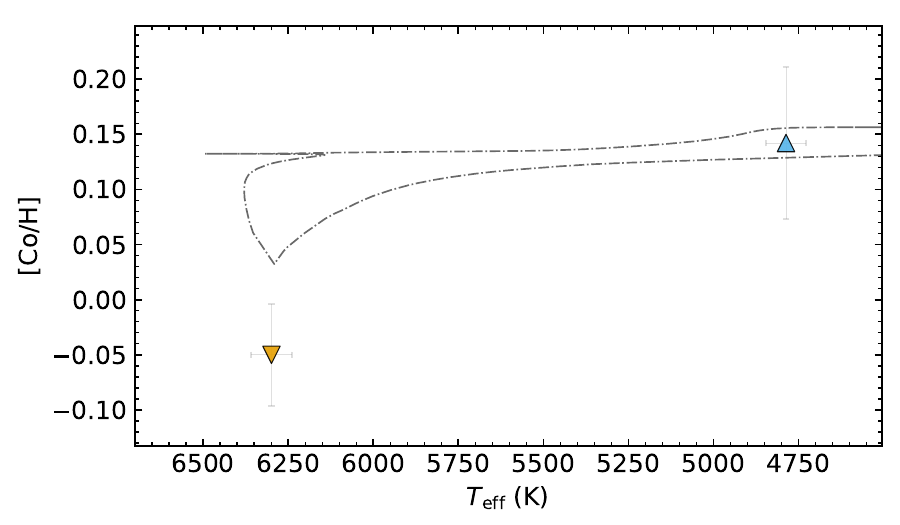}}
   { \includegraphics[width=0.32\textwidth]{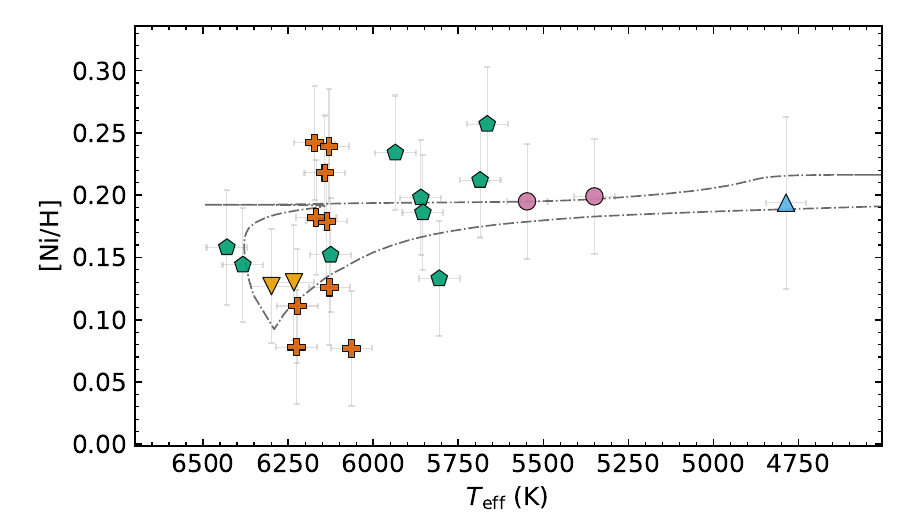}}
   \caption{Abundances, [X/H], as a function of effective temperature of the studied elements for the open cluster Ruprecht 147. Each panel corresponds to a different chemical species. The different symbols indicate the evolutionary stages: pink triangles represent red giant (RG) stars, green squares indicate main-sequence (MS) stars, yellow circles show stars in the main-sequence turnoff transition region (MS--TO), and blue plus symbols represent turnoff (TO) stars. The dot-dashed curves show the MIST atomic diffusion model \citep{MIST_Choi2016} for an age of 2.5~Gyr and [Fe/H] = 0.18.}
    \label{fig:RUP147_teff_elems}
\end{figure*}

In Figures \ref{fig:NGC752_teff_elems} and \ref{fig:RUP147_teff_elems}, we show chemical abundances of the studied elements as a function of effective temperature for sample stars from the clusters NGC 752 and Ruprecht 147, respectively, with each panel representing a studied element. 
Markers are the same as Figure \ref{fig:mass_analysis} bottom panel, where pink circles are the coolest stars (CS), green pentagons the main sequence (MS), yellow downward triangles are most diffused stars (Diff), orange pluses are the main sequence turnoff transition (MS-TO), and blue upward triangles are the red giants sample (RG).
Gray error bars show the uncertainties in both $T_{\rm eff}$ and [Fe/H].

\subsubsection{Carbon}

Carbon abundances in the APOGEE spectra are primarily derived from two sets of spectral features: atomic C~I lines in the warmer MS and MS--TO stars, and molecular CO lines in the cooler CS and RG stars (\citealt{Smith2013_method,souto_2018}).
The carbon abundances in NGC 752 reveal a remarkably clean signature of stellar evolution: atomic diffusion in the Diff stars, and the onset of the first dredge-up in the cooler red giants.
Our derived abundances indicate a mean $\langle$[C/H]$\rangle$ of –0.07$\pm$0.02 dex for the warmer Diff stars, –0.04 for the MS-TO, –0.05$\pm$0.03 dex for the main-sequence end (MS), 0.01$\pm$0.01 for the coolest stars (CS) and a significantly lower abundance for RG, with $\langle$[C/H]$\rangle$ = –0.24$\pm$0.04 dex.
The abundance difference between Diff-CS stars ($\Delta$(Diff-CS)) is -0.08. This difference is $>3$ times the typical abundance scatter in our measurements ($\sigma$), providing evidence for carbon depletion as a signature of atomic diffusion.

As shown in Figure \ref{fig:NGC752_teff_elems}, the determined carbon abundances in NGC 752 align closely with the model predictions, reinforcing the presence of both atomic diffusion in warmer stars and convective mixing in more evolved giants. At the cooler end, the behavior is consistent with the effects of the first dredge-up, a well-understood mixing process that reduces surface carbon abundances by bringing CNO-processed material to the surface \citep{Lagarde2012}. Together, these abundance patterns provide compelling observational support for the combined action of atomic diffusion and mixing in the chemical evolution of NGC 752 stars.

For Ruprecht~147, we also find evidence that atomic diffusion and mixing affect the surface abundances of Diff and red giant stars, respectively. 
The CS stars yield $\langle$[C/H]$\rangle$ = $0.08 \pm 0.01$ dex, which we adopt as representative of the cluster’s pristine carbon abundance. In comparison, the Diff stars show lower carbon abundances, with $\langle$[C/H]$\rangle$ = -$0.04 \pm 0.02$ dex, whereas the MS-TO has $\langle$[C/H]$\rangle$ = $0.01 \pm 0.05$ dex.
The RG star has a carbon abundance of $\langle$[C/H]$\rangle$ = $-0.05 \pm 0.00$ dex,\footnote{We report a standard deviation of zero to indicate that either only one star has a precise carbon abundance measurement or all the derived abundances converged to the same value.} in the same scale as consistent with carbon depletion caused by the first dredge-up.
The derived $\Delta$(Diff-CS) is -0.13 with $>3\sigma$ confidence level. It indicates that atomic diffusion operates in the carbon abundances of Ruprecht~147.

Figure~\ref{fig:RUP147_teff_elems} follows the same format as Figure~\ref{fig:NGC752_teff_elems}, but for the Ruprecht~147 open cluster. The carbon abundances of the Diff and the MS--TO population lie near the depletion level predicted by the MIST isochrones, consistent with the expected signature of atomic diffusion. The lower carbon abundance measured for the RG star is also consistent with carbon depletion associated with first dredge-up.

\subsubsection{Nitrogen}

Nitrogen abundances in APOGEE spectra are derived from several well-defined CN molecular lines. Because these molecular features are highly sensitive to $T_{\rm eff}$, we cannot reliably determine nitrogen abundances for the warmer MS-TO and Diff stars. Therefore, we cannot directly probe the signature of atomic diffusion in nitrogen for either cluster.

Figure~\ref{fig:NGC752_teff_elems} shows a clear signature of extra mixing in NGC~752, as indicated by the enhanced nitrogen abundance of the red giant stars. We obtain $\langle$[N/H]$\rangle$ = $-0.12 \pm 0.04$ dex for the CS stars and $\langle$[N/H]$\rangle$ = $0.26 \pm 0.07$ dex for the RG stars. A similar behavior is seen in Figure~\ref{fig:RUP147_teff_elems} for Ruprecht~147. However, in this cluster we lack nitrogen abundances for the MS stars and have only one RG star with a nitrogen measurement, yielding [N/H] = $0.35 \pm 0.04$ dex.

\subsubsection{Sodium}

Sodium abundances are derived from two weak neutral Na~I lines in the APOGEE spectra ($\lambda$16373.850~\AA{} and $\lambda$16388.860~\AA{}). In warmer stars, both features become increasingly weak, making precise sodium abundance determinations challenging. As a result, we could not derive reliable Na abundances for the Diff stars in NGC 752 and therefore cannot directly probe atomic diffusion of sodium in that cluster.

The previous work by \cite{Loaiza-Tacuri2023MNRAS.526.2378L} has shown that the Na abundances of red giants belonging to the young open cluster NGC 6705 (age=316; turnoff mass = 3.3 M$\odot$) show sodium enhancements ([Na/Fe]=+0.29), in agreement with expectations from first-dredge-up on the red-giant branch of models by \cite{Lagarde2012}. The open clusters in this study, NGC 752 and Ruprecht 147, have significantly older ages and lower turnoff masses, and are expected to show much less Na enhancement due to mixing.
For NGC~752, we obtain $\langle$[Na/H]$\rangle$ = $-0.00 \pm 0.04$, $0.01 \pm 0.02$, and $0.03 \pm 0.05$ dex for the CS, MS, and RG stars, respectively. 
For Ruprecht~147, we determine $\langle$[Na/H]$\rangle$ = $0.21 \pm 0.01$, $0.19 \pm 0.00$, and $0.24 \pm 0.00$ dex for the same evolutionary groups. In both clusters, the CS and MS abundances agree within the uncertainties, while the RG stars show slightly higher Na abundances. This behavior is visible in Figures~\ref{fig:NGC752_teff_elems} and~\ref{fig:RUP147_teff_elems}, where the Na enhancement in the red giants is generally consistent with the MIST predictions and with the effects of extra mixing operating in the convective envelopes of red giant stars \citep[e.g.,][]{Lagarde2012}. We find a clear signature of atomic diffusion in Ruprecht 147, with $\Delta$(Diff-CS) of -0.26. This is the largest difference in abundance among the stellar classes analyzed in this work.

\subsubsection{Magnesium}\label{sec:mg}

The APOGEE spectra contain six well-defined Mg~I lines that can be used to derive magnesium abundances. However, three of the strongest features, at $\lambda$15740.72, $\lambda$15748.99, and $\lambda$15765.84~\AA{}, which are shown in the right panel of Figure \ref{fig:CMD_Spec}, are known to exhibit strong systematic abundance trends with effective temperature in the dwarf stars, being particularly biased for the K dwarfs \citep{Grilo2024}. We excluded these three Mg I lines from the present analysis. Instead, we used the Mg~I lines at $\lambda$15879.50, $\lambda$15886.20, and $\lambda$15954.48~\AA{}, which provide more reliable diagnostics for investigating abundance trends associated with stellar evolution.

For the NGC~752 open cluster, our analysis yields $\langle$[Mg/H]$\rangle$ = $0.01 \pm 0.03$, $-0.01 \pm 0.06$, $-0.02 \pm 0.05$, $-0.015 \pm 0.00$, and $-0.06 \pm 0.03$ dex for the CS, MS, Diff, MS-TO, and RG stars, respectively. 
The abundance difference between the Diff and CS stars is $\Delta({\rm Diff-CS}) = -0.03$ dex, significant at the $1.5\sigma$ level. As shown in Figure~\ref{fig:NGC752_teff_elems}, [Mg/H] follows a temperature-dependent trend broadly consistent with the predictions of diffusion models, supporting the operation of atomic diffusion in NGC~752.

A similar behavior is found for Ruprecht~147. We obtain $\langle$[Mg/H]$\rangle$ = $0.10 \pm 0.00$, $0.07 \pm 0.04$, $0.03 \pm 0.04$, $0.02 \pm 0.05$, and $0.08 \pm 0.00$ dex for the CS, MS, Diff, MS-TO, and RG stars, respectively. In this case, $\Delta({\rm Diff-CS}) = -0.07$ dex, significant at the $2.2\sigma$ level, providing additional evidence for atomic diffusion in Ruprecht~147. Figure~\ref{fig:RUP147_teff_elems} shows this behavior, with most Diff and MS-TO stars overall following the expected abundance dip associated with diffusion.

We note, however, that the RG [Mg/H] values fall slightly below the MIST evolutionary tracks in both clusters, as seen in Figures~\ref{fig:NGC752_teff_elems} and~\ref{fig:RUP147_teff_elems}. 
Magnesium also exhibits some of the largest abundance uncertainties in our sample, ranging from $\sim$0.06 dex for RG stars to $\sim$0.10 dex for TO and MS stars. These larger uncertainties likely contribute to the increased dispersion in abundance observed in the figures, but cannot explain systematic differences.

\subsubsection{Aluminium}

Aluminum abundances are derived from three well-defined Al~I lines in the redder region of the APOGEE spectra. However, as with the Mg~I features discussed above, two of these Al~I lines, at $\lambda$16719 and $\lambda$16750.6~\AA{}, show systematic abundance trends with effective temperature in dwarfs and are therefore excluded from our analysis. We use only one Al~I line at $\lambda$16763.4~\AA{}. Although this leaves a single diagnostic feature, it provides the most reliable aluminum indicator for investigating abundance trends associated with stellar evolution and for searching for signatures of atomic diffusion.

For NGC~752, the diagnostic Al~I line is not sufficiently well defined in the MS-TO stars to allow precise abundance determinations. For the CS, MS, Diff, and RG stars, we obtain $\langle$[Al/H]$\rangle$ = $-0.02 \pm 0.05$, $-0.04 \pm 0.07$, $-0.11 \pm 0.14$, and $-0.07 \pm 0.05$ dex, respectively. Similar to Mg, the Al abundances show a relatively large scatter, likely driven by their larger typical uncertainties ($\sim$0.06--0.10 dex), as seen in Figure~\ref{fig:NGC752_teff_elems}. 
The $\Delta({\rm Diff-CS})$ is -0.09 dex, significant at the $\sim$1$\sigma$ level. 
Figure~\ref{fig:NGC752_teff_elems} shows the overall behavior of the Diff stars following the expected diffusion-induced abundance dip.

We find a clear signature of atomic diffusion in Ruprecht~147, with $\Delta({\rm Diff-CS}) = -0.22$ dex, significant at the $>3\sigma$ level. We derive $\langle$[Al/H]$\rangle$ = $0.21 \pm 0.06$, $0.16 \pm 0.06$, $-0.2 \pm 0.06$, $0.16 \pm 0.09$, and $0.23 \pm 0.00$ dex for the CS, MS, Diff, MS-TO, and RG stars, respectively. Figure~\ref{fig:RUP147_teff_elems} shows [Al/H] as a function of the effective temperature, where the Al abundances also exhibit larger scatter than most other elements.

\subsubsection{Silicon}

Nine Si~I lines are available for deriving silicon abundances from APOGEE spectra. As discussed by \citet{Grilo2024}, four of these lines, at $\lambda$15888.4, $\lambda$15960.1, $\lambda$16060.0, and $\lambda$16094.8~\AA{}, show strong systematic trends with $T_{\rm eff}$ for the dwarfs. We therefore exclude these features from the present analysis and use only five Si~I lines at $\lambda$15361.2, $\lambda$15376.8, $\lambda$16215.7, $\lambda$16680.8, and $\lambda$16828.2~\AA{}.

We find strong evidence for atomic diffusion affecting silicon abundances in the Diff stars of NGC~752, with $\Delta({\rm Diff-CS}) = -0.06$ dex at a significance level of $2.1\sigma$. The CS, MS, MS-TO, and RG stars have similar abundances, with $\langle$[Si/H]$\rangle$ = $0.02 \pm 0.03$, $-0.01 \pm 0.04$, $0.05 \pm 0.00$, and $-0.01 \pm 0.01$ dex, respectively. In contrast, the Diff stars show a systematically lower abundance, [Si/H] = $-0.04 \pm 0.07$ dex, as expected from atomic diffusion. This behavior is shown in Figure~\ref{fig:NGC752_teff_elems}. We also note that the Si abundances display a smaller typical scatter than the Mg and Al abundances discussed above.

A similar result is obtained for Ruprecht~147. For this cluster, we derive $\langle$[Si/H]$\rangle$ = $0.15 \pm 0.05$, $0.15 \pm 0.05$, $0.05 \pm 0.06$, $0.15 \pm 0.06$, and $0.17 \pm 0.00$ dex for the CS, MS, Diff, MS--TO, and RG stars, respectively. We obtain $\Delta({\rm Diff-CS}) = -0.10$ dex at a significance level of $1.8\sigma$, suggesting that atomic diffusion also affects the silicon abundances in Ruprecht~147.

\subsubsection{Sulfur}

For sulfur, we identify eight well-defined S~I lines in the APOGEE spectra, at $\lambda$15400.1, $\lambda$15402.3, $\lambda$15403.8, $\lambda$15406.0, $\lambda$15422.3, $\lambda$15469.8, $\lambda$15475.6, and $\lambda$15478.5~\AA{}. These features are stronger in warmer stars and become progressively weaker toward lower effective temperatures.

For NGC~752, we obtain $\langle$[S/H]$\rangle$ = $0.06 \pm 0.03$, $0.01 \pm 0.04$, $0.03 \pm 0.04$, $-0.05 \pm 0.00$, and $0.03 \pm 0.02$ dex for the CS, MS, Diff, MS--TO, and RG stars, respectively. The abundance difference between the Diff and CS stars is $\Delta({\rm Diff-CS}) = -0.03$ dex, significant at the $1.1\sigma$ level. As shown in Figure~\ref{fig:NGC752_teff_elems}, the overall sulfur abundance pattern follows the MIST isochrone for the cluster, with the lowest abundances occurring near the turnoff.

For Ruprecht~147, we obtain $\langle$[S/H]$\rangle$ = $0.18 \pm 0.01$, $0.17 \pm 0.02$, $0.09 \pm 0.03$, $0.16 \pm 0.04$, and $0.18 \pm 0.00$ dex for the CS, MS, Diff, MS--TO, and RG stars, respectively. The abundance difference between the Diff and CS stars is $\Delta({\rm Diff-CS}) = -0.09$ dex, significant at the $>3\sigma$ level. As shown in Figure~\ref{fig:RUP147_teff_elems}, for the two studied open clusters, the sulfur abundances display a smooth expected depletion toward the TO region, broadly consistent with the behavior predicted by the MIST diffusion models.

\subsubsection{Potassium}

Potassium abundances are derived from two K~I lines located near the beginning of the blue APOGEE detector, at $\lambda$15163.1 and $\lambda$15168.4~\AA{}. This spectral region can be affected by persistence effects, as discussed by \citet{holtzman_2018}. However, we find no evidence that persistence significantly affected the spectra analyzed in this work.

For NGC~752, we derive $\langle$[K/H]$\rangle$ = $-0.11 \pm 0.01$, $-0.11 \pm 0.03$, and $-0.10 \pm 0.04$ dex for the CS, MS, and RG stars, respectively. 
We are unable to derive reliable potassium abundances for the Diff and MS-TO stars because the K~I lines are too weak in the warmer stellar spectra; therefore, it is expected that our sample cannot properly probe the effects of diffusion for this element. 
Within the uncertainties, the potassium abundances remain nearly constant across the different evolutionary stages, showing no statistically significant abundance trend associated with atomic diffusion or extra mixing. As shown in Figure~\ref{fig:NGC752_teff_elems}, the [K/H] abundances exhibit relatively small scatter and remain broadly consistent with a constant abundance pattern across the cluster sequence.

For Ruprecht~147, on the other hand, the MIST model shown in Figure~\ref{fig:RUP147_teff_elems} indicates that the diffusion dip occurs at a cooler effective temperature of around 6250 K. 
We derive $\langle$[K/H]$\rangle$ = $0.08 \pm 0.01$, $0.12 \pm 0.04$, $0.04 \pm 0.04$, $0.12 \pm 0.04$, and $0.11 \pm 0.00$ dex for the CS, MS, Diff, MS--TO, and RG stars, respectively. 
The Diff stars exhibit systematically lower potassium abundances relative to the other evolutionary stages, with $\Delta({\rm Diff-CS}) = -0.04$ dex at a significance level of $1.3\sigma$. This is evidence of atomic diffusion operating in the potassium abundances of Ruprecht~147 stars. As shown in Figure~\ref{fig:RUP147_teff_elems}, the overall abundance pattern follows the expected depletion near the turnoff predicted by the MIST diffusion models, while the RG abundance returns to values comparable to those observed in the CS and MS stars.

\subsubsection{Calcium}

There are four well-defined Ca~I lines in the APOGEE spectra, at $\lambda$16136.8, $\lambda$16150.8, $\lambda$16155.236, and $\lambda$16157.4~\AA{}, which can be used to derive precise calcium abundances. These features are located in a relatively narrow spectral region. However, the Ca~I line at $\lambda$16155.236~\AA{} is often blended with other species and is excluded from our analysis.

For NGC~752, calcium shows one of the clearest abundance separations for the Diff stars. We obtain $\langle$[Ca/H]$\rangle$ = $0.04 \pm 0.03$, $0.01 \pm 0.05$, $-0.09 \pm 0.01$, and $0.01 \pm 0.03$ dex for the CS, MS, Diff, and RG stars, respectively. The Diff abundance is substantially lower than the CS abundance, yielding $\Delta({\rm Diff-CS}) = -0.13$ dex with a significance of $>3\sigma$. This strong offset indicates a pronounced calcium depletion at the 1.2 M$\odot$, consistent with the expected effect of atomic diffusion. 
In contrast, the RG abundance is comparable to the CS value, suggesting that the surface calcium depletion is largely erased after the deepening of the convective envelope during post-main-sequence evolution. Figure~\ref{fig:NGC752_teff_elems} shows this behavior, and we can see a group of Diff stars falling at the bottom of the MIST isochrone where we expect the stronger signature of atomic diffusion.

For the open cluster Ruprecht~147, the calcium abundances show a similar evolutionary pattern. We derive $\langle$[Ca/H]$\rangle$ = $0.18 \pm 0.01$, $0.15 \pm 0.05$, $0.10 \pm 0.03$, $0.11 \pm 0.05$, and $0.15 \pm 0.00$ dex for the CS, MS, Diff, MS--TO, and RG stars, respectively. 
The Diff stars are systematically more Ca-poor than the CS stars, giving $\Delta({\rm Diff-CS}) = -0.09$ dex at a significance level of $>3\sigma$. Although less significant than in NGC~752, this offset still supports the presence of atomic diffusion in Ruprecht~147, from the smooth agreement between the derived abundance pattern and the MIST isochrone, as shown in Figure~\ref{fig:RUP147_teff_elems}.

\subsubsection{Chromium}

Chromium abundances are derived from a single weak Cr~I line in the APOGEE spectra, at $\lambda$15680.1~\AA{}. 
For NGC~752, we obtain $\langle$[Cr/H]$\rangle$ = $0.06 \pm 0.04$, $0.02 \pm 0.09$, and $0.01 \pm 0.03$ dex for the CS, MS, and RG stars, respectively. 
The CS and MS abundances agree within the uncertainties, and the RG abundance is also consistent with the same abundance scale. We note that the typical scatter in Cr is generally larger than for several other species, likely because the abundance determination relies on a single weak Cr~I feature. Since reliable Cr abundances are unavailable for the Diff stars, where this line is too weak, we cannot directly evaluate the expected turnoff depletion of this element in NGC~752. Figure~\ref{fig:NGC752_teff_elems} shows that the Cr abundances are more scattered within the evolutionary stages, particularly among the MS stars.

For Ruprecht~147, as previously, the model diffusion dip is cooler, and chromium provides a strong indication of abundance depletion at the Diff stars. 
We derive $\langle$[Cr/H]$\rangle$ = $0.19 \pm 0.01$, $0.16 \pm 0.04$, $0.09 \pm 0.00$, $0.13 \pm 0.06$, and $0.16 \pm 0.00$ dex for the CS, MS, Diff, MS--TO, and RG stars, respectively. The Diff abundance is substantially lower than the CS value, yielding $\Delta({\rm Diff-CS}) = -0.10$ dex with a significance of $>3\sigma$. This large offset suggests pronounced chromium depletion in the Diff stars, consistent with the expected signature of atomic diffusion. The MS--TO stars remain closer to the MS abundance scale, as shown in Figure~\ref{fig:RUP147_teff_elems}.

\subsubsection{Manganese}

The APOGEE spectral coverage includes three Mn~I lines in the blue region of the detector, at $\lambda$15159.0, $\lambda$15217.8, and $\lambda$15262.4~\AA{}. As discussed for the potassium abundances, this spectral region can be affected by persistence. However, we do not find evidence that persistence significantly impacted the spectra analyzed in this work.

For NGC~752, manganese shows a clear decrease in abundance toward the TO region. We obtain $\langle$[Mn/H]$\rangle$ = $-0.15 \pm 0.02$, $-0.18 \pm 0.04$, $-0.23 \pm 0.06$, and $-0.16 \pm 0.03$ dex for the CS, MS, Diff, and RG stars, respectively. No MS-TO abundance could be derived for Mn. 
The Diff abundance is significantly lower than the CS abundance, yielding $\Delta({\rm Diff-CS}) = -0.08$ dex at the $1.8\sigma$ level. This offset suggests substantial manganese depletion in the Diff star, consistent with the expected signature of atomic diffusion. Figure~\ref{fig:NGC752_teff_elems} summarizes this behavior. 

For Ruprecht~147, the manganese abundances also show a clear depletion at the Diff point. We obtain $\langle$[Mn/H]$\rangle$ = $0.01 \pm 0.02$(CS), $0.02 \pm 0.06$ (MS), $-0.11 \pm 0.02$ (Diff), $-0.06 \pm 0.05$ (MS-TO), and $0.01 \pm 0.00$ dex (RG). 
The abundance difference between the Diff and CS stars is $\Delta({\rm Diff-CS}) = -0.12$ dex, significant at the $>3\sigma$ level. This offset indicates a strong manganese depletion in the Diff stars, consistent with the expected signature of atomic diffusion. The MS-TO stars show intermediate abundances, while the RG abundance is comparable to the MS value, suggesting that the diffusion signature is largely reduced after post-main-sequence evolution, as shown in Figure~\ref{fig:RUP147_teff_elems}.

\subsubsection{Titanium, vanadium, and cobalt}

Titanium, vanadium, and cobalt are grouped together in this analysis because we are unable to determine reliable abundances for the warmer stars in NGC 752, and for the CS sample in Ruprecht 147, where we derive abundances for one Diff and one RG star.
For Ti, six well-defined Ti~I lines are available in the APOGEE spectra. Following the recommendation of \citet{Grilo2024}, we use only the three lines that do not show systematic abundance trends with effective temperature, at $\lambda$15602.8, $\lambda$15699.0, and $\lambda$16635.2~\AA{}. The remaining Ti~I lines, at $\lambda$15334.8, $\lambda$15543.8, and $\lambda$15715.6~\AA{}, are excluded from this work. For V and Co, only one weak line is available in the APOGEE spectral coverage, at $\lambda$15924.9 and $\lambda$16757.7~\AA{}, respectively.

In NGC~752, we derive titanium abundances of $\langle$[Ti/H]$\rangle$ = $0.05 \pm 0.03$ and $0.01 \pm 0.02$ dex for the CS and RG stars, respectively. Titanium is a challenging element to study in APOGEE spectra because its abundances often show larger scatter and/or systematic offsets when compared with benchmark stars (see discussion in \citealt{Jonsson2020AJ....160..120J,Souto2016,souto_19}. 
In this work, however, we find a relatively small scatter for both the CS and RG samples, with standard deviations of 0.03 and 0.02 dex, respectively. This improvement may be related to our more restrictive line selection, in which we exclude Ti~I features known to display systematic abundance trends. 
For vanadium, both the CS and RG stars yield $\langle$[V/H]$\rangle$ = $-0.01$ dex, with standard deviations of 0.00 and 0.08 dex, respectively. Cobalt abundances could only be derived for the red giant stars, for which we obtain $\langle$[Co/H]$\rangle$ = $-0.11 \pm 0.02$ dex.

In Ruprecht~147, abundances for these elements could only be determined for the Diff and red giant stars, for which we obtain $\langle$[Ti/H]$\rangle$ = -0.05 and 0.13, [V/H] = -0.05 and 0.28, and $\langle$[Co/H]$\rangle$ = -0.05 and 0.14 dex, respectively 
These measurements are shown in Figure~\ref{fig:RUP147_teff_elems}, and if we consider the RG star to indicate the pristine cluster composition, we can indicate that atomic diffusion is operating in these elements.

\subsubsection{Nickel}

We used seven well-defined Ni~I lines ($\lambda$15605.7, $\lambda$15632.7, $\lambda$16584.4, $\lambda$16589.3, $\lambda$16673.7, $\lambda$16815.5, and $\lambda$16818.8~\AA{}) to derive nickel abundances from the APOGEE spectra of the sample stars.

For NGC~752, we were unable to derive Ni abundances for Diff stars to probe the signature of atomic diffusion. For the other stellar classes, we obtained $\langle$[Ni/H]$\rangle$ = $0.07 \pm 0.02$ (CS), $0.02 \pm 0.03$ (MS), and -$0.01 \pm 0.01$ (RG). The CS and MS abundances are consistent within the uncertainties, and the RG abundance is only slightly lower. As shown in Figure~\ref{fig:NGC752_teff_elems}, the Ni abundance distribution remains relatively flat across the available evolutionary stages, as we also lack MS-TO stars with higher $T_{\rm eff}$, suggesting that nickel does not provide a strong diagnostic of atomic diffusion in NGC~752 with the present sample.

For Ruprecht~147, on the other hand, the nickel abundances show a clear depletion in the Diff sample. We obtain $\langle$[Ni/H]$\rangle$ = $0.20 \pm 0.01$ (CS), $0.19 \pm 0.04$ (MS), $0.13 \pm 0.00$ (Diff), $0.16 \pm 0.06$ (MS-TO), and $0.19 \pm 0.00$ (RG). The Diff stars are systematically more Ni-poor than the CS stars, yielding $\Delta({\rm Diff-CS}) = -0.07$ dex at $>3\sigma$ significance. This offset suggests that nickel is also affected by atomic diffusion in Ruprecht~147. The MS and MS--TO stars show intermediate abundances, while the RG abundance is closer to the CS abundance scale, as shown in Figure~\ref{fig:RUP147_teff_elems}.

\subsubsection{Iron}

Several Fe~I lines are present in the APOGEE spectra. In this work, we analyzed 77 Fe~I features but excluded 25 lines from the final abundance determination, following the recommendation of \citet{Grilo2024}, to avoid transitions that exhibit systematic abundance trends with $T_{\rm eff}$. We note that for cooler stars such as late-K and M dwarfs, roughly 30 well-defined FeH lines can also serve as iron abundance diagnostics \citep{Souto2017ApJ...835..239S,Souto2018ApJ...860L..15S,Souto2026ApJ..1001L..28S}. However, these FeH transitions are not detected in the K dwarfs analyzed in this work.

For NGC~752, iron shows a clear depletion pattern in the Diff sample. We obtain $\langle$[Fe/H]$\rangle$ = $0.01 \pm 0.02$, $-0.01 \pm 0.03$, $-0.07 \pm 0.04$, $-0.07 \pm 0.00$, and $-0.02 \pm 0.02$ dex for the CS, MS, Diff, MS-TO, and RG stars, respectively. The Diff stars are systematically more Fe-poor than the CS stars, yielding $\Delta({\rm Diff-CS}) = -0.09$ dex at a significance level of $>3\sigma$. 
This strong offset provides one of the clearest signatures of atomic diffusion in NGC~752, as shown in Figure~\ref{fig:NGC752_teff_elems}, where the Diff sample falls right at the expected atomic diffusion model.

A similar conclusion is obtained for Ruprecht~147, as iron also shows a clear abundance depletion in the Diff sample. We obtain $\langle$[Fe/H]$\rangle$ = $0.17 \pm 0.00$ (CS), $0.15 \pm 0.03$ (MS), $0.08 \pm 0.01$ (Diff), $0.13 \pm 0.04$ (MS-TO), and $0.19 \pm 0.00$ dex (RG).
We obtain a $\Delta({\rm Diff-CS}) = -0.09$ dex at a significance level of $>3\sigma$. This strong offset provides robust evidence that atomic diffusion affects the surface iron abundances in Ruprecht~147. 
Figure~\ref{fig:RUP147_teff_elems} shows a clear trend where these Diff stars are right at the dip of atomic diffusion in Ruprecht~147.

\begin{figure*}
\epsscale{1.1}
\plottwo{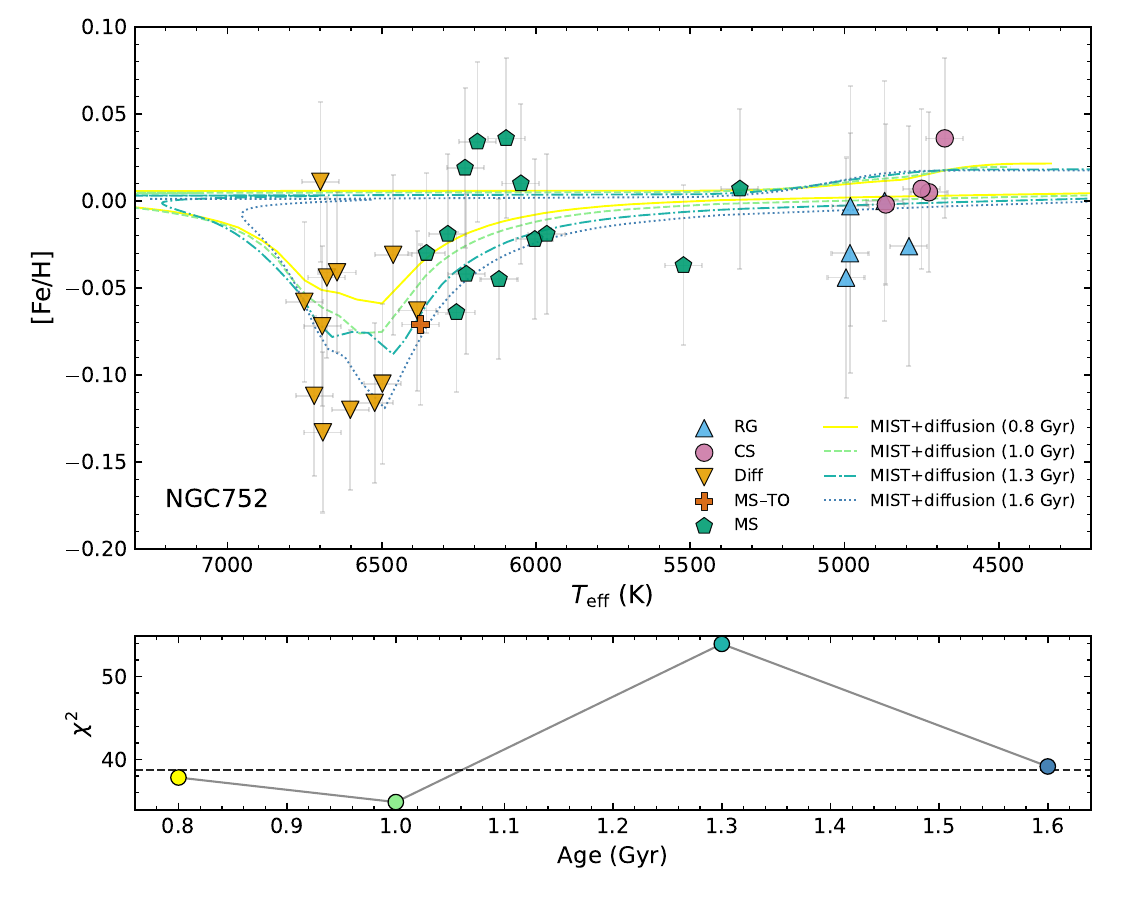}{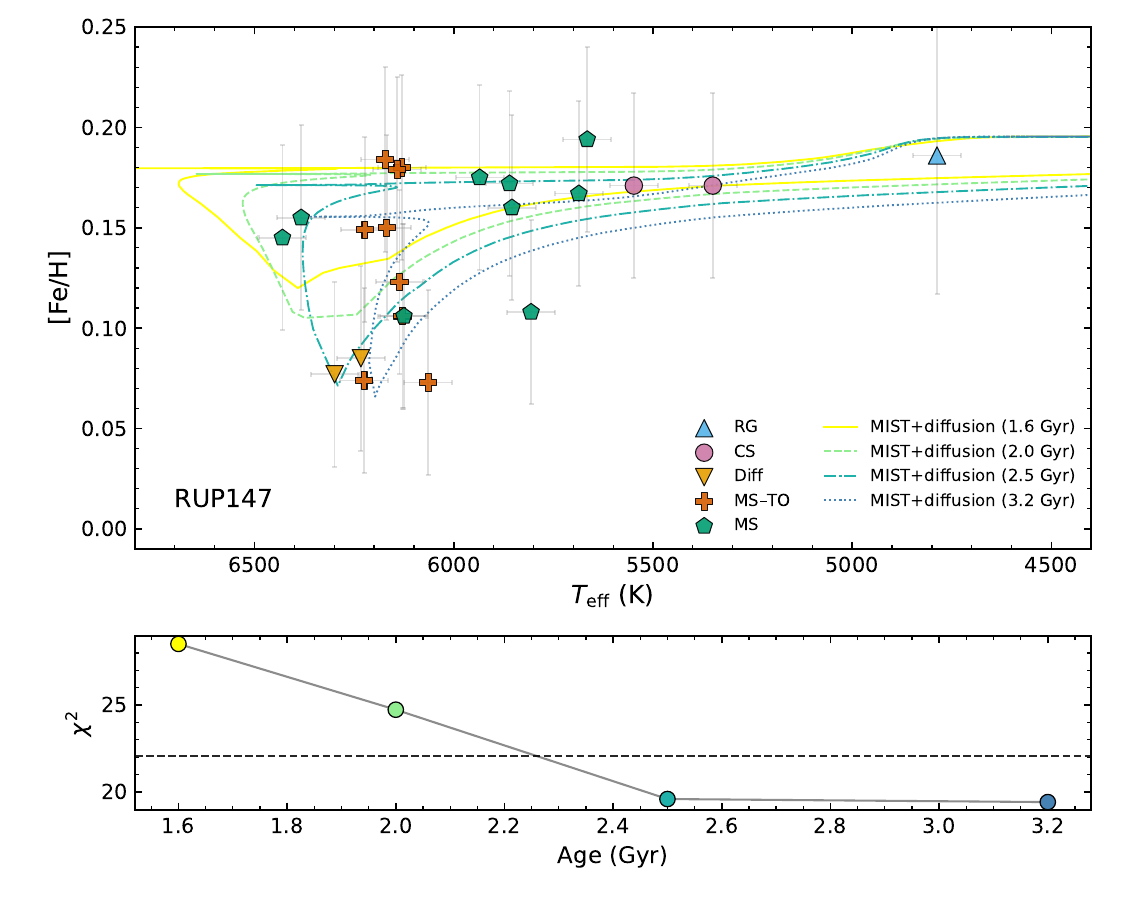}
\caption{{\it Top left}: Metallicity as a function of effective temperature for NGC 752. The solid and dashed lines represent MIST isochrones assuming [Fe/H]=-0.05 and ages = 0.8, 1.0, 1.3, 1.6 Gyr. {\it Bottom left}: A chi-squared minimization comparing the derived metallicities with MIST isochrones. The grey dashed line is a chi-squared comparison assuming the mean metallicity is constant. The color of the circle markers corresponds to the five models shown above, which are used to calculate their chi-squared value. The right panels are similar to the left panels, but for Ruprecht 147, where the MIST isochrones assume [Fe/H]=0.12 and ages = 1.6, 2.0, 2.5, 3.2 Gyrs.}
\label{fig:chi2}
\end{figure*}

We also performed a quantitative comparison between the observed [Fe/H] abundances derived in this study and the MIST isochrone predictions for different ages, using a chi-squared minimization approach. 
The results for NGC 752 and Ruprecht 147 are presented in Figure \ref{fig:chi2}, in the left and right panels, respectively.
The atomic diffusion models used for the comparison with the NGC 752 stars (top-left panel of Figure \ref{fig:chi2}) 
correspond to ages of 800 Myr, 1.0 Gyr, 1.3 Gyr, and 1.6 Gyr. For Ruprecht 147, the models (top-right panel of Figure \ref{fig:chi2}) cover ages of 1.6 Gyr, 2.0 Gyr, 2.5 Gyr, and 3.2 Gyr. 
The model points used in our $\chi^2$ comparison were carefully selected to match the observed stellar parameters ($T_{\rm eff}$, log $g$, and [Fe/H]) as closely as possible. This ensures that stars are compared at the correct stage of their evolution. Failure to constrain all three parameters can lead to incorrect matches between observed stars and model points, particularly since stars in different evolutionary phases (e.g., main sequence and red giant branch) may share similar effective temperatures and metallicities, as illustrated in Figures \ref{fig:NGC752_teff_elems} and \ref{fig:RUP147_teff_elems}.

The dashed gray horizontal lines in the bottom panels of Figure \ref{fig:chi2} represent the $\chi^2$ values obtained by assuming a constant mean [Fe/H] for all evolutionary phases: –0.03 dex for NGC 752 (left) and +0.14 dex for Ruprecht 147 (right). This baseline model assumes no variation in surface [Fe/H] with evolutionary phase, serving as a reference to assess whether atomic diffusion improves the agreement between observations and theory. 
To perform this comparison consistently, all metallicities were converted from logarithmic to linear space before computing the $\chi^2$ values.

For NGC~752, the MIST models including atomic diffusion provide a better description of the observed [Fe/H] pattern than a constant-abundance model. In particular, the diffusion models reproduce the abundance depression observed near $T{\rm eff}$ = 6500K, and the partial recovery toward cooler and more evolved stars. This is reflected in the lower $\chi^2$ values relative to the constant [Fe/H] model, with the best agreement obtained for the 1.0~Gyr isochrone. The improvement in $\chi^2$ indicates that the observed abundance variations are not simply random scatter around a single cluster metallicity, but instead trace a temperature-dependent pattern consistent with atomic diffusion.

A similar result is obtained for Ruprecht~147. In this case, the diffusion models also yield lower $\chi^2$ values than the constant-abundance model, with the best fit obtained for the 2.5--3.2~Gyr isochrone. The agreement between the observed depletion and the predicted diffusion pattern provides an independent way to probe the cluster's evolutionary state, and potentially its age, using chemical abundances alone. For comparison, photometric isochrone fitting based on {\it Gaia} data yields ages of 1.3~Gyr for NGC~752 and 2.5~Gyr for Ruprecht~147 \citep{cg20}, in good agreement with the ages favored by our abundance-based $\chi^2$ comparison.

These results suggest that atomic diffusion signatures in open clusters are not only useful for testing the physical processes operating during stellar evolution, but may also provide an additional chemical clock for constraining cluster ages. This is particularly valuable because the method is based on the relative abundance pattern along the cluster sequence, rather than on photometry alone. However, we emphasize that this approach depends on the precision of the abundance measurements, the adopted stellar models, and the treatment of additional mixing processes that can moderate the effects of atomic diffusion.

\subsection{Consequences of Diffusion}

Combining the results of this study for the open clusters NGC 752 and Ruprecht 147 with those, also from analyses of APOGEE spectra, from \citet{souto_2018} for the older M 67 open cluster (age $\sim$ 4.0 Gyr \citealt{Sarajedini2009ApJ...698.1872S}) and \citet{Souto2021_ComaBer} for the younger Coma Berenices open cluster (age $\sim$ 600 Myr, \citealt{Agueros2025ApJ...993..144A}), we find that atomic diffusion can have a significant impact on the measured surface abundances of stars near the main-sequence turnoff and along the subgiant branch.
The results from this work reinforce previous findings and the importance of considering atomic diffusion.

In particular, care should be taken when interpreting observed metallicities of stars around the turnoff without accounting for atomic diffusion. This omission may result in the selection of isochrones that are too metal-poor compared to the star’s birth composition, leading to a systematic overestimation of stellar ages.
The earlier conclusions of \citet{Vandenberg2002} indicated that neglecting atomic diffusion can bias age estimates by up to 10\%. As more clusters exhibit evidence of atomic diffusion signatures, these effects must be taken into account in both stellar modeling and age determinations, particularly for field stars.
\citet{Dotter2017} further showed that assuming a constant metallicity in isochrone fitting, rather than allowing for metallicity variations due to diffusion, can lead to age overestimates of up to 20\%. Failing to account for this process introduces a systematic offset in derived ages, which can propagate into errors in Galactic archaeology studies and reconstructions of the Milky Way’s formation history.

While this study does not yet provide a complete framework for correcting surface abundances back to their primordial values, it highlights the need for future work aimed at quantifying the impact of diffusion across a broad range of cluster ages and metallicities. Achieving this goal will require high-signal-to-noise, high-precision abundance measurements in open clusters with well-sampled evolutionary sequences, including stars on the main sequence, TO, SGB, and RGB. Such studies will be essential for mapping diffusion-induced abundance changes and recovering more accurate stellar ages in both cluster and field star populations.

\section{Conclusions}

In this work, we determine stellar parameters ($T_{\rm eff}$, $\log g$, $\xi$) and elemental abundances for Fe, C, N, Na, Mg, Al, Si, S, K, Ca, Ti, V, Cr, Mn, Co, and Ni for stars at different evolutionary stages in two open clusters: NGC~752 and Ruprecht~147. 
This work sample is designed to investigate whether there are signatures of atomic diffusion in these open clusters. 
For each cluster, we divide the sample into five evolutionary groups: the coolest stars in the sample (CS), the main-sequence (MS), the stars most likely to present an atomic diffusion signature based on their mass (Diff), the main-sequence--turnoff transition (MS--TO), and red giant (RG) stars.

Atomic diffusion has often been treated as a second-order effect in stellar abundance studies, but our results show that it produces measurable signatures in both NGC~752 and Ruprecht~147. For every element with reliable Diff abundances, the Diff stars are systematically depleted relative to the CS reference population, with offsets significantly above the $2\sigma$ level in both clusters. 
For example, averaging over all elements with abundances measured in both stellar classes off Diff and CS, we find $\langle \Delta({\rm Diff-CS})[X/H] \rangle = -0.08 \pm 0.01$ dex for NGC~752 and $-0.12 \pm 0.01$ dex for Ruprecht~147. In both clusters, these offsets are significant at more than the $3\sigma$ level, confirming that stars in the diffusion-dominated regime are systematically depleted relative to the cool-star reference population.
We also find evidence for post-main-sequence mixing in the carbon and sodium abundances of red giant stars. Carbon is especially diagnostic, as it traces both the depletion caused by atomic diffusion and the abundance changes produced by first dredge-up and extra mixing in evolved stars. These results demonstrate that stellar evolution effects must be accounted for when interpreting precise chemical abundances in open clusters.

Using the atomic diffusion signature in the Fe abundances, we modeled the $T_{\rm eff}$--[Fe/H] distribution with MIST isochrones, including diffusion over a range of cluster ages. A $\chi^2$ comparison between the observed abundances and the model predictions gives the best agreement for the 1.0~Gyr isochrone in NGC~752 and the 2.5--3.2~Gyr isochrone in Ruprecht~147. These age estimates are based solely on the stellar chemical abundance patterns and are in excellent agreement with the cluster ages derived from {\it Gaia} photometry by \citet{cg20}. This result suggests that atomic diffusion signatures in open clusters are not only valuable probes of the physical processes operating during stellar evolution, but may also provide an independent chemical clock for constraining cluster ages.

With the advent of ESA’s {\em Gaia} mission, we are now able to determine precise absolute magnitudes for stars near the main-sequence turnoff — precisely where the effects of atomic diffusion are expected to be strongest. Ignoring atomic diffusion in this regime leads to systematically underestimated metallicities and, consequently, biased age determinations. As large-scale spectroscopic surveys increasingly target these evolutionary phases \citep[e.g., GALAH;][]{Hayden2022_GALAHchemclock}, neglecting the effects of atomic diffusion will introduce systematics that may compromise our understanding of Galactic evolution.
A broader sample of open clusters with full evolutionary coverage (including the main sequence, turnoff, subgiant branch, and red giant branch) is essential to calibrate the surface abundance variations induced by atomic diffusion. We aim to address this need in future work by expanding the number and diversity of open clusters analyzed.


\begin{acknowledgments}
We thank the referee for the helpful suggestions that improved the paper. D.S. acknowledges support from the Foundation for Research and Technological Innovation Support of the State of Sergipe (FAPITEC/SE) and the National Council for Scientific and Technological Development (CNPq), under grant numbers 404056/2021-0, 794017/2013, and 444372/2024-5.
TS, PMF, KC, AIW, and NM acknowledge support for this research from the National Science Foundation Astronomy and Astrophysics grants AST-1715662, AST-2206541, and AST-2206543. 
PMF and JT acknowledge this work was performed at the Aspen Center for Physics, which is supported by National Science Foundation grant PHY-1607611.

Funding for the Sloan Digital Sky 
Survey IV has been provided by the 
Alfred P. Sloan Foundation, the U.S. 
Department of Energy Office of 
Science, and the Participating 
Institutions. 

SDSS-IV acknowledges support and 
resources from the Center for High 
Performance Computing  at the 
University of Utah. The SDSS 
website is www.sdss.org.

SDSS-IV is managed by the 
Astrophysical Research Consortium 
for the Participating Institutions 
of the SDSS Collaboration including 
the Brazilian Participation Group, 
the Carnegie Institution for Science, 
Carnegie Mellon University, Center for 
Astrophysics | Harvard \& 
Smithsonian, the Chilean Participation 
Group, the French Participation Group, 
Instituto de Astrof\'isica de 
Canarias, The Johns Hopkins 
University, Kavli Institute for the 
Physics and Mathematics of the 
Universe (IPMU) / University of 
Tokyo, the Korean Participation Group, 
Lawrence Berkeley National Laboratory, 
Leibniz Institut f\"ur Astrophysik 
Potsdam (AIP),  Max-Planck-Institut 
f\"ur Astronomie (MPIA Heidelberg), 
Max-Planck-Institut f\"ur 
Astrophysik (MPA Garching), 
Max-Planck-Institut f\"ur 
Extraterrestrische Physik (MPE), 
National Astronomical Observatories of 
China, New Mexico State University, 
New York University, University of 
Notre Dame, Observat\'ario 
Nacional / MCTI, The Ohio State 
University, Pennsylvania State 
University, Shanghai 
Astronomical Observatory, United 
Kingdom Participation Group, 
Universidad Nacional Aut\'onoma 
de M\'exico, University of Arizona, 
University of Colorado Boulder, 
University of Oxford, University of 
Portsmouth, University of Utah, 
University of Virginia, University 
of Washington, University of 
Wisconsin, Vanderbilt University, 
and Yale University.

Funding for the Sloan Digital Sky Survey IV has been provided by the Alfred P. Sloan Foundation, the U.S. Department of Energy Office of Science, and the Participating Institutions. SDSS-IV acknowledges support and resources from the Center for High-Performance Computing at the University of Utah. The SDSS website is www.sdss.org.
SDSS-IV is managed by the Astrophysical Research consortium for the Participating Institutions of the SDSS Collaboration including the Brazilian Participation Group, the Carnegie Institution for Science, Carnegie Mellon University, the Chilean Participation Group, the French Participation Group, Harvard-Smithsonian Center for Astrophysics, Instituto de Astrof\'isica de Canarias, The Johns Hopkins University, Kavli Institute for the Physics and Mathematics of the Universe (IPMU) /  University of Tokyo, Lawrence Berkeley National Laboratory, Leibniz Institut f\"ur Astrophysik Potsdam (AIP),  Max-Planck-Institut f\"ur Astronomie (MPIA Heidelberg), Max-Planck-Institut f\"ur Astrophysik (MPA Garching), Max-Planck-Institut f\"ur Extraterrestrische Physik (MPE), National Astronomical Observatory of China, New Mexico State University, New York University, University of Notre Dame, Observat\'orio Nacional / MCTI, The Ohio State University, Pennsylvania State University, Shanghai Astronomical Observatory, United Kingdom Participation Group, Universidad Nacional Aut\'onoma de M\'exico, University of Arizona, University of Colorado Boulder, University of Oxford, University of Portsmouth, University of Utah, University of Virginia, University of Washington, University of Wisconsin, Vanderbilt University, and Yale University.

\end{acknowledgments}

This work has made use of data from the European Space Agency (ESA) mission {\it Gaia} (\url{https://www.cosmos.esa.int/gaia}), processed by the {\it Gaia} Data Processing and Analysis Consortium (DPAC, \url{https://www.cosmos.esa.int/web/gaia/dpac/consortium}). Funding for the DPAC has been provided by national institutions, in particular the institutions participating in the {\it Gaia} Multilateral Agreement.

This research made use of Astropy, a community-developed core Python package for Astronomy (Astropy Collaboration, 2018).

%

\vspace{5mm}
\facilities{Du Pont (APOGEE), Sloan (APOGEE), Gaia}


\software{BACCHUS (\citealt{Masseron2016}), Turbospectrum (\citealt{AlvarezPlez1998}; \citealt{turbospec}), Astropy \citep{2013A&A...558A..33A,2018AJ....156..123A, astropy:2022}, Numpy \citep{numpy}, Matplotlib \citep{matplotlib}, Scipy \citep{scipy}.
}

\bibliography{sample7}{}
\bibliographystyle{aasjournalv7}



\appendix
\section{Comparison of the effective temperature}\label{app:teff}

As discussed in Section~\ref{sec:atm_par}, the effective temperatures adopted in this work are derived from the well-established photometric calibrations of \citet{GHB2009_PhotCal}, using five color indices: $B-V$, $V-J$, $V-H$, $V-K_s$, and $J-K_s$. In this section, we provide a one-to-one comparison between our adopted $T_{\rm eff}$ scale and several alternative temperature estimates: the \citet{GHB2009_PhotCal} calibration using only the $J-K_s$ color, the APOGEE DR17 temperatures \citep{dr17}, and the APOGEE DR19 temperatures \citep{DR19_2025arXiv250707093S}.

\begin{figure*}
    \centering
   { \includegraphics[width=0.95\textwidth]{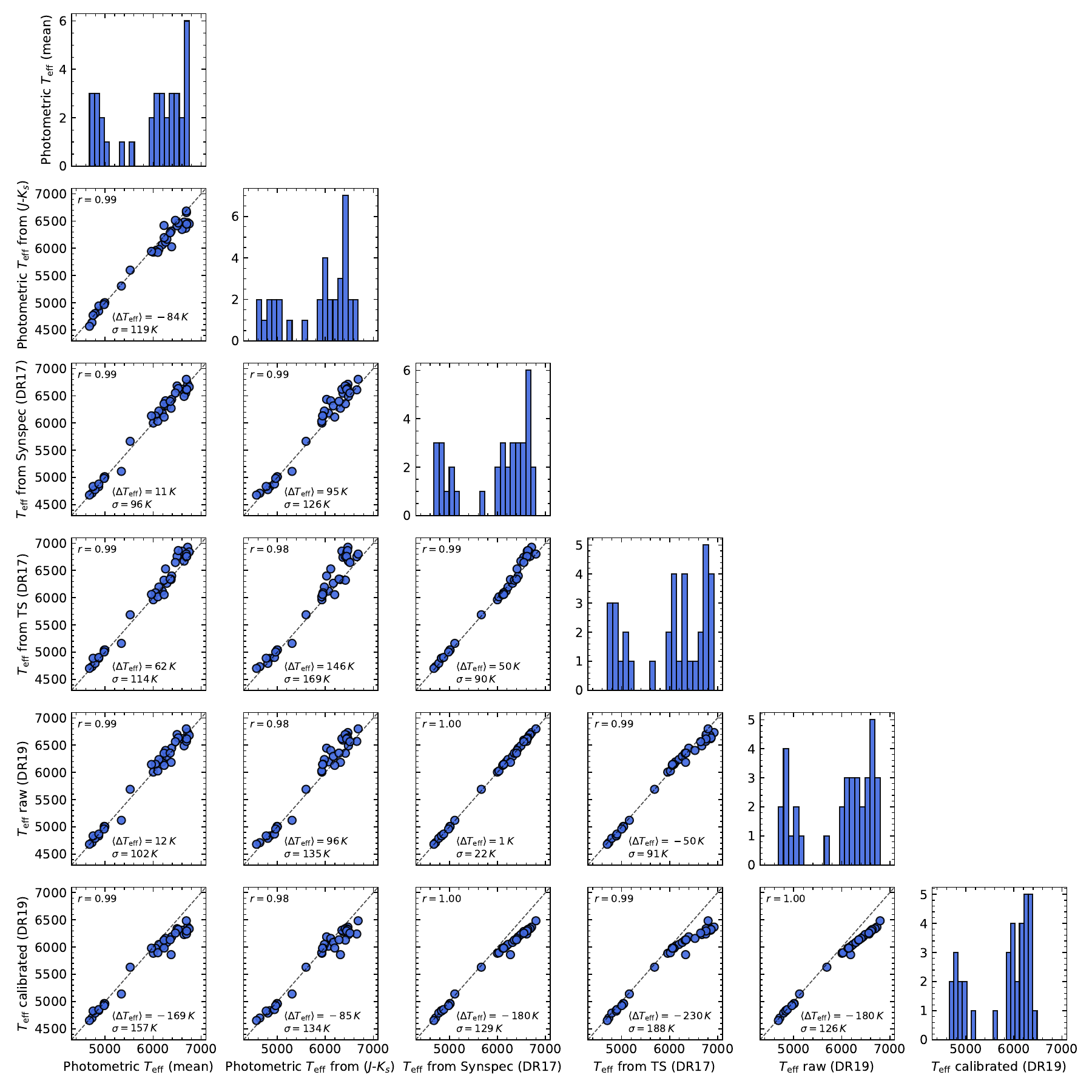}}
    \caption{
    Comparison between the different effective temperature scales considered for NGC~752. The diagonal panels show the temperature distributions for each scale, while the lower panels show the pairwise comparisons. The temperature scales include the photometric $T_{\rm eff}$ adopted in this work, computed as the mean of five color calibrations; the photometric $T_{\rm eff}$ derived using only the $J-K_s$ color; the raw DR17 Synspec and Turbospectrum temperatures; and the raw and calibrated DR19 temperatures. The dashed line in each comparison panel indicates the one-to-one relation. Each panel also reports the Pearson correlation coefficient, $r$, the mean temperature difference, $\langle \Delta T_{\rm eff} \rangle$, and the corresponding dispersion, $\sigma$. Overall, the different temperature scales are strongly correlated, with small systematic offsets between them.}
    \label{fig:teff_comp_NGC752}
\end{figure*}

\begin{figure*}
    \centering
   { \includegraphics[width=0.95\textwidth]{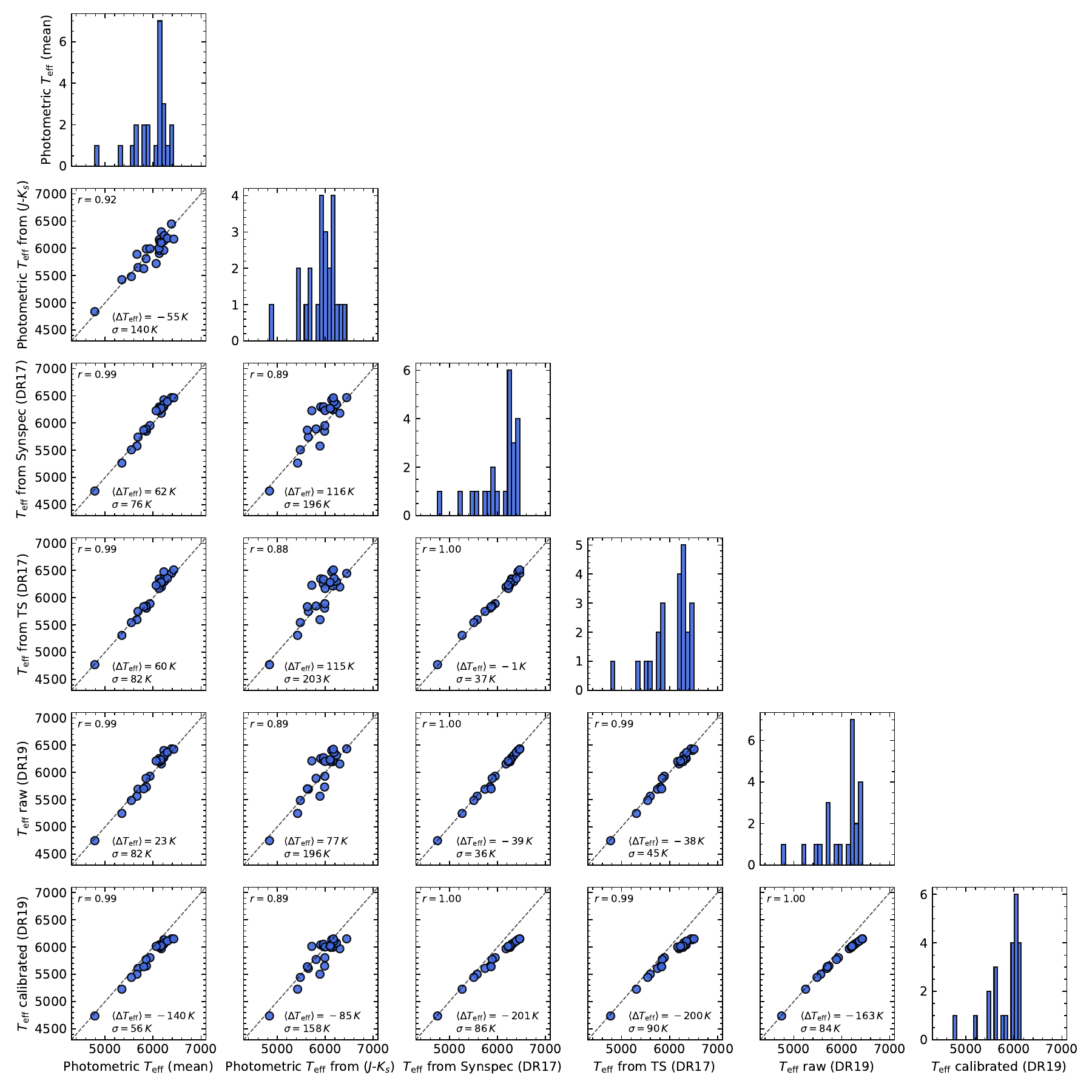}}
    \caption{Same as Figure \ref{fig:teff_comp_NGC752}, but for Ruprecht 147.}
    \label{fig:teff_comp_Rup147}
\end{figure*}

Figures~\ref{fig:teff_comp_NGC752} and~\ref{fig:teff_comp_Rup147} present a comparison among the effective-temperature scales used or considered in this work for the NGC~752 and Ruprecht~147 open clusters, respectively. These include the photometric temperatures derived from different color combinations, the raw DR17 Synspec and Turbospectrum estimates obtained directly from spectral fitting, the raw DR19 effective temperatures, and the calibrated DR19 effective temperatures. The diagonal panels show the temperature distributions for each scale, while the lower panels show the pairwise comparisons. In each comparison panel, the dashed line marks the one-to-one relation. We also report the Pearson correlation coefficient, the mean temperature difference, and the dispersion around this difference.

Overall, the different temperature scales are strongly correlated, with most comparisons yielding correlation coefficients close to unity. The photometric effective temperatures adopted in this work are generally consistent with the spectroscopic and calibrated APOGEE temperature scales, although small systematic offsets are present. These offsets are typically of order $\sim$50--150~K, depending on the comparison scale, and are comparable to the expected uncertainties in the temperature determinations. 
The typical effective temperature uncertainty is about 50--70 K, comparable to the marker sizes in the figure. The raw DR19 and DR17 Synspec effective temperatures agree most closely with our adopted photometric scale, whereas the calibrated DR19 temperatures and the $J-K_s$-based photometric effective temperatures show the largest offsets. We also note that the raw DR17 Synspec and raw DR19 temperatures yield virtually the same $T_{\rm eff}$ scale for NGC~752.

\clearpage
\newpage
\section{Comparison with automated abundance pipelines}\label{app:abundances}

This appendix presents abundance comparisons between the values derived in this work and those available from APOGEE DR19, APOGEE DR17 Turbospectrum, and APOGEE DR17 Synspec. The presented figures are intended to quantify possible zero-point offsets and systematic differences among abundance scales, and to verify whether the abundance trends discussed in the main text are specific to our analysis or also present in independent APOGEE determinations.

For all Figures in the Appendix, we show a corner-plot comparison of [X/H] abundances for NGC~752 (top panel) and Ruprecht (bottom panel) derived in this work and from different APOGEE abundance scales. The diagonal panels show the abundance distributions for each source: this work, APOGEE DR19, APOGEE DR17 Turbospectrum (TS), and APOGEE DR17 Synspec. The numbers in each diagonal panel indicate the mean abundance and standard deviation for the full sample and for each evolutionary stage studied. The value of $\Delta({\rm Diff-CS})$ gives the abundance difference between the stars most likely to have atomic diffusion depletion and those at the cooler end of the main-sequence population studied in this work, and $S$ gives the significance of this difference in units of $\sigma$.

For each element, we quantify the abundance difference between the sample most likely to present an atomic diffusion signature (Diff) and the coolest stellar sample (CS) in what we assume to have the pristine cluster composition as
\begin{equation}
\Delta({\rm Diff-CS}) = \mu_{\rm Diff} - \mu_{\rm CS},
\end{equation}
where $\mu_{\rm Diff}$ and $\mu_{\rm CS}$ represent the mean abundances. The uncertainty on this difference is estimated from the dispersion of each evolutionary group as
\begin{equation}
\sigma_{\Delta} =
\left(
\frac{\sigma_{\rm TO}^{2}}{N_{\rm TO}} +
\frac{\sigma_{\rm MS}^{2}}{N_{\rm MS}}
\right)^{1/2},
\end{equation}
where $\sigma_{\rm Diff}$ and $\sigma_{\rm CS}$ are the abundance dispersions within the samples, and $N_{\rm Diff}$ and $N_{\rm CS}$ are the corresponding number of stars. We then define the significance of the abundance offset as
\begin{equation}
S = \frac{\Delta({\rm Diff-CS})}{\sigma_{\Delta}}.
\end{equation}
Negative values of $\Delta({\rm Diff-CS})$ indicate that the Diff stars are depleted relative to the CS reference population. We consider abundance differences with $|S| > 1$ statistically significant (\textit{likely} in the Figures), and those with $|S| > 2$ as strong evidence for a depletion signature (\textit{diff} in the Figures).

The lower panels of this appendix figure show the pairwise abundance comparisons between different sources. The dashed line indicates the one-to-one relation, and the inset text gives the mean abundance offset, $\langle \Delta{\rm [Fe/H]} \rangle$, and standard deviation for all stars and for each evolutionary group. Symbols indicate the evolutionary stage: cyan upwards triangles are red giants (RG), pink circles are the cooler main-sequence (CS), gold inverted triangles are stars at the expected atomic diffusion dip (Diff), red pluses are the stars close to the turnoff region (MS-TO), and the green pentagon are main-sequence stars with masses close to the Sun (MS).

\begin{figure*}
    \centering
   { \includegraphics[width=0.65\textwidth]{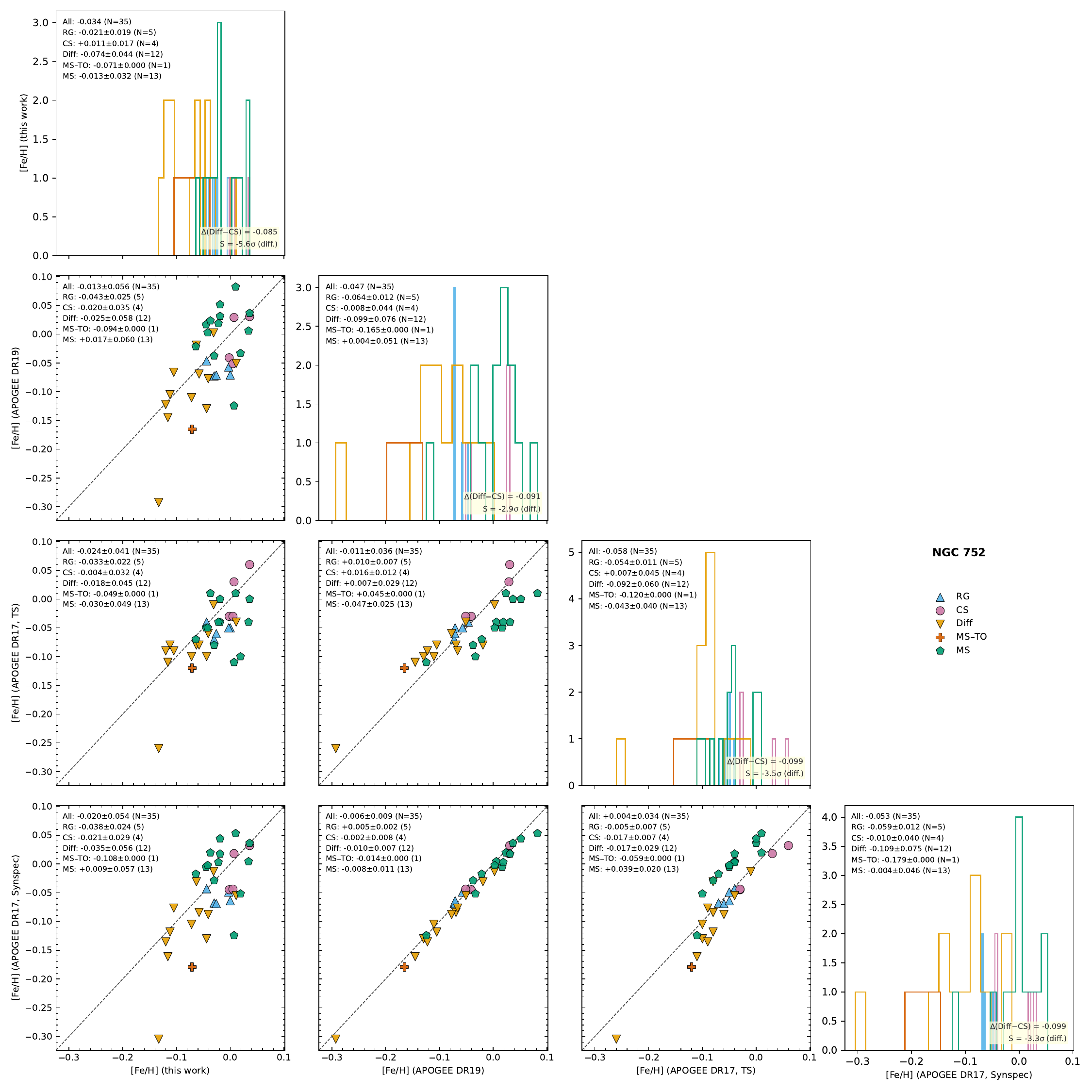}}
   { \includegraphics[width=0.65\textwidth]{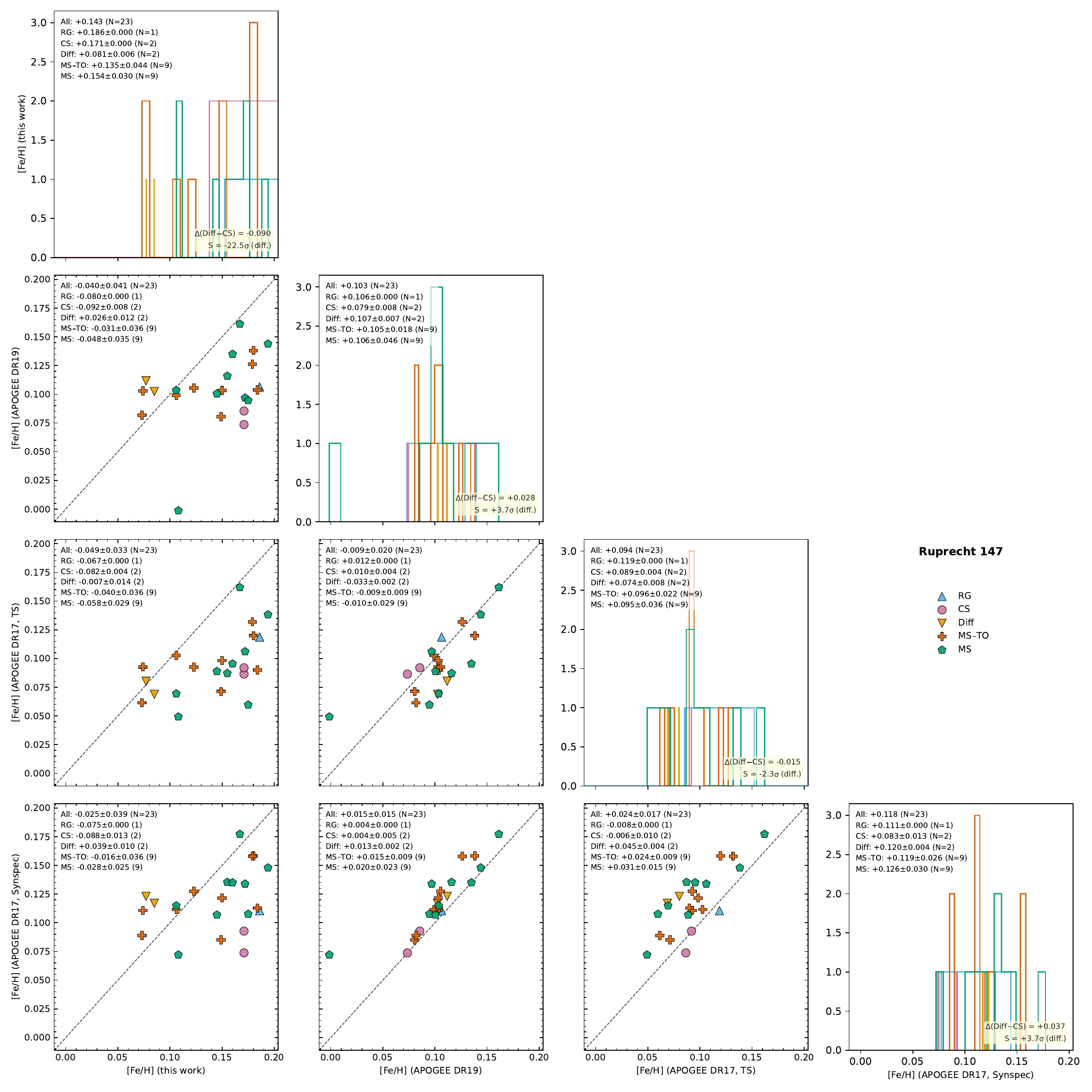}}
   \caption{
    Top panel: Abundance scale comparison for Fe in the NGC 752 open cluster. Bottom panel: the same as top, but for Ruprecht 147.}
   \label{AP_abu_Fe}
\end{figure*}

\begin{figure*}
    \centering
   { \includegraphics[width=0.65\textwidth]{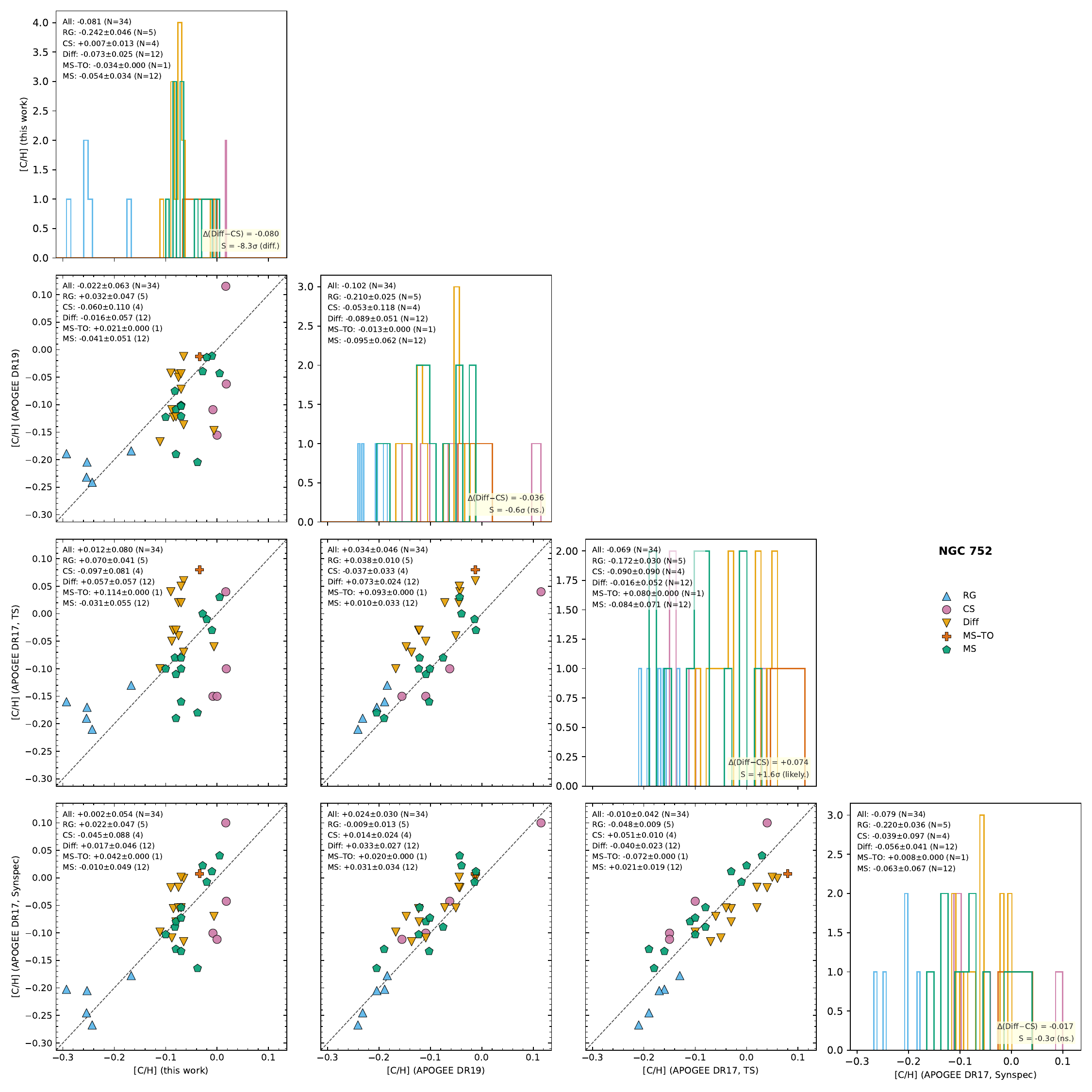}}
   { \includegraphics[width=0.65\textwidth]{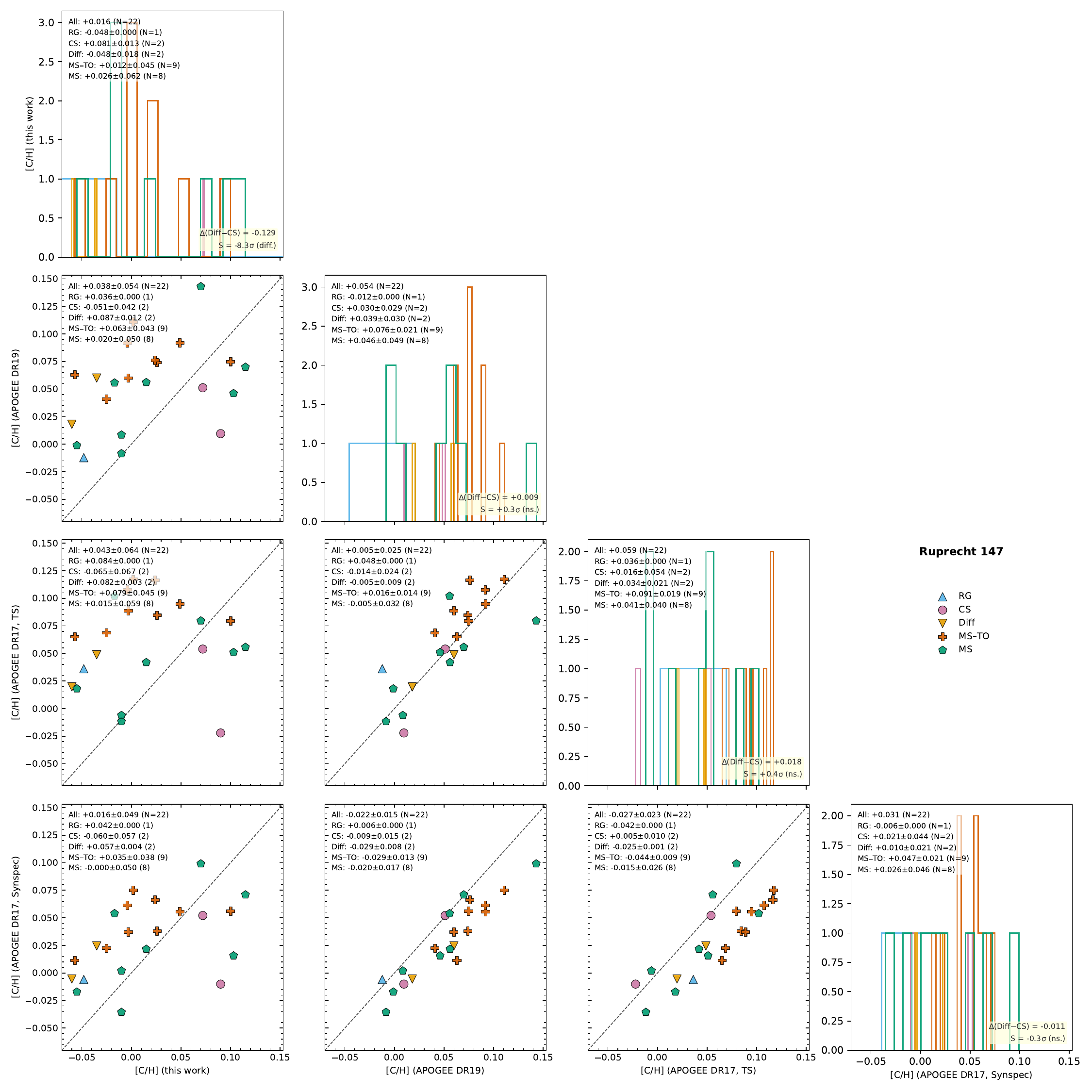}}
   \caption{Same as Figure \ref{AP_abu_Fe}, but for C.}
   \label{AP_abu_C}
\end{figure*}

\begin{figure*}
    \centering
   { \includegraphics[width=0.65\textwidth]{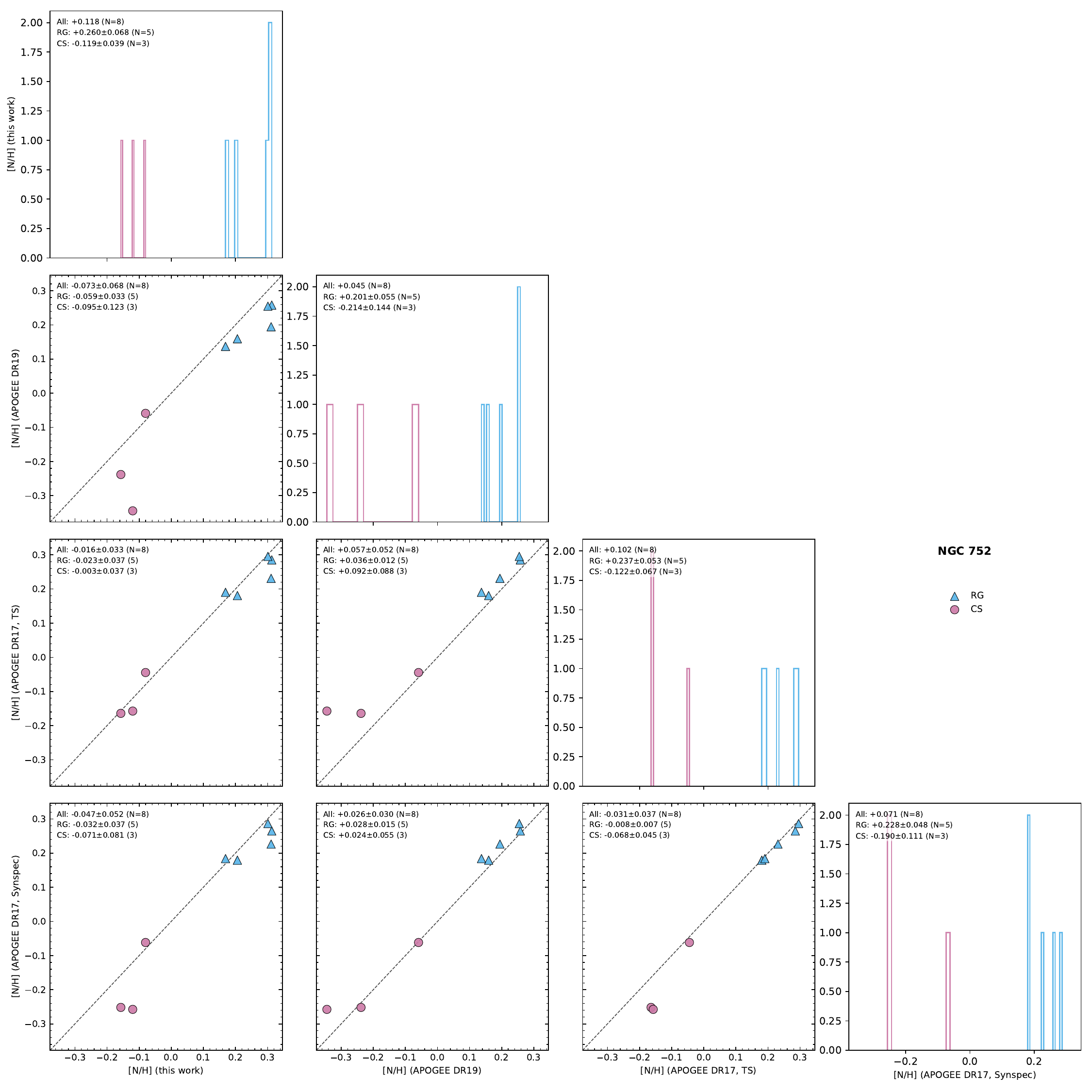}}
   { \includegraphics[width=0.65\textwidth]{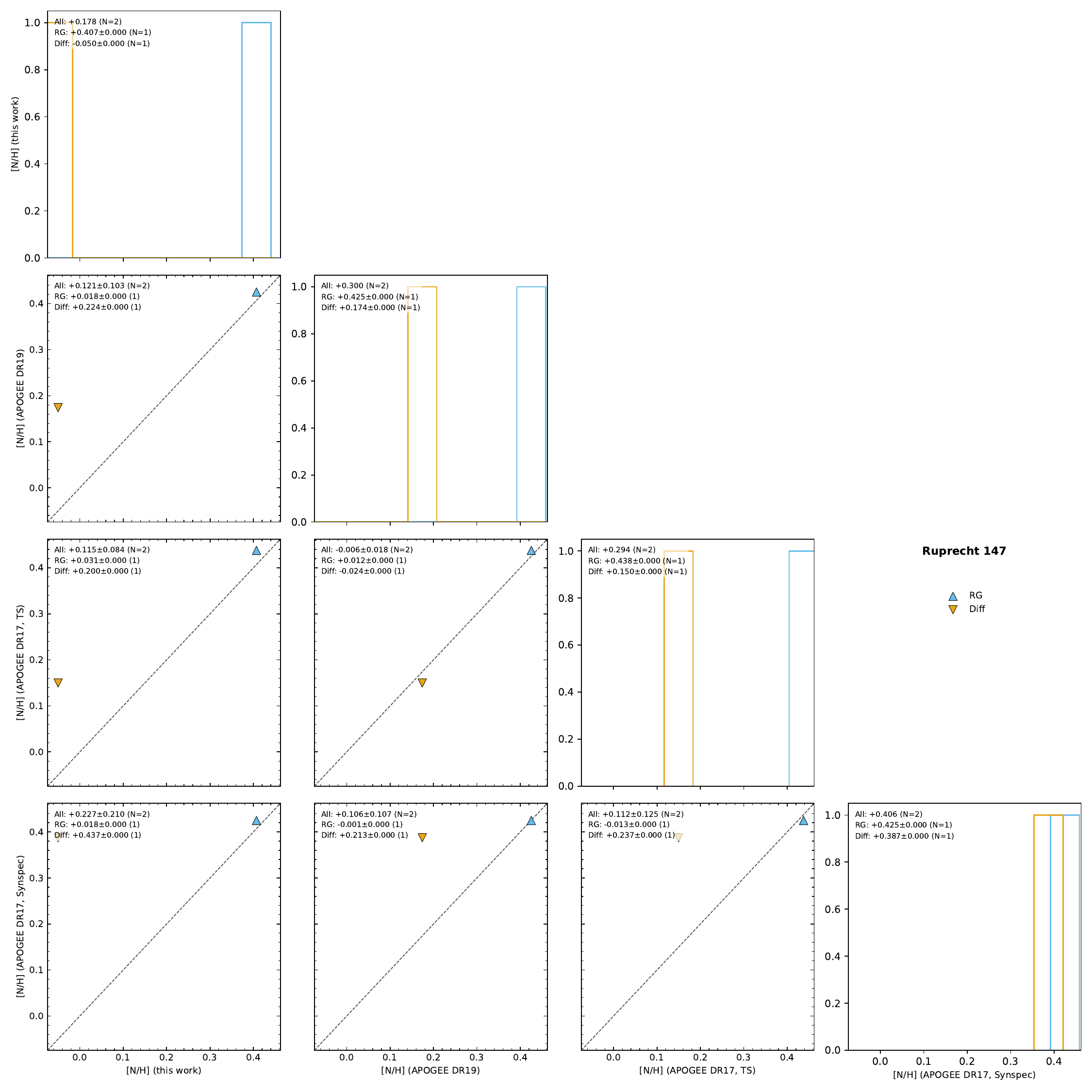}}   
   \caption{Same as Figure \ref{AP_abu_Fe}, but for N. }
   \label{AP_abu_N}
\end{figure*}

\begin{figure*}
    \centering
   { \includegraphics[width=0.65\textwidth]{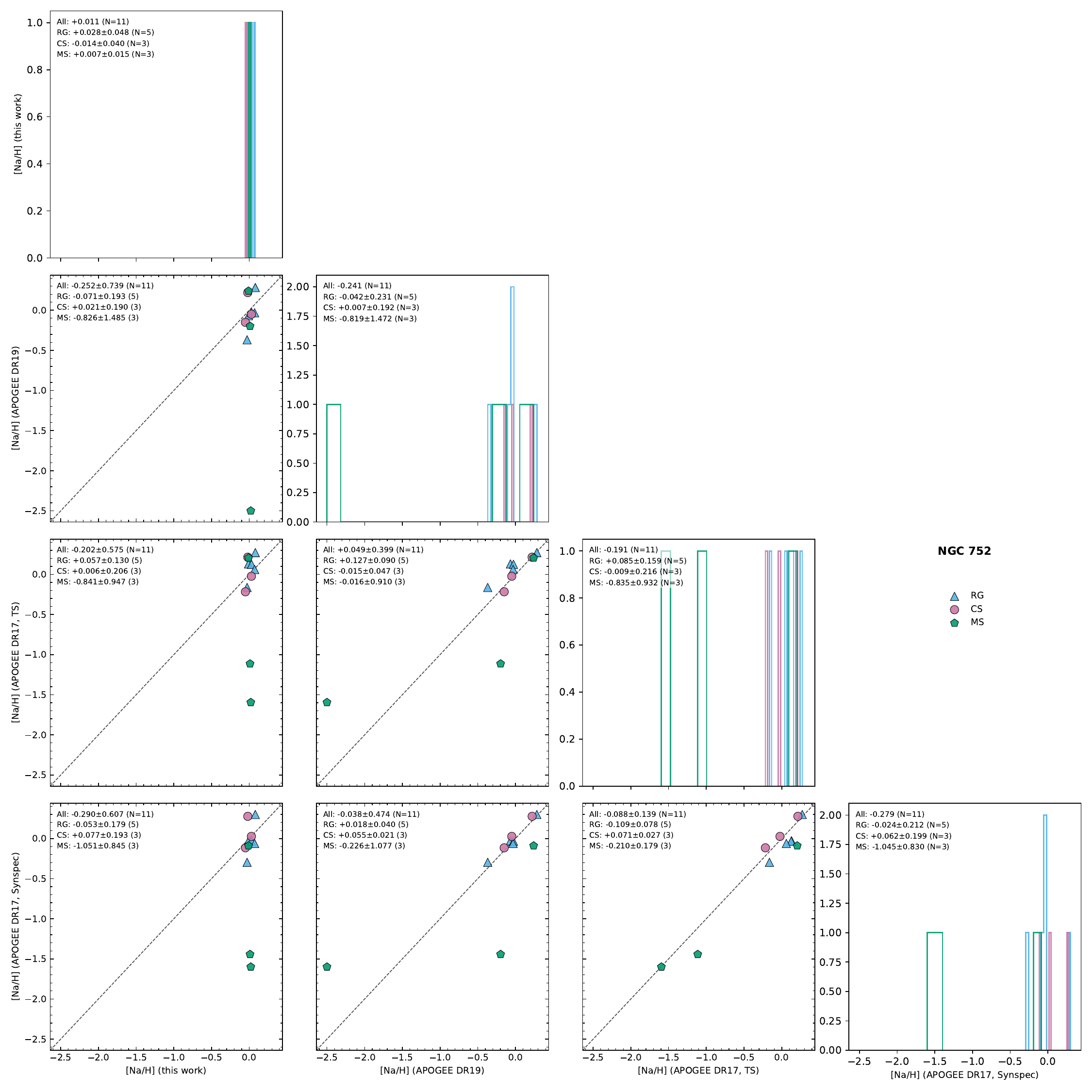}}
   { \includegraphics[width=0.65\textwidth]{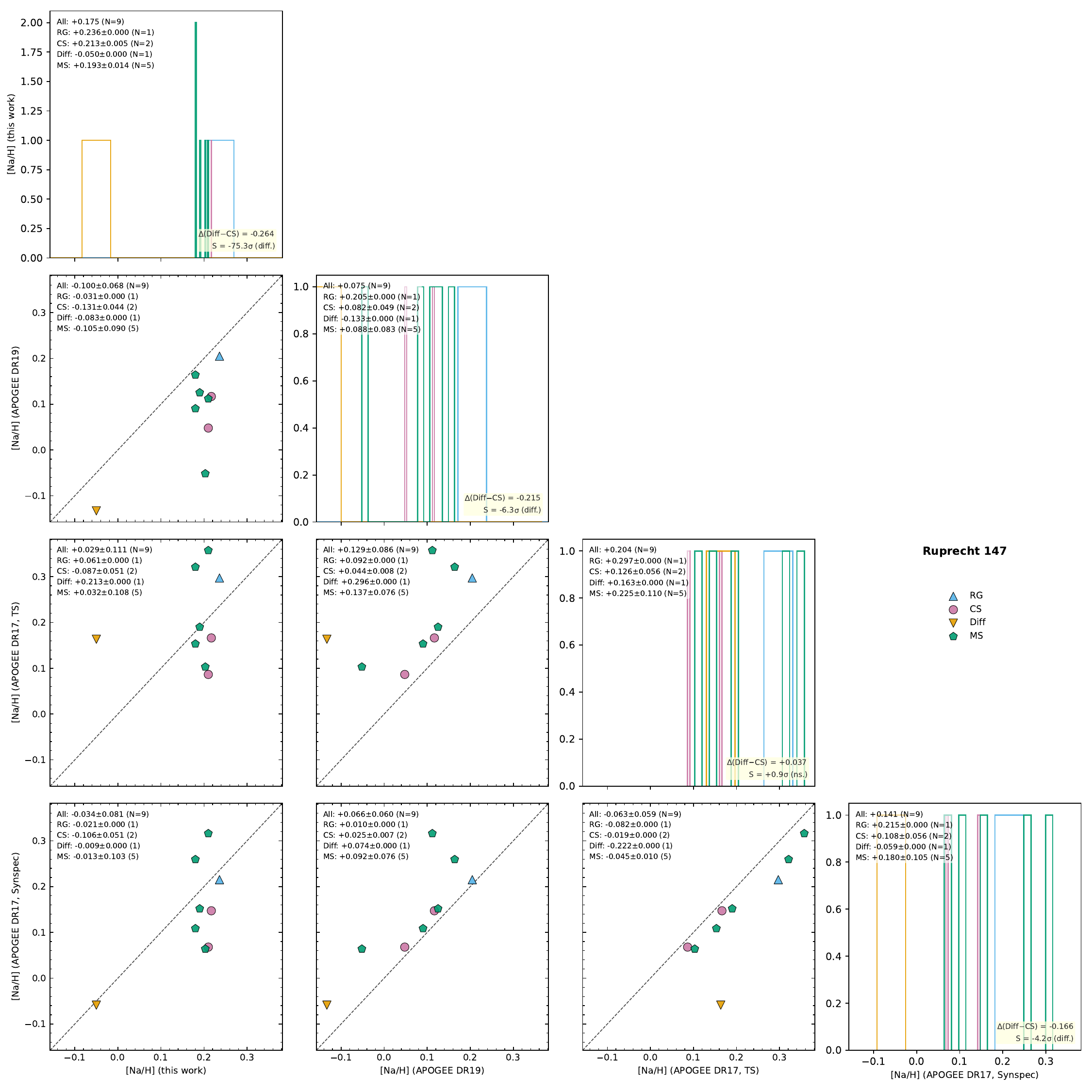}}
   \caption{Same as Figure \ref{AP_abu_Fe}, but for Na.}
   \label{AP_abu_Na}
\end{figure*}

\begin{figure*}
    \centering
   { \includegraphics[width=0.65\textwidth]{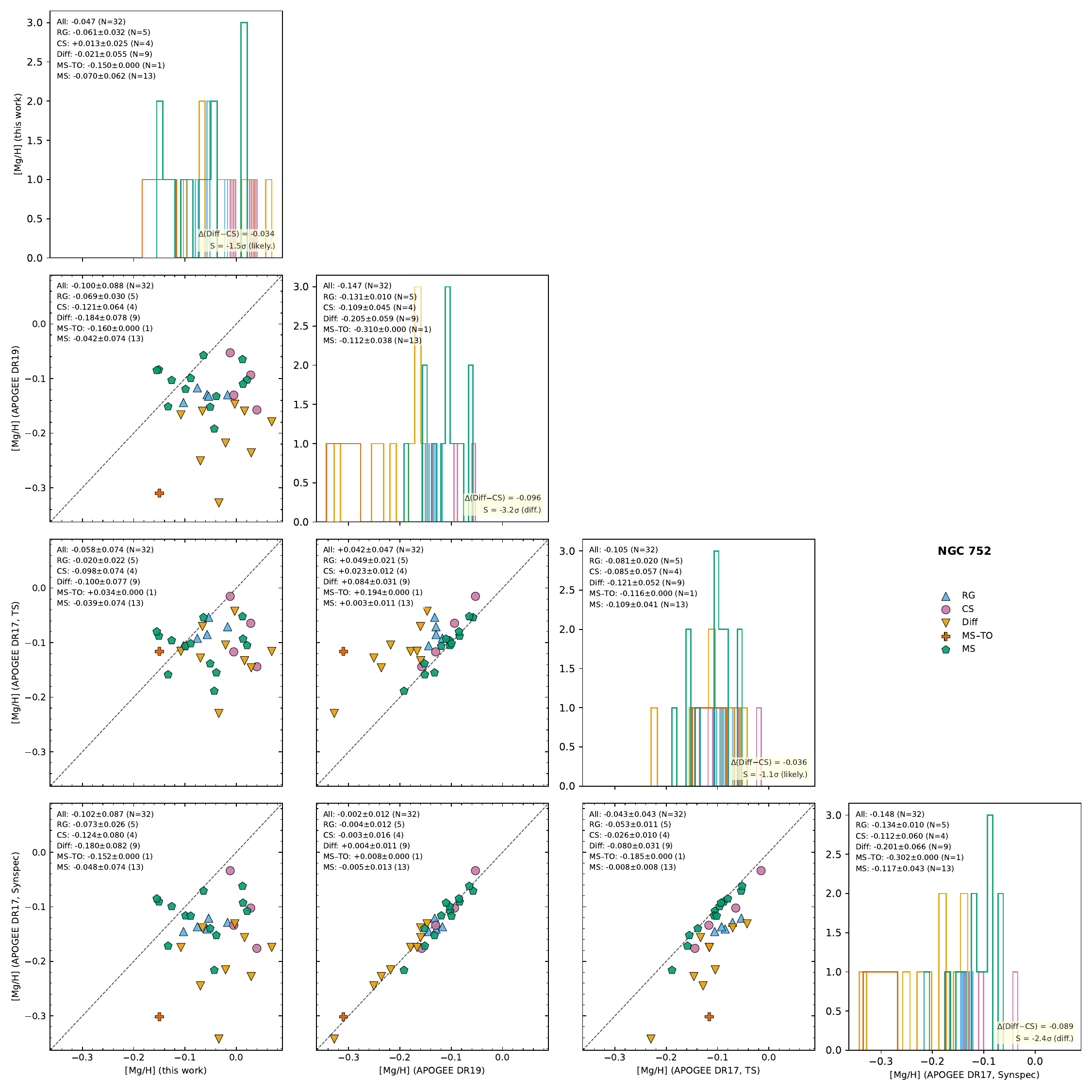}}
   { \includegraphics[width=0.65\textwidth]{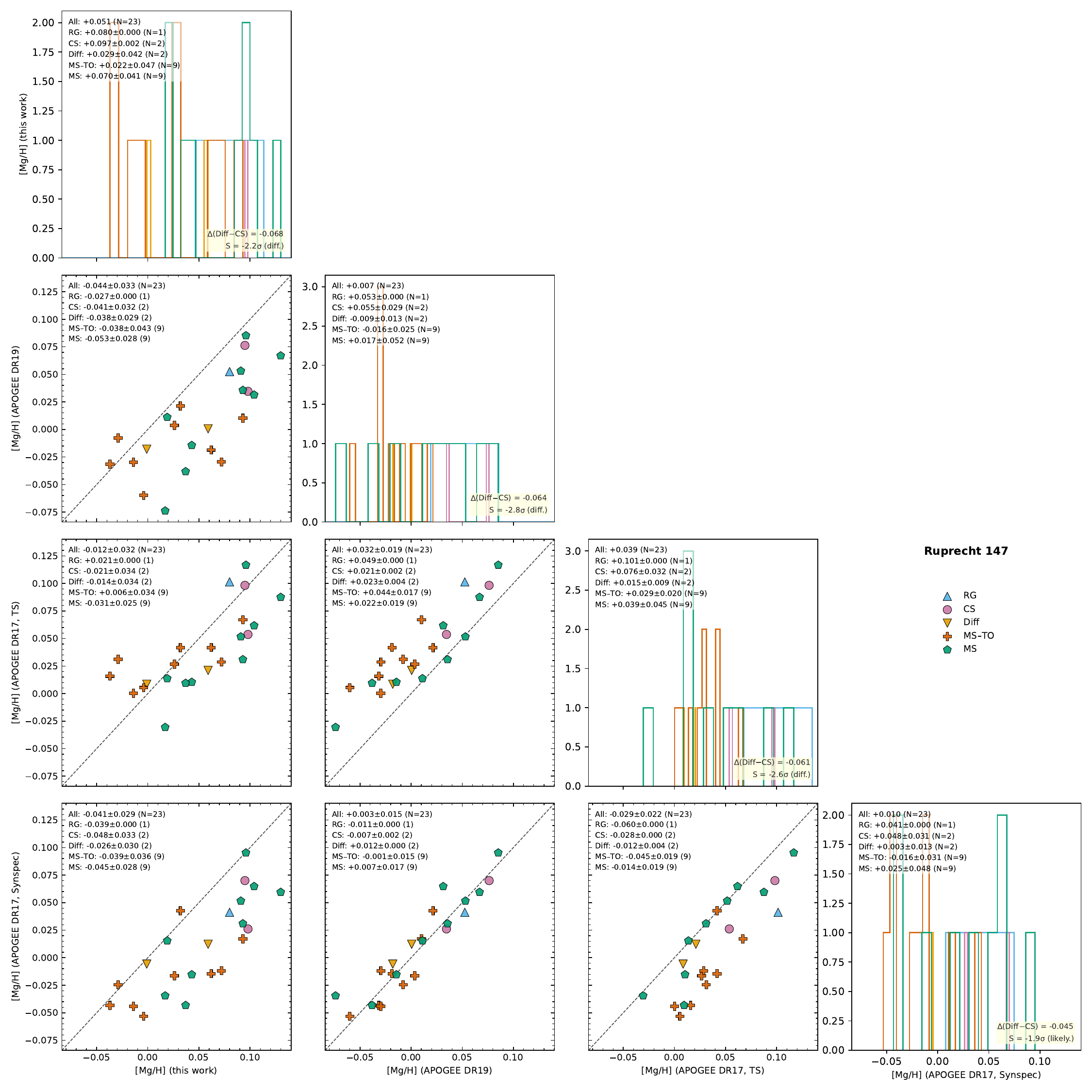}}
   \caption{Same as Figure \ref{AP_abu_Fe}, but for Mg.}
   \label{AP_abu_Mg}
\end{figure*}

\begin{figure*}
    \centering
   { \includegraphics[width=0.65\textwidth]{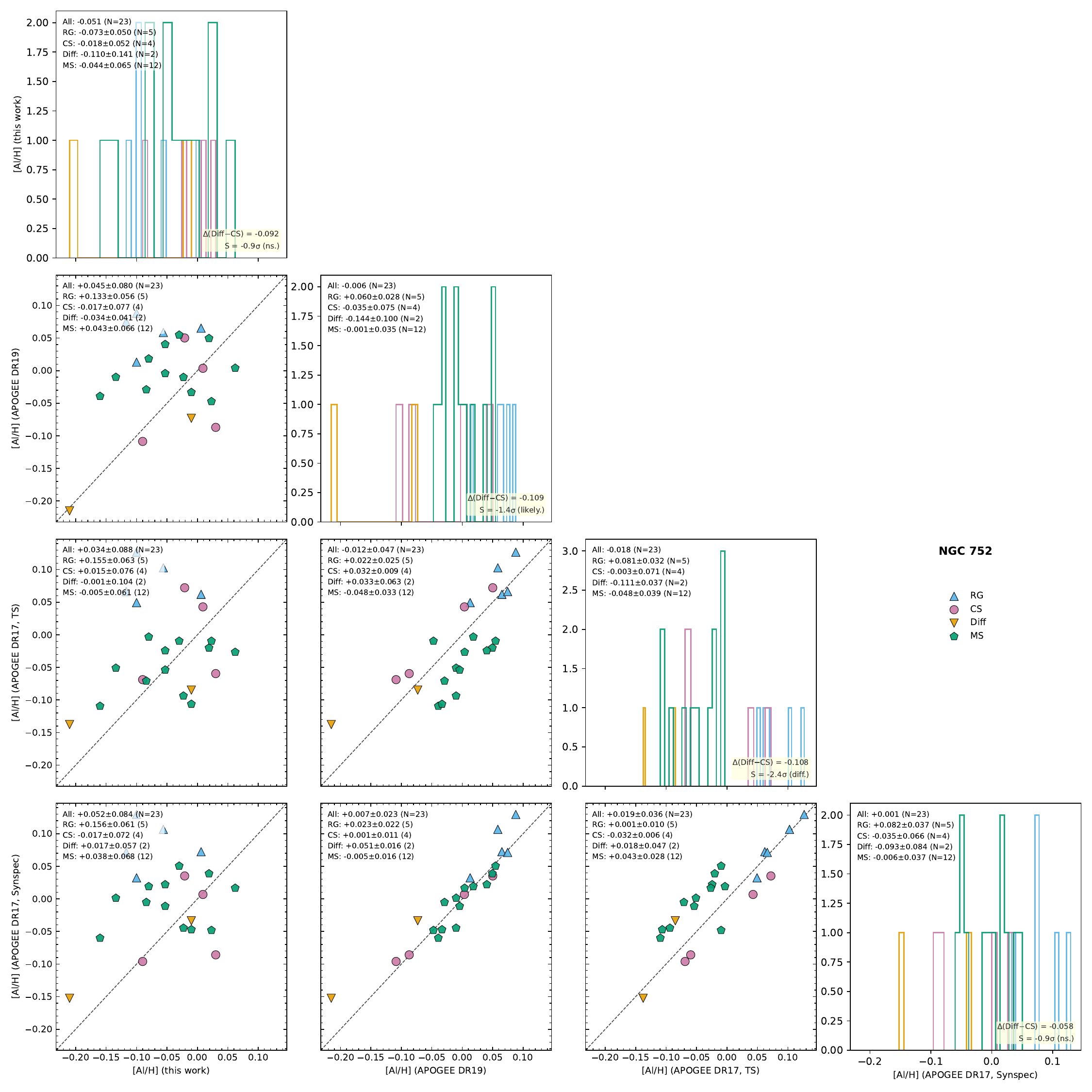}}
   { \includegraphics[width=0.65\textwidth]{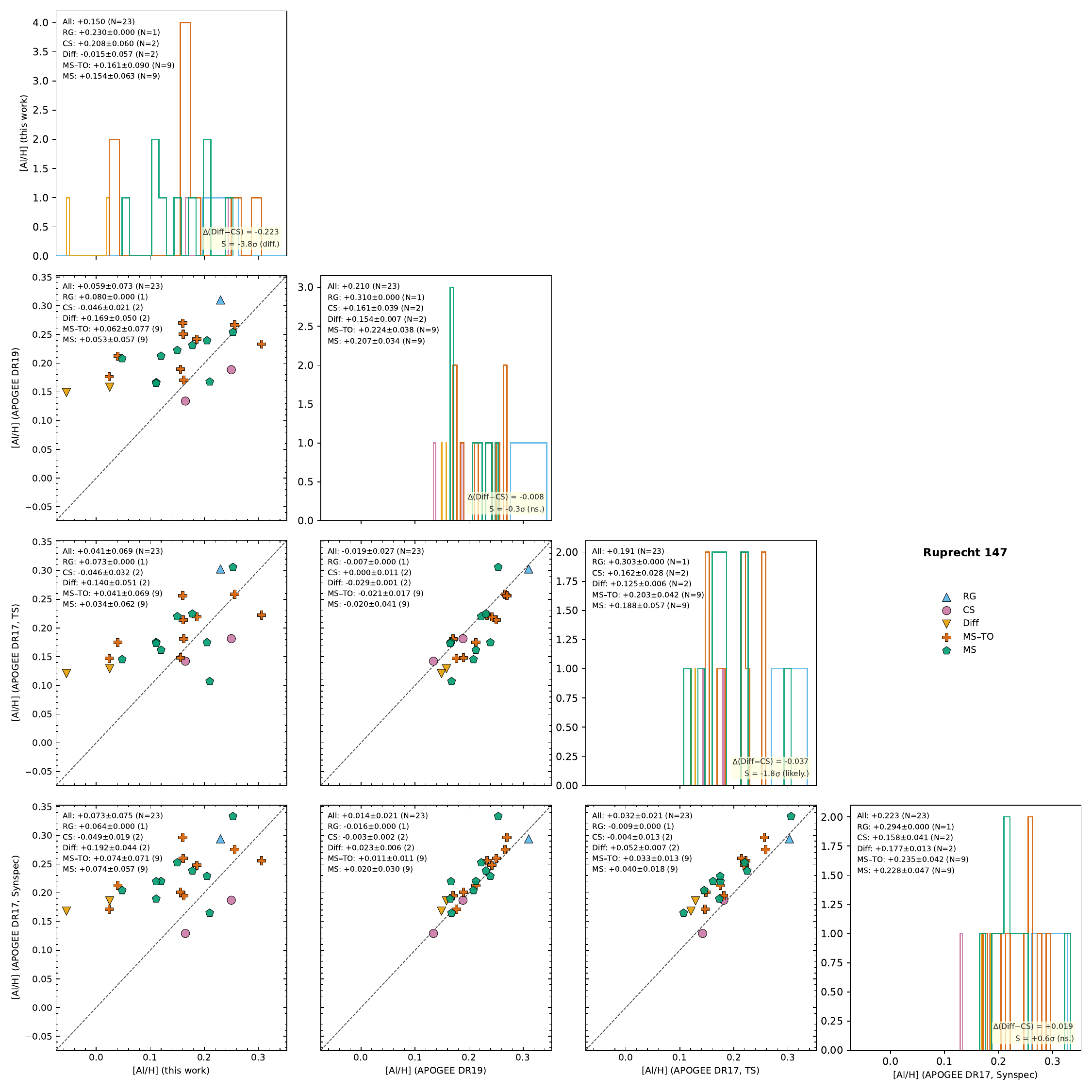}}
   \caption{Same as Figure \ref{AP_abu_Fe}, but for Al.}
   \label{AP_abu_Al}
\end{figure*}

\begin{figure*}
    \centering
   { \includegraphics[width=0.65\textwidth]{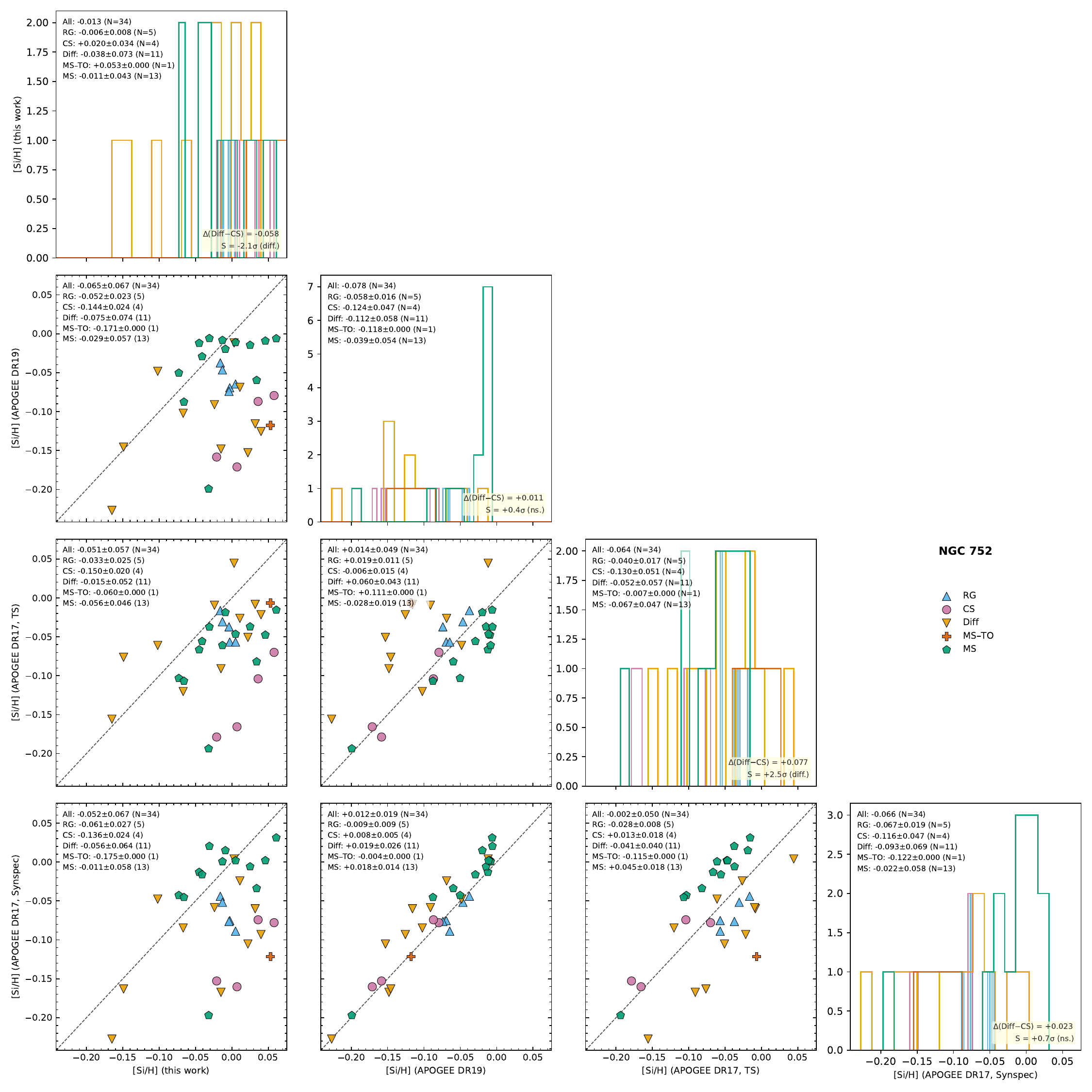}}
   { \includegraphics[width=0.65\textwidth]{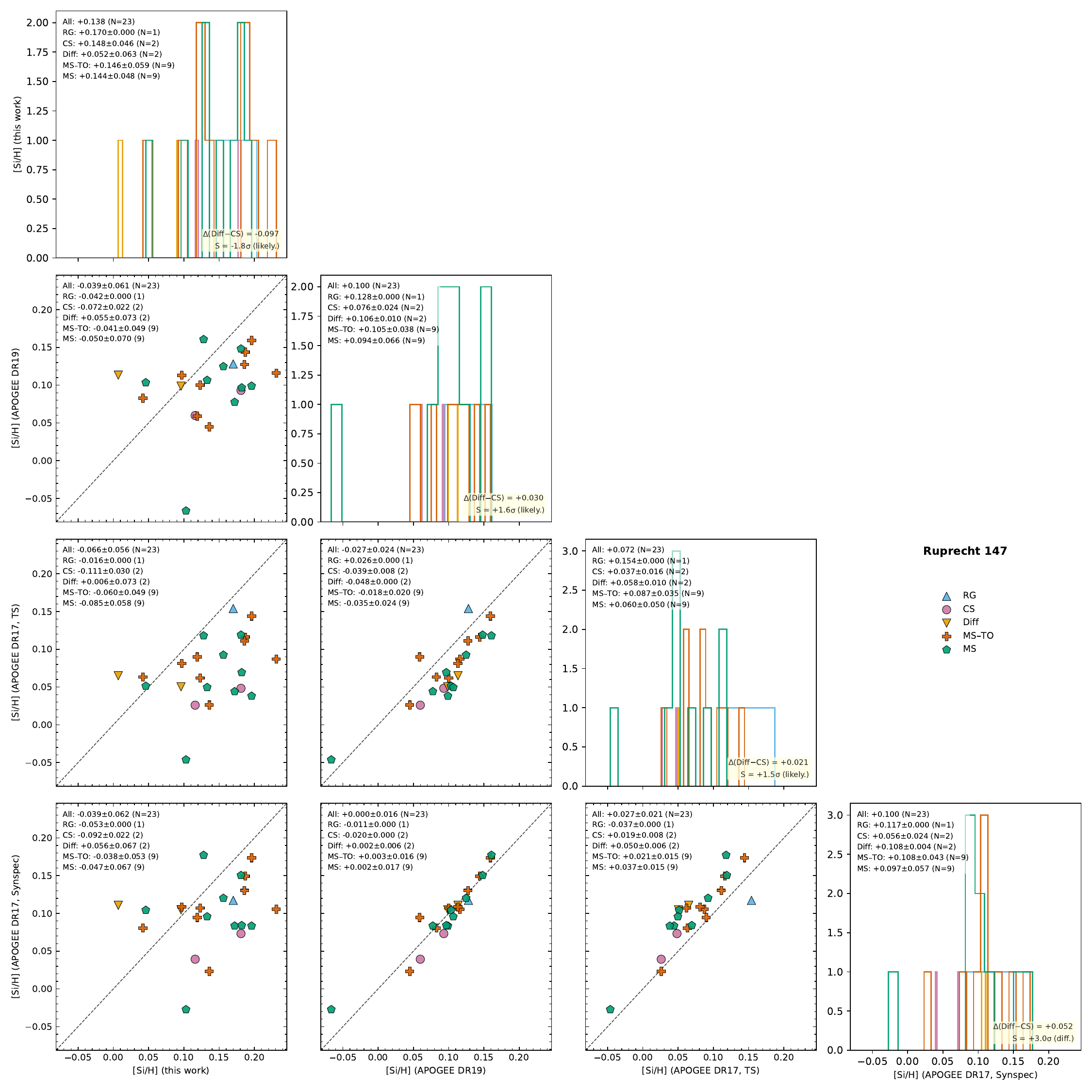}}
   \caption{Same as Figure \ref{AP_abu_Fe}, but for Si.}
   \label{AP_abu_Si}
\end{figure*}

\begin{figure*}
    \centering
   { \includegraphics[width=0.65\textwidth]{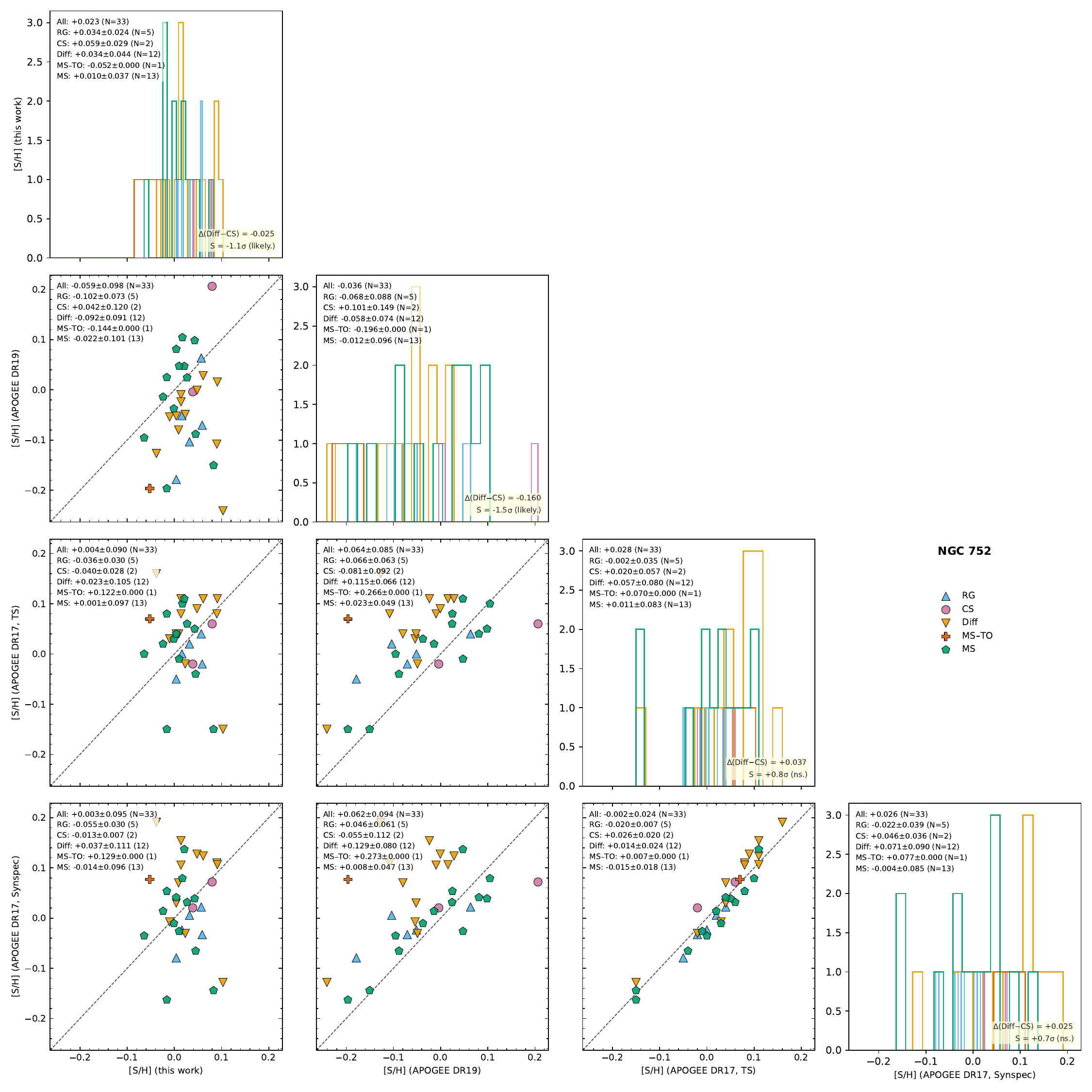}}
   { \includegraphics[width=0.65\textwidth]{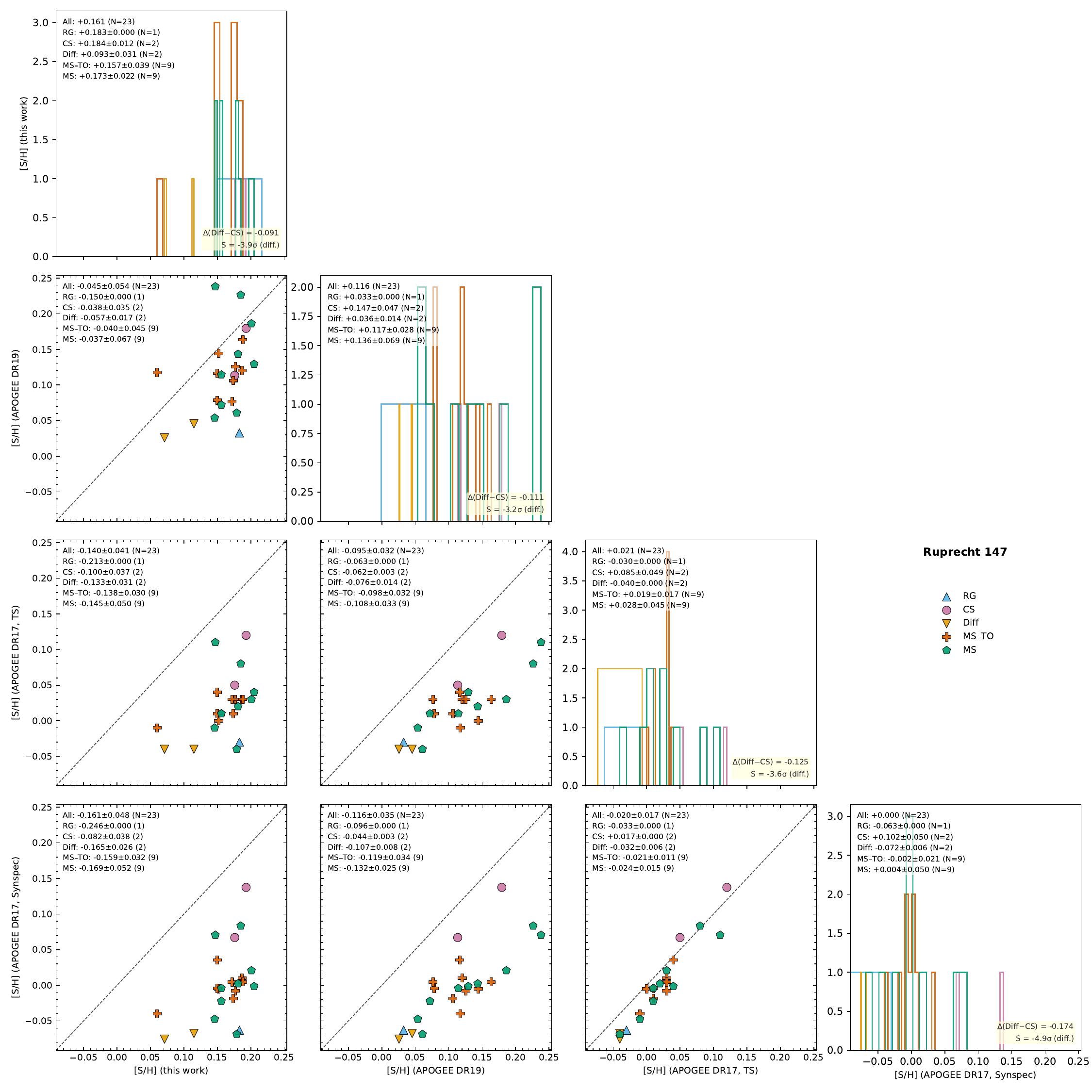}}
   \caption{Same as Figure \ref{AP_abu_Fe}, but for S.}
   \label{AP_abu_S}
\end{figure*}

\begin{figure*}
    \centering
   { \includegraphics[width=0.65\textwidth]{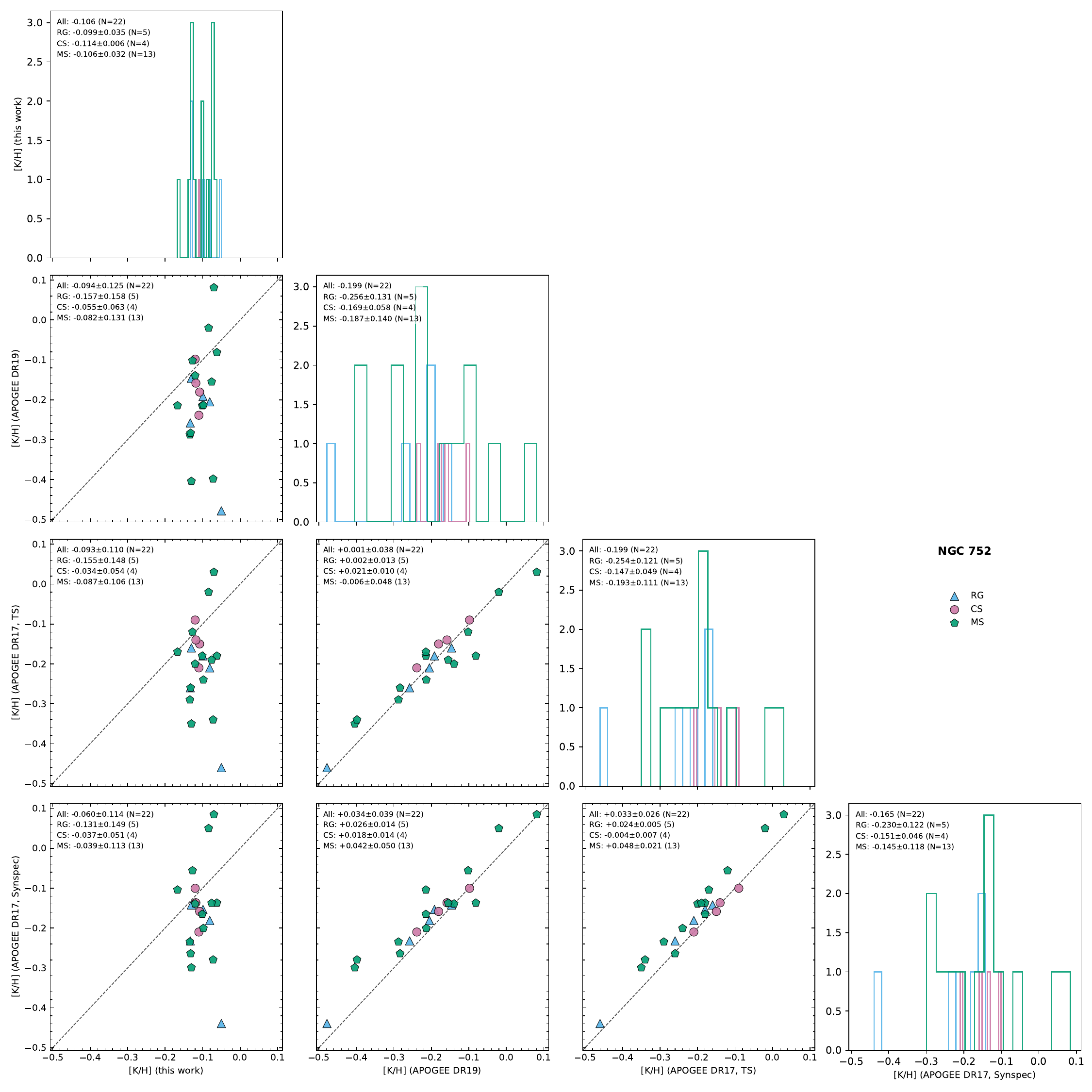}}
   { \includegraphics[width=0.65\textwidth]{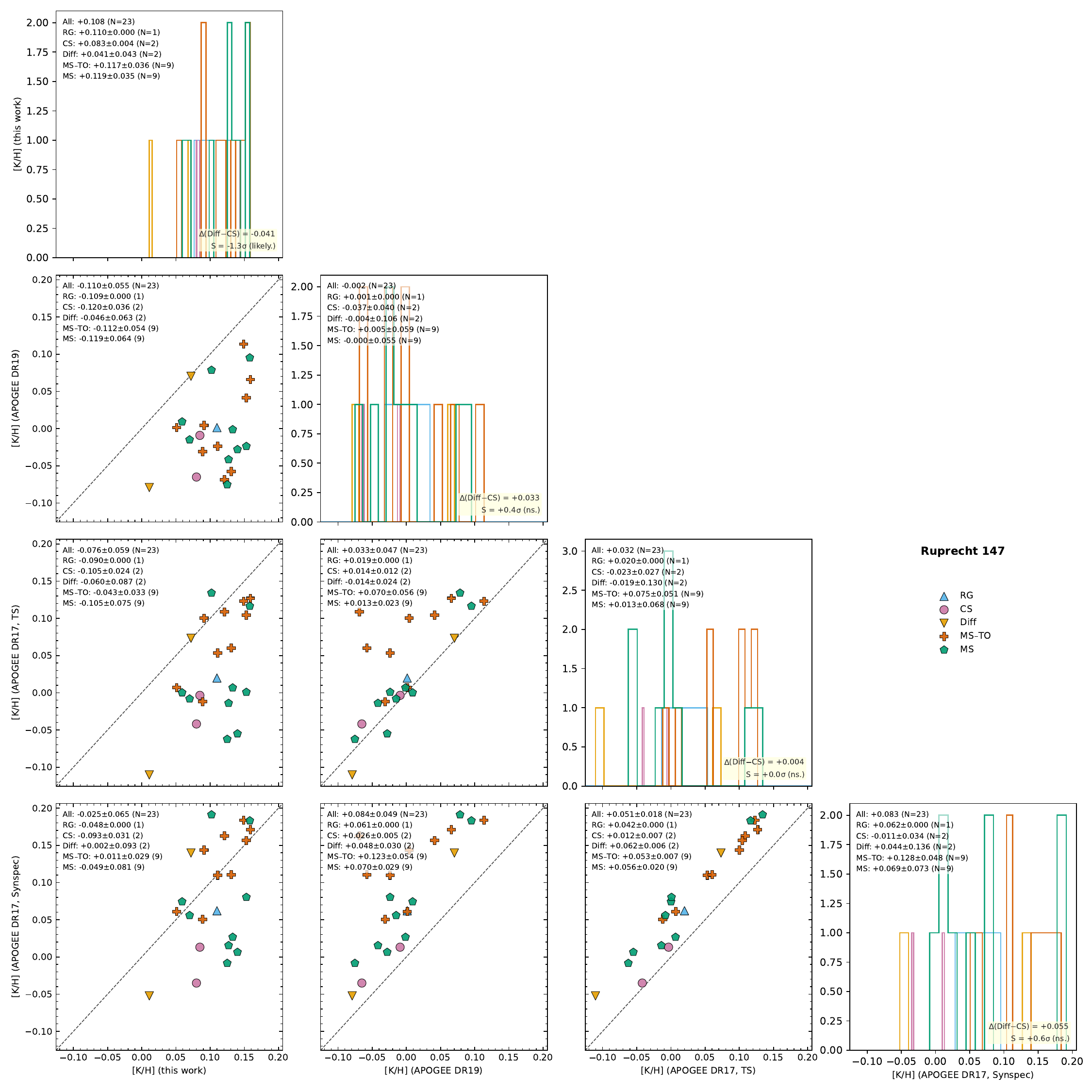}}
   \caption{Same as Figure \ref{AP_abu_Fe}, but for K.}
   \label{AP_abu_K}
\end{figure*}

\begin{figure*}
    \centering
   { \includegraphics[width=0.65\textwidth]{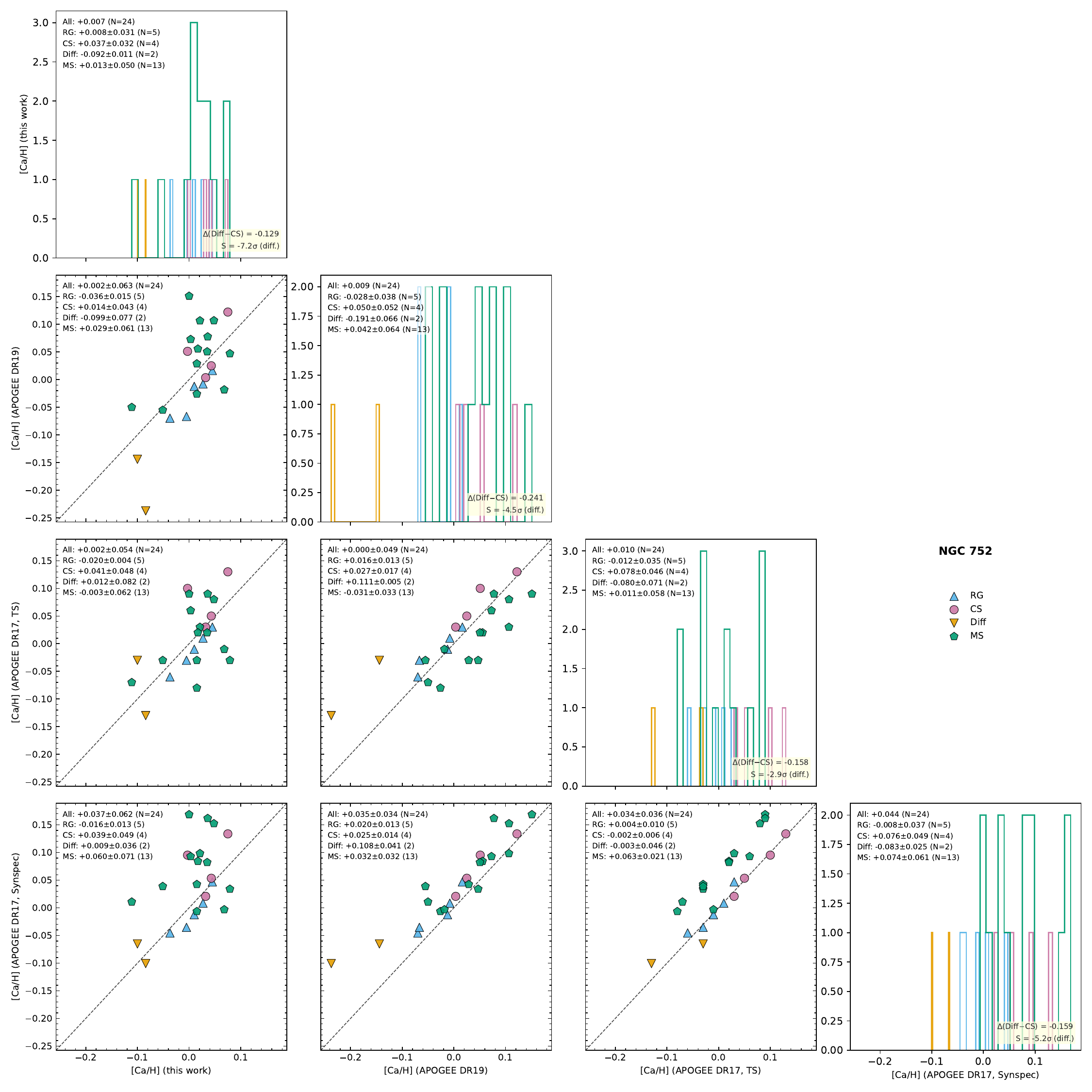}}
   { \includegraphics[width=0.65\textwidth]{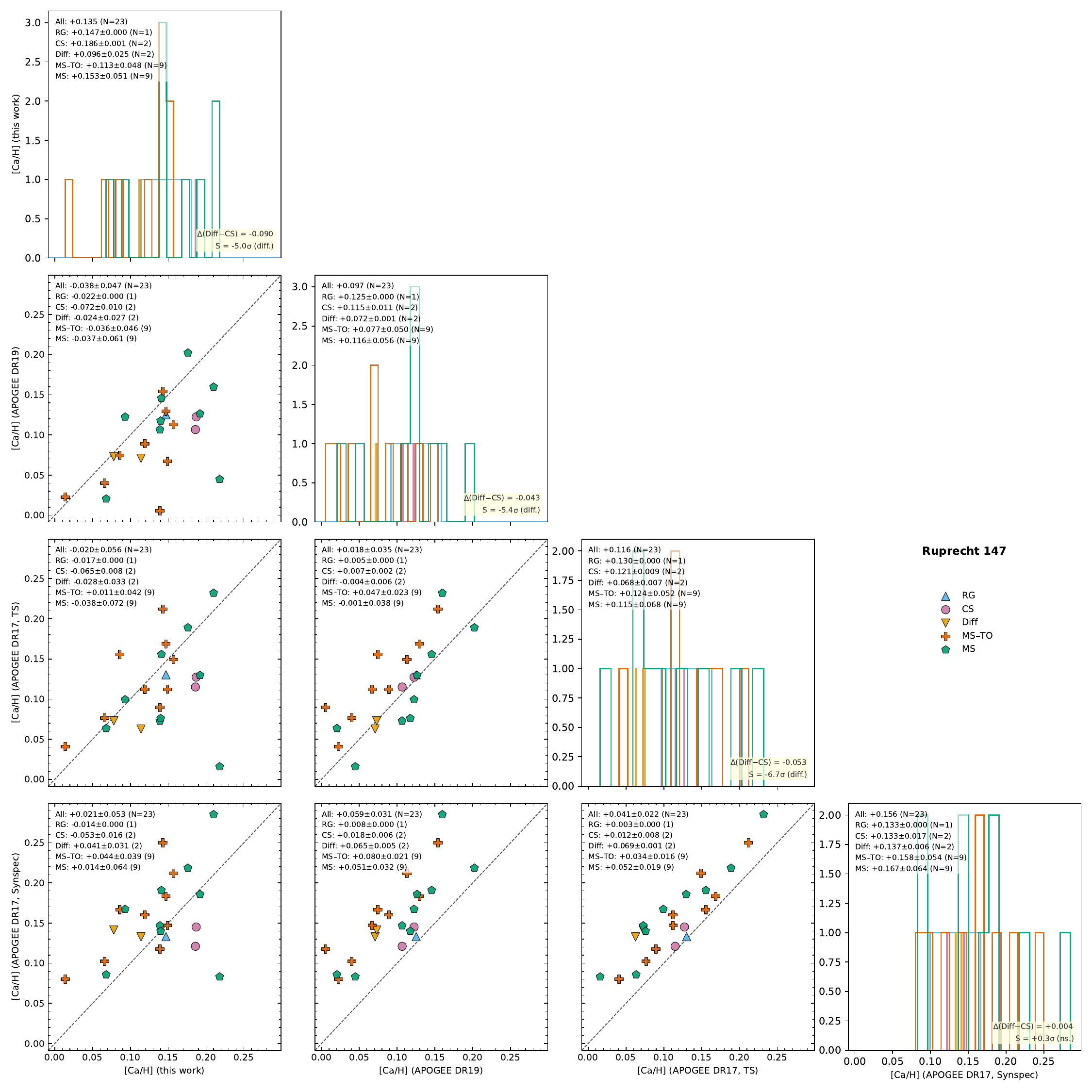}}
   \caption{Same as Figure \ref{AP_abu_Fe}, but for Ca.}
   \label{AP_abu_Ca}
\end{figure*}

\begin{figure*}
    \centering
   { \includegraphics[width=0.65\textwidth]{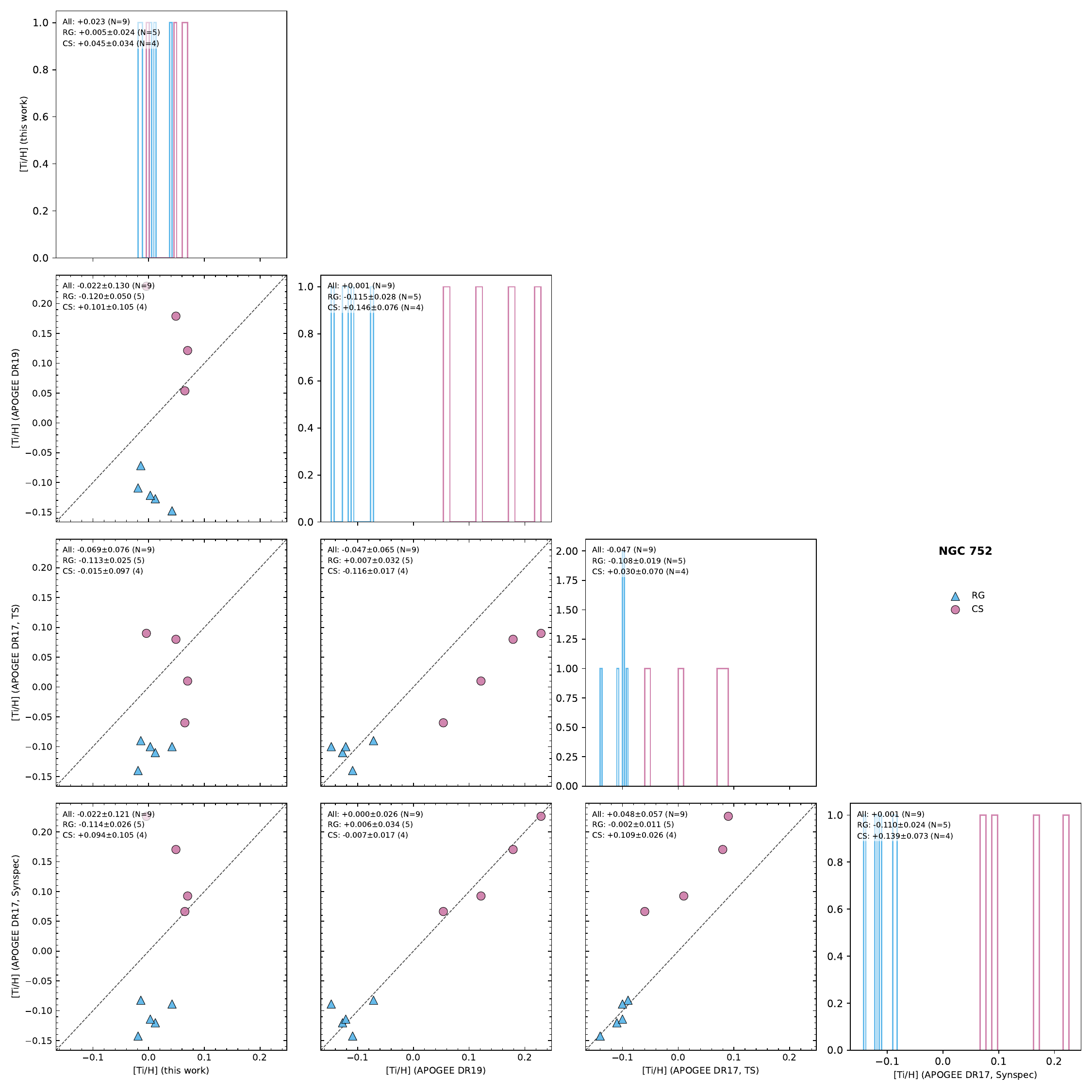}}
   { \includegraphics[width=0.65\textwidth]{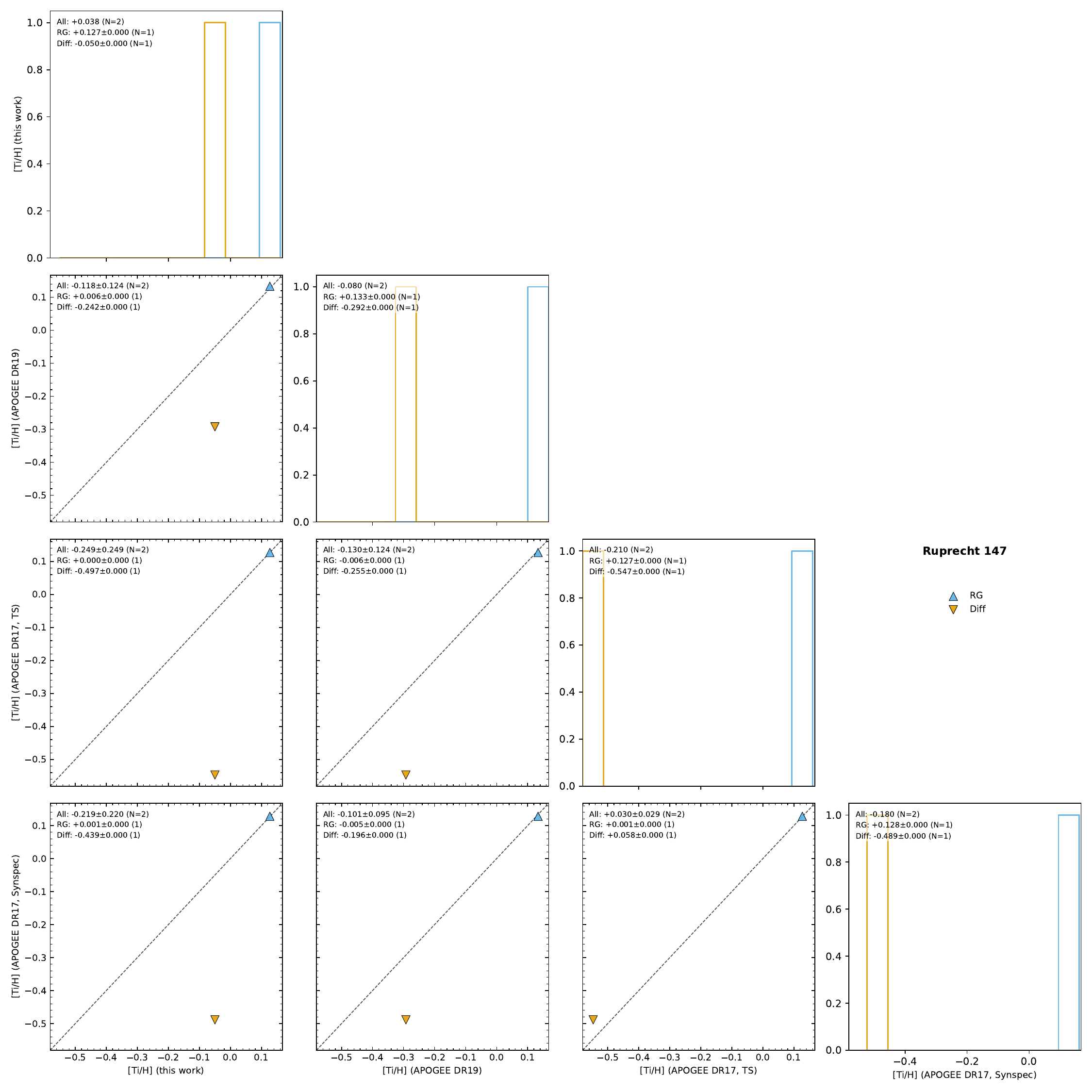}}
     \caption{Same as Figure \ref{AP_abu_Fe}, but for Ti.}
   \label{AP_abu_Ti}
\end{figure*}

\begin{figure*}
    \centering
   { \includegraphics[width=0.65\textwidth]{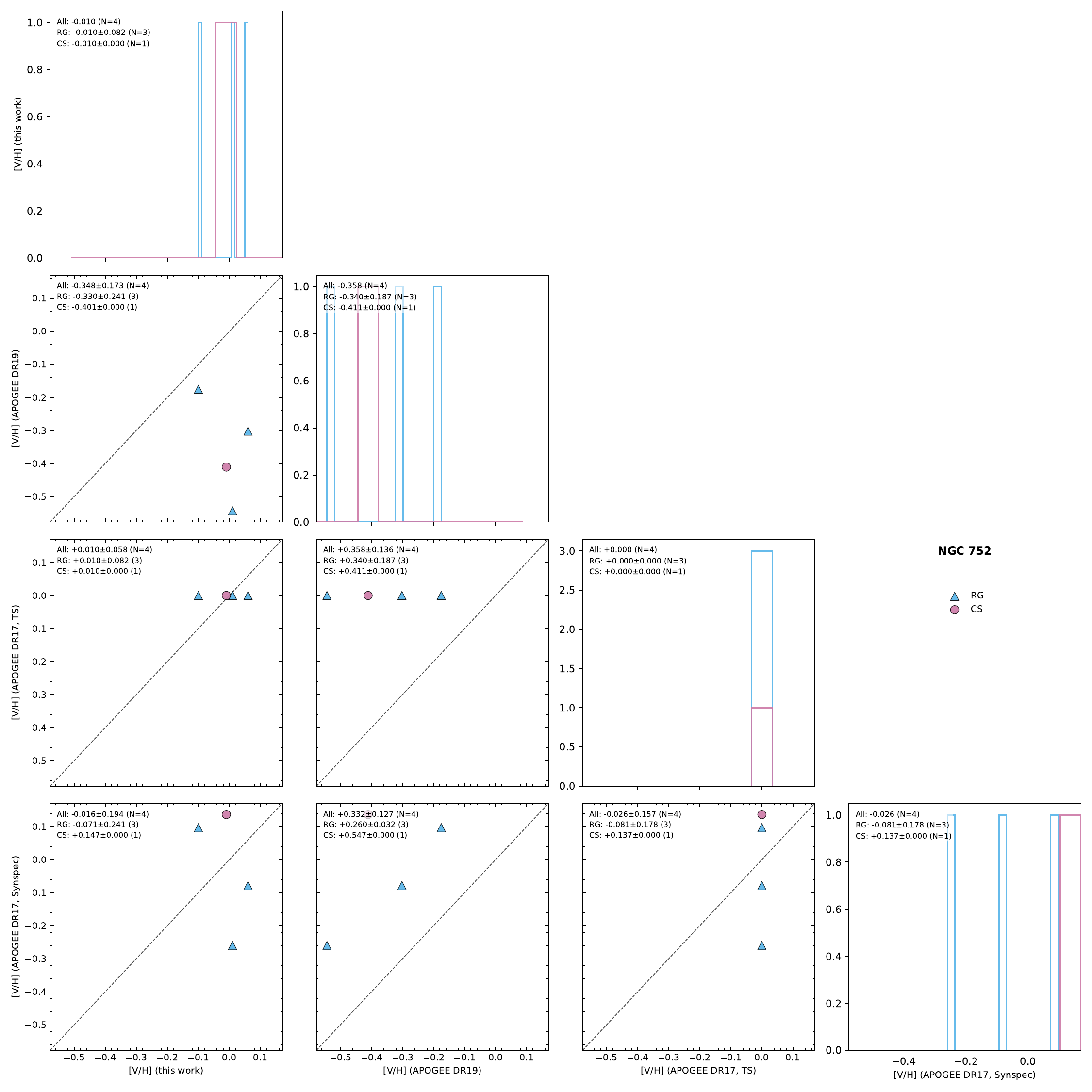}}
   { \includegraphics[width=0.65\textwidth]{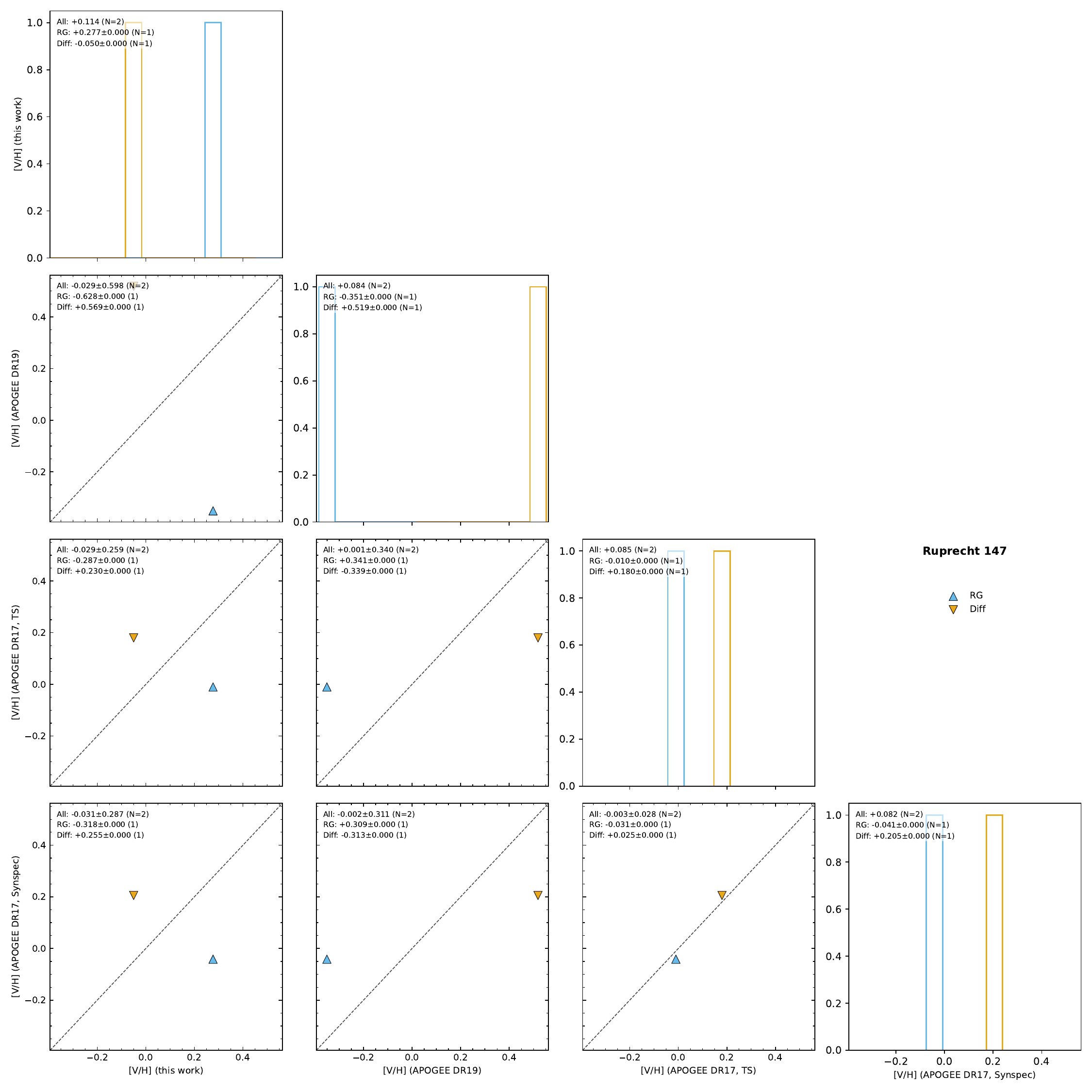}} 
   \caption{Same as Figure \ref{AP_abu_Fe}, but for V.}
   \label{AP_abu_V}
\end{figure*}

\begin{figure*}
    \centering
   { \includegraphics[width=0.65\textwidth]{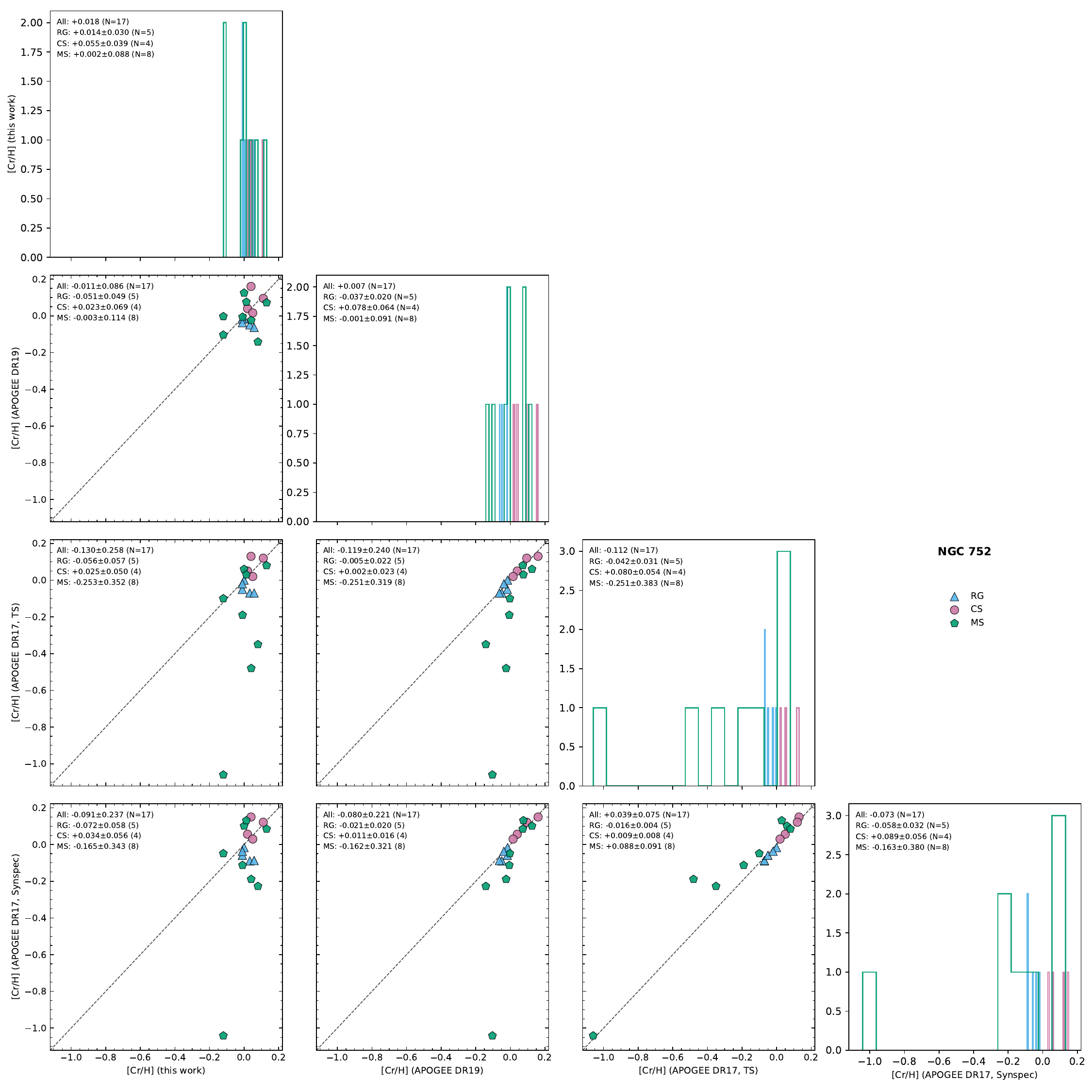}}
   { \includegraphics[width=0.65\textwidth]{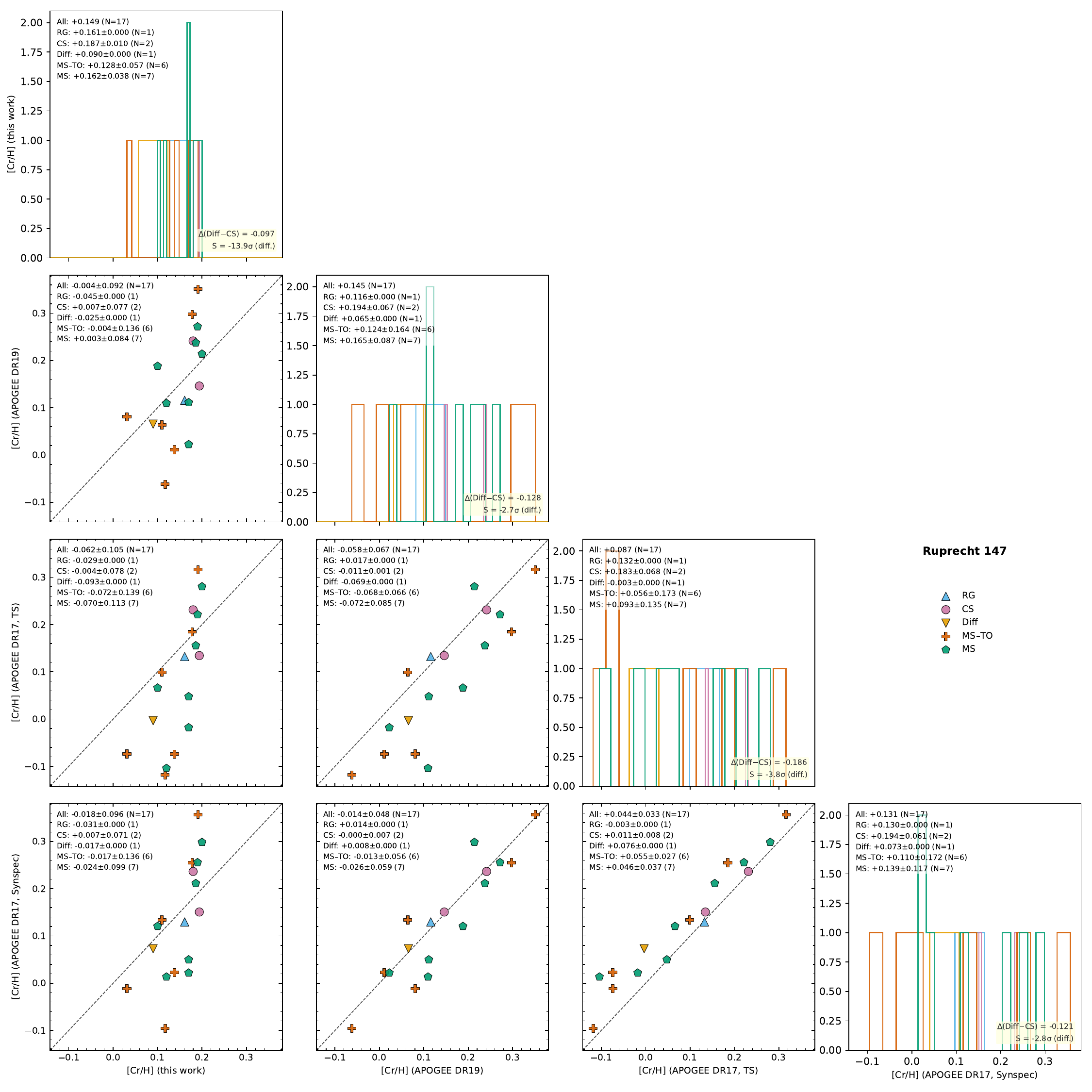}}
   \caption{Same as Figure \ref{AP_abu_Fe}, but for Cr.}
   \label{AP_abu_Cr}
\end{figure*}

\begin{figure*}
    \centering
   { \includegraphics[width=0.65\textwidth]{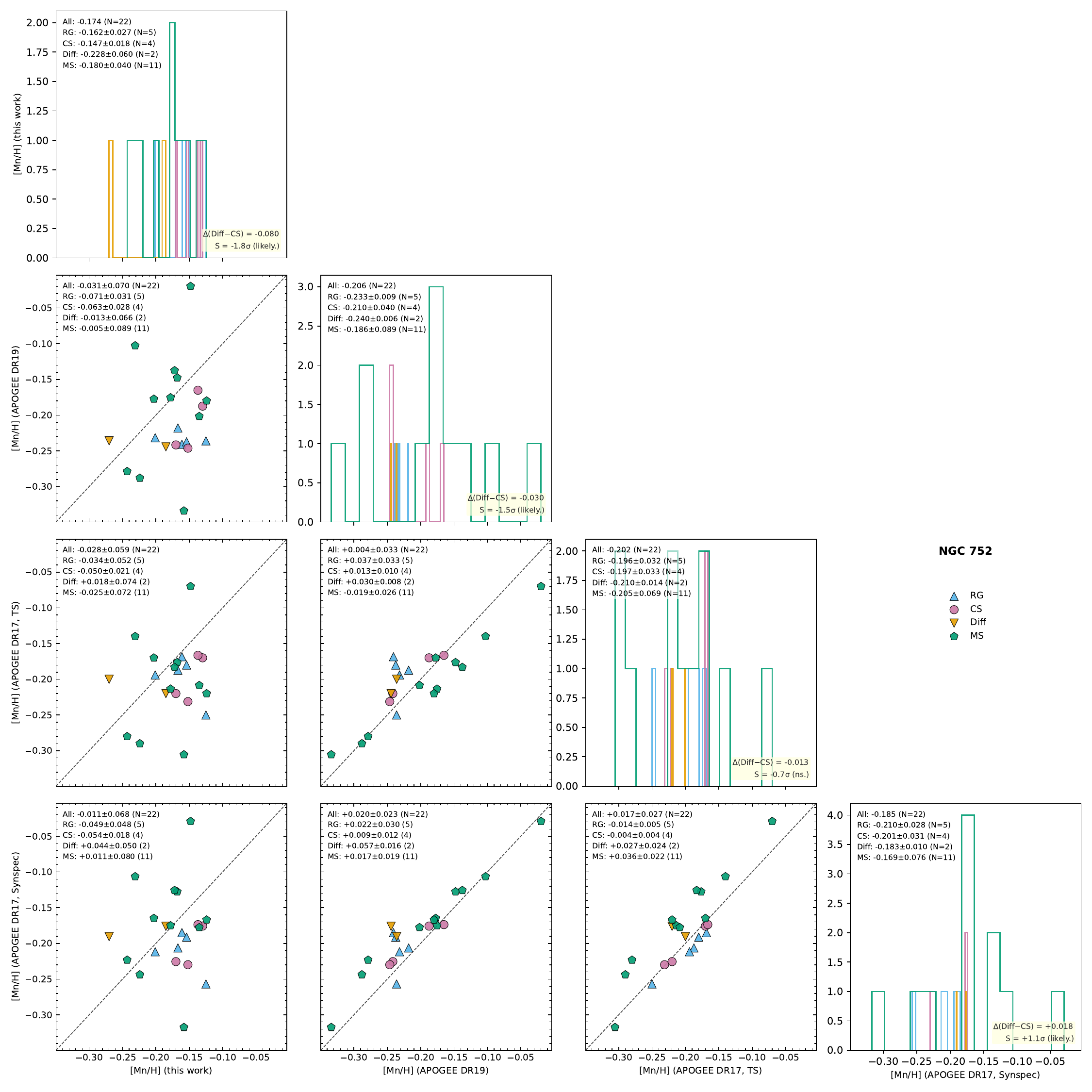}}
   { \includegraphics[width=0.65\textwidth]{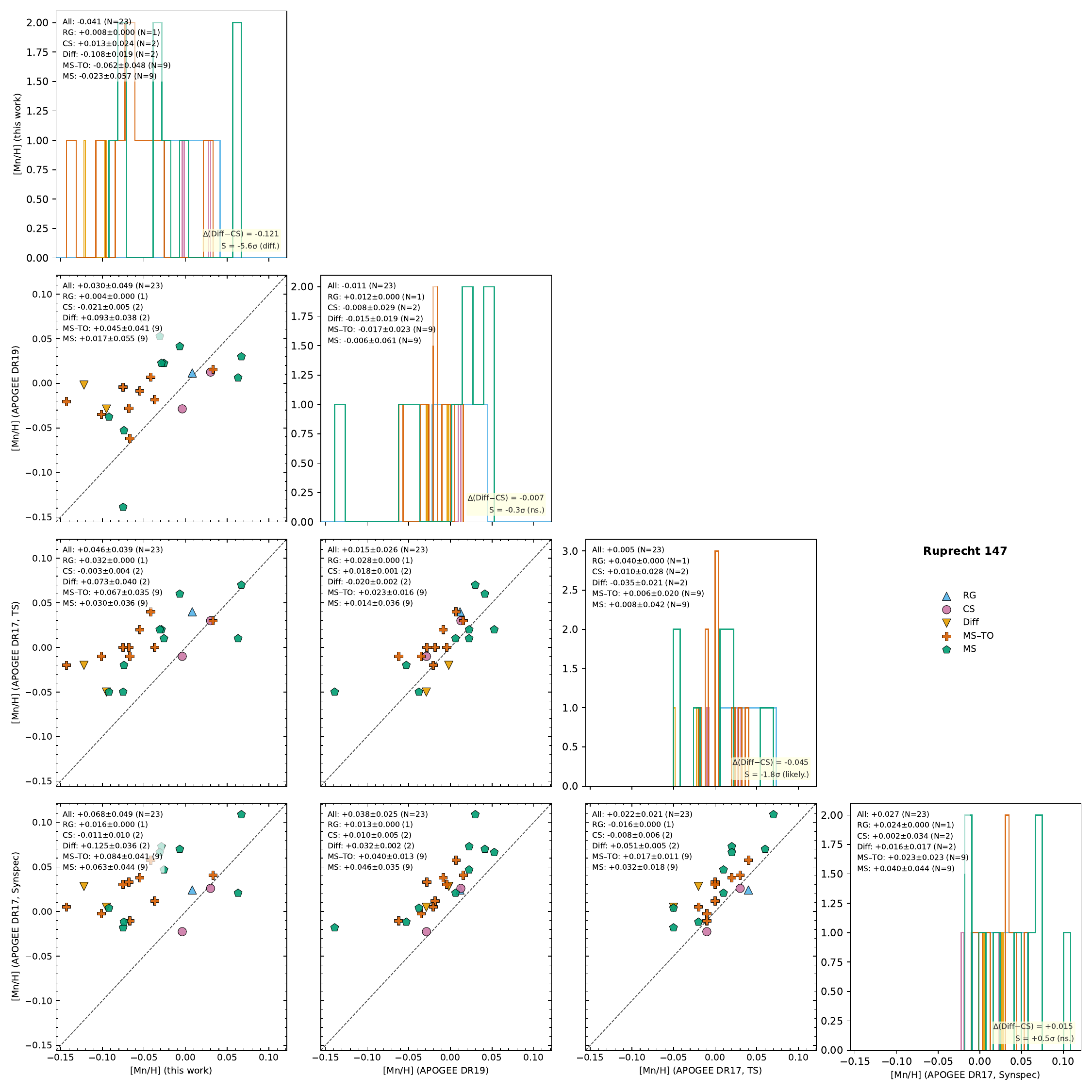}}
   \caption{Same as Figure \ref{AP_abu_Fe}, but for Mn.}
   \label{AP_abu_Mn}
\end{figure*}

\begin{figure*}
    \centering
   { \includegraphics[width=0.65\textwidth]{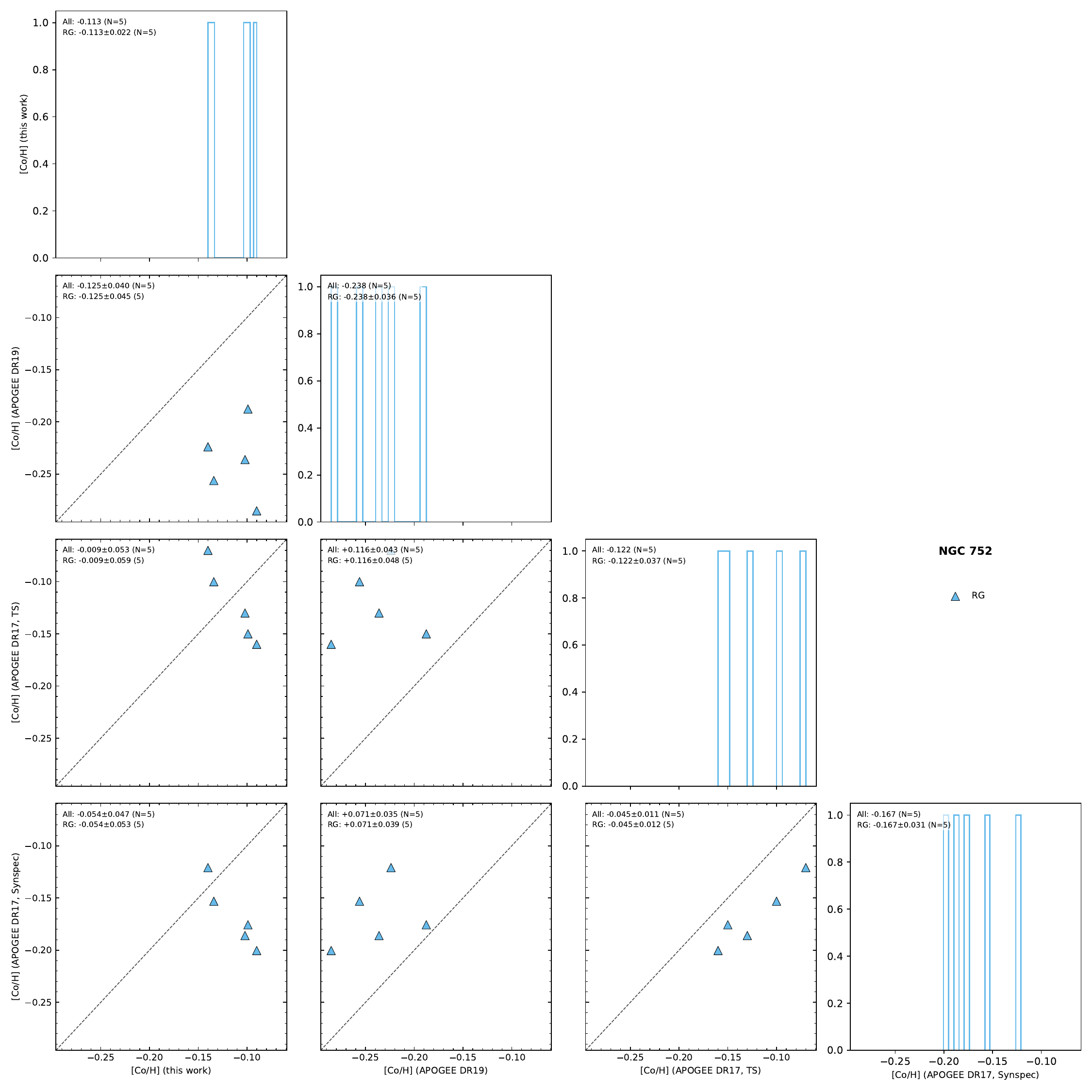}}
   { \includegraphics[width=0.65\textwidth]{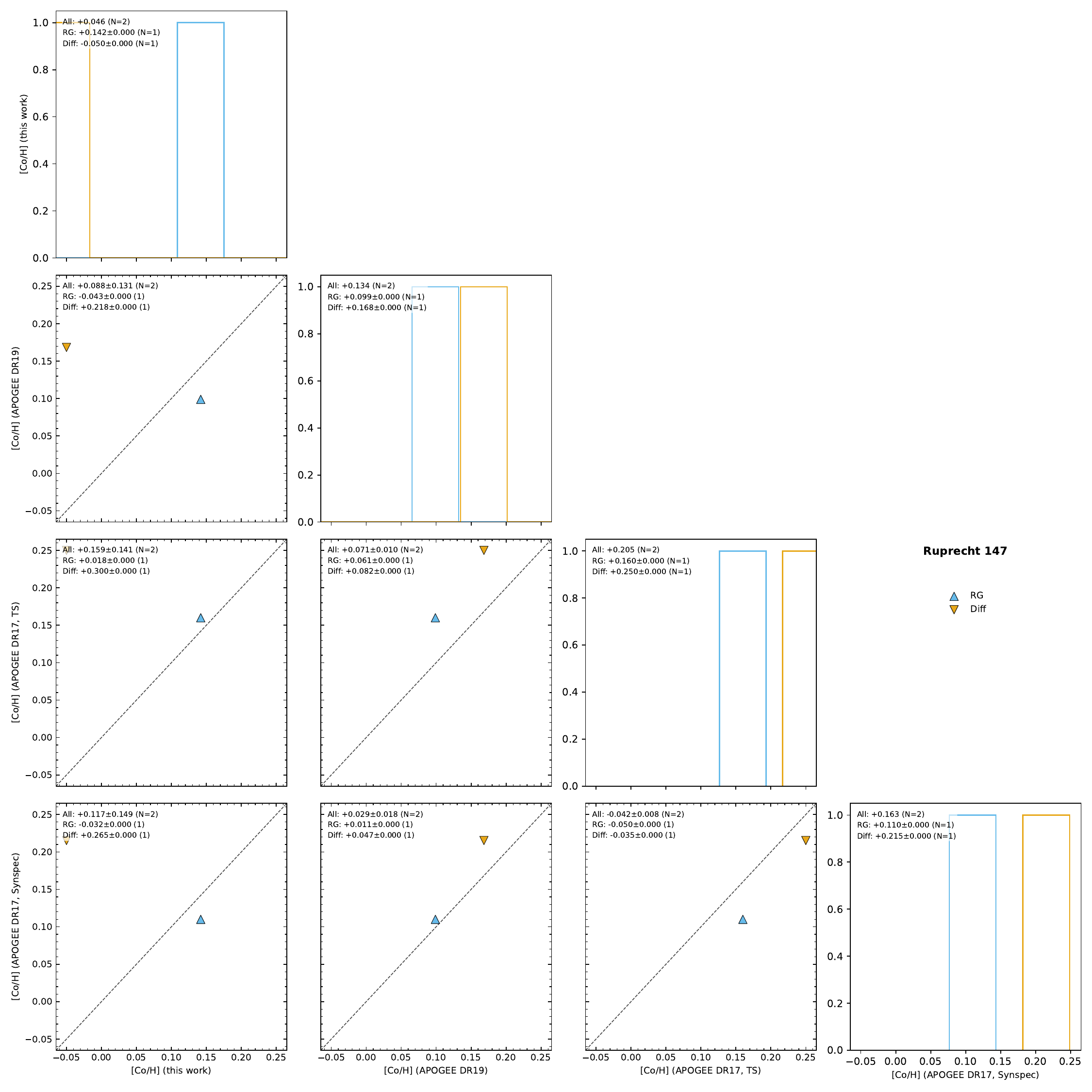}}   
   \caption{Same as Figure \ref{AP_abu_Fe}, but for Co.}
   \label{AP_abu_Co}
\end{figure*}

\begin{figure*}
    \centering
   { \includegraphics[width=0.65\textwidth]{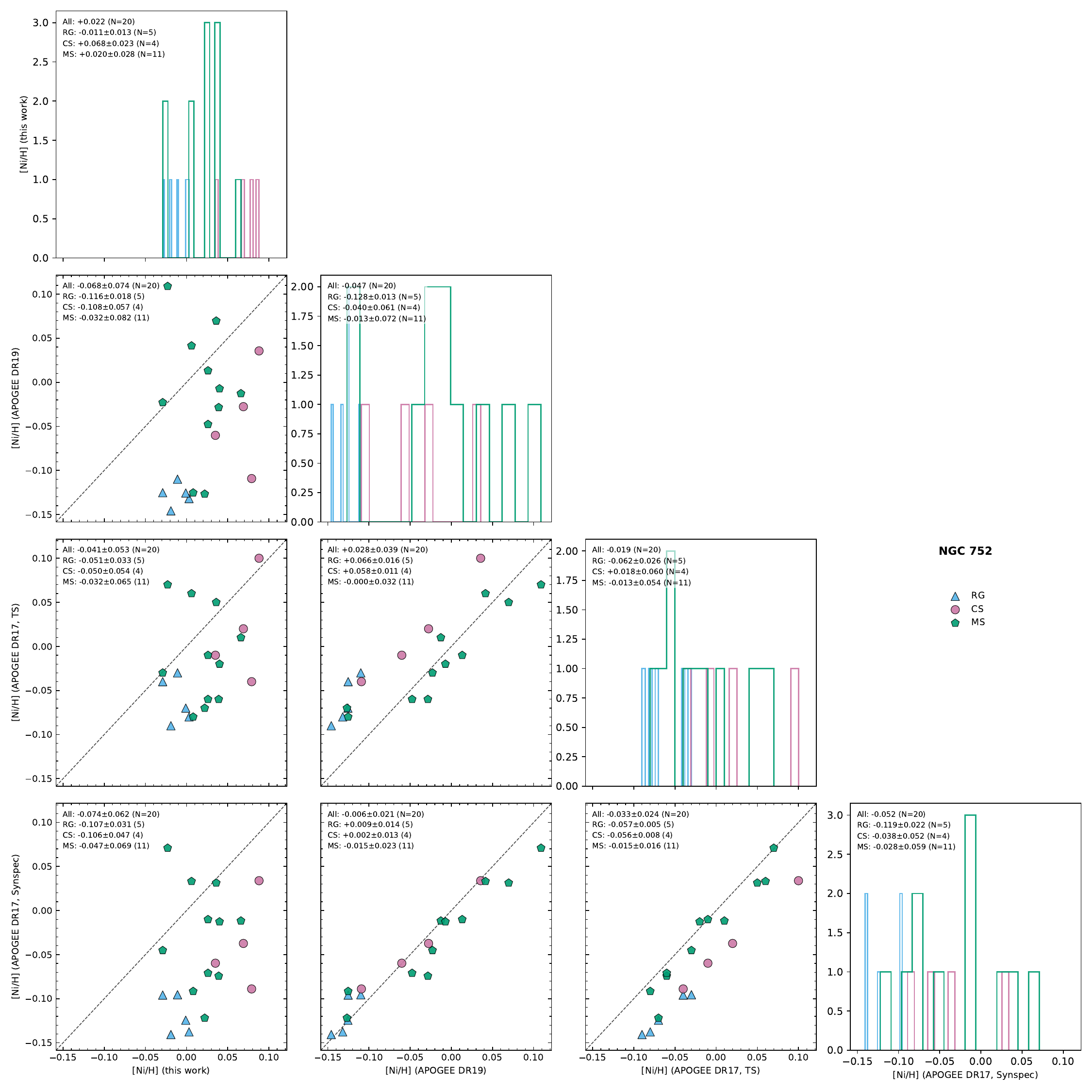}}
   { \includegraphics[width=0.65\textwidth]{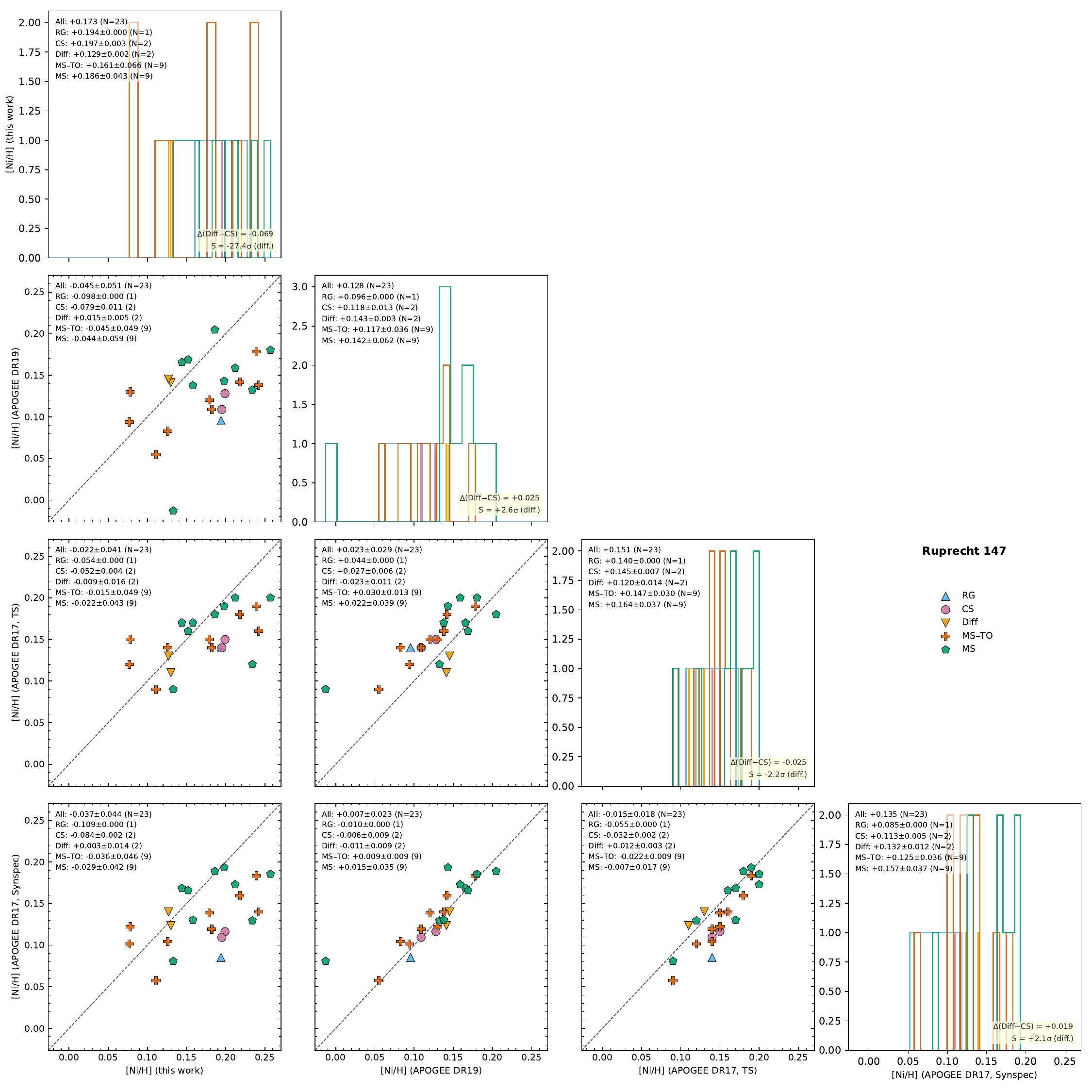}}
   \caption{Same as Figure \ref{AP_abu_Fe}, but for Ni.}
   \label{AP_abu_Ni}
\end{figure*}

\end{document}